\documentclass[11pt]{article}
\usepackage{graphics,epsfig, psfrag}
\usepackage{pstricks}
\usepackage{pst-node}
\usepackage{amscd}
\usepackage{amssymb}
\usepackage{amsfonts}
\usepackage{amsmath}
\usepackage{mathrsfs}
\makeatother
\makeatletter
\@addtoreset{equation}{section}
\makeatother
\renewcommand{\theequation}{\arabic{section}.\arabic{equation}}
\begin{document}

\setcounter{page}{1}
\vfil

\pagestyle{plain}

\begin{center}
{\LARGE {\bf  Koopman-von Neumann Formulation of Classical Yang-Mills Theories: I}}

\bigskip
\bigskip
\bigskip

{\large P. Carta}

The Royal Bank of Scotland, \\
Financial Markets \\
135 Bishopsgate \\
London EC2M 3UR \\
e-mail: {\it Paolo.Carta@rbos.com}

\bigskip 

{\large E. Gozzi} and {\large D. Mauro}

             Department of Theoretical Physics, 
	     University of Trieste, \\
	     Strada Costiera 11, Miramare-Grignano, 34014 Trieste, Italy,\\
	     and INFN, Trieste, Italy\\
	     e-mail: {\it gozzi@ts.infn.it} and {\it mauro@ts.infn.it}

\end{center}

\bigskip
\bigskip
\begin{abstract}
\noindent In this paper we present the Koopman-von Neumann (KvN) formulation of classical non-Abelian gauge field theories. In particular we shall explore the functional (or classical path integral) counterpart of the KvN method. In the quantum path integral quantization of Yang-Mills theories concepts like gauge-fixing and Faddeev-Popov determinant appear in a quite natural way. We will prove that these same objects are needed also in this classical path integral formulation for Yang-Mills theories. We shall also explore the classical path integral counterpart of the BFV formalism and build all the associated universal and gauge charges. These last are quite different from the analog quantum ones and we shall show the relation between the two. This paper lays the foundation of this formalism which, due to the many auxiliary fields present, is rather heavy. Applications to specific topics outlined in the paper will appear in later publications. 

\end{abstract}

\section{Introduction}
In 1931, soon after the appearance of {\it quantum mechanics} (QM) in its operatorial formulation, Koopman and von Neumann (KvN) proposed \cite{Koopman} an operatorial version also for {\it classical mechanics} (CM). It is well known that, given any operatorial formalism, it is always possible to build an associated path integral, like Feynman did in 1942 for QM. The same was done in 1989 \cite{martin} for the KvN operatorial formalism of CM.
We will indicate this path integral with the acronym CPI for Classical Path Integral in order to distinguish it from the Feynman's one associated to QM which we will indicate with QPI for Quantum Path Integral.  
For a brief review of the CPI of a point particle we refer the interested reader to Ref. \cite{martin} and later papers \cite{math}. In this section we want to briefly review the extension of the CPI to the case of a system with an infinite number of degrees of freedom, i.e. the scalar field theory with a $\varphi^4$ interaction already introduced in \cite{carta}-\cite{field}. This case is free from the technical complications due to the constraints present in a field theory with gauge invariance. The Lagrangian density describing a scalar field $\varphi$ with a $\varphi^4$ self-interaction is given by:
\begin{equation}
\displaystyle {\cal{L}}=\frac{1}{2}(\partial_{\mu}\varphi)(\partial^{\mu}\varphi)-\frac{1}{2}m^2\varphi^2-\frac{1}{24}g\varphi^4. \label{lag}
\end{equation}
Indicating with $\pi$ the conjugate momentum of $\varphi$ we can build from the Lagrangian (\ref{lag}) the following Hamiltonian:
\begin{displaymath}
\displaystyle H=\int \textrm{d} {\bf x} \biggl[\frac{1}{2}\pi^2+\frac{1}{2}(\partial_r \varphi)^2+
\frac{1}{2}m^2\varphi^2+\frac{1}{24}g\varphi^{4}\biggr], 
\end{displaymath}
where $\textrm{d} {\bf x}$ means the integration over all the space variables and $\displaystyle \partial_r \varphi \equiv \frac{\partial \varphi}{\partial {\bf x}^r}$ stands for the space derivative of the field $\varphi$. The equations of motion of the system, which are
\begin{displaymath}
\displaystyle \partial_{\mu}\partial^{\mu}\varphi+m^2\varphi^2+\frac{1}{6}g\, \varphi^3=0
\end{displaymath}
can be written as a couple of first order equations in terms of the Hamiltonian $H$ as follows:
\begin{equation}
\displaystyle \dot{\varphi}(x)=\frac{\partial H}{\partial \pi(x)}, \qquad \dot{\pi}(x)=-\frac{\partial H}{\partial \varphi(x)}, \label{eqq}
\end{equation}
where $\partial$ indicates a functional derivative and $x$ collects together all the space-time variables $x\equiv (t, {\bf x})$. Using the compact notation $\xi^a=(\varphi,\pi), \; a=1,2$ and the $2\times 2$ symplectic matrix $\omega^{ab}$ we can write (\ref{eqq}) as:
\begin{equation}
\displaystyle \dot{\xi}^a=\omega^{ab}\frac{\partial H}{\partial \xi^b}. \label{eqcomp}
\end{equation}
We shall now implement the standard steps of the CPI procedure \cite{martin}. Let us ask ourselves, if we start from an initial field $\xi^a_i$ at time $t_i$, which is the probability amplitude of finding the system in $\xi^a$ at time $t$? This probability amplitude is one if the field $\xi^a$ lies on the solutions of the classical equations of motion and zero otherwise. To write this probability amplitude as a path integral we must integrate over all the paths in phase space with a weight which is given by a functional Dirac delta:
\begin{equation}
\displaystyle K(\xi^a({\bf x}), t|\xi^a_i({\bf x}),t_i)= \int {\mathcal D}\xi^a\, \delta\left[ \xi^a(x)-\xi^a_{\textrm{cl}}(x;\xi^a_i({\bf x}),t_i) \right], \label{1.3bis}
\end{equation}
where $\xi^a_{\textrm{cl}}(x;\xi^a_i({\bf x}),t_i)$ is the solution of the equations of motion (\ref{eqcomp}) with initial condition $\xi^a(t_i, {\bf x})=\xi^a_i({\bf x})$. The functional Dirac delta on the solutions 
of the equations of motion can be converted into a functional Dirac delta on the associated equations, provided we introduce also a suitable functional determinant
\begin{equation}
\displaystyle \delta\left[\xi^a(x)-\xi^a_{\textrm{cl}}(x;\xi^a_i({\bf x}),t_i)\right]=\delta \left[\dot{\xi}^a-\omega^{ab}\partial_b H\right]\, \textrm{det}\left[\delta({\bf x}-{\bf y})\delta^a_c\partial_t -\omega^{ab}\frac{\partial^2 H}{\partial \xi^b({\bf x})\partial \xi^c({\bf y})}\right]. \label{trasf}
\end{equation}
If we use the integral representation of the Dirac delta we can rewrite the Dirac delta on the RHS of (\ref{trasf}) as:
\begin{displaymath}
\delta\left[\dot{\xi}^a-\omega^{ab}\partial_bH\right]=\int {\mathcal D}\lambda_a \, \exp \left[ i\int {\textrm d}x\, \lambda_a(x)\left[\dot{\xi}^a(x)-\omega^{ab}\partial_bH(\xi)\right] \right].
\end{displaymath}
The functional determinant in (\ref{trasf}) can be exponentiated via suitable Grassmann fields $c^a(x)$ and $\overline{c}_a(x)$ using the formula 
\begin{eqnarray*}
&& \displaystyle \textrm{det}\left[\delta({\bf x}-{\bf y})\delta^a_c\partial_t -\omega^{ab}\frac{\partial^2 H}{\partial \xi^b({\bf x})\partial \xi^c({\bf y})}\right]\nonumber\\
&& \qquad =\int {\mathcal D}c{\mathcal D}\overline{c}\, \exp \left[ i \int {\textrm d}x \, \overline{c}_a(x)\left[\delta^a_c\partial_t -\omega^{ab}\partial_b\partial_c H(\xi)\right]c^c(x) \right].
\end{eqnarray*}
The final result of these formal manipulations is that the probability amplitude (\ref{1.3bis}) can be rewritten as the following path integral over an {\it extended phase space} given not only by the fields $\xi$ but also by the auxiliary fields $(\lambda, c,\overline{c})$:
\begin{equation}
\displaystyle K(\xi^a,t| \xi^a_i,t_i)= \int {\mathcal D}\xi {\mathcal D}\lambda {\mathcal D}c 
{\mathcal D}\overline{c} \, \exp \left[ i\int {\textrm d}x \, \widetilde{\cal L} \right], \label{kernelphi}
\end{equation}
where $\widetilde{\cal L}=\lambda_a\dot{\xi}^a+i \overline{c}_a\dot{c}^a-\widetilde{\cal H}$ and 
$\widetilde{\cal H}$ is the following Hamiltonian density:
\begin{equation}
\displaystyle \widetilde{\cal H}=\lambda_a\omega^{ab}\partial_bH+i\overline{c}_a\omega^{ab}\partial_b\partial_d Hc^d. \label{1.5bis}
\end{equation}
From a formal point of view the kernel of propagation (\ref{kernelphi}) is identical to the one given in \cite{martin} for a point particle. The main differences are due to the fact that a field is a system with an infinite number of degrees of freedom. This implies that the derivatives $\partial_b$ must be intended as functional derivatives and that in the weight of the path integral the Lagrangian density $\widetilde{\cal L}$ is integrated over all the space-time variables $x$. Another difference is in the functional measure: for example in the expression ${\mathcal D}\xi^a=\prod_{k, {\bf x}}d\xi^a(k, {\bf x})$, besides a slicing in $t$, there is also a slicing in the space variables ${\bf x}$. The CPI for the point particle has been thoroughly studied \cite{martin}-\cite{math}-\cite{topological} from a geometrical point of view and contact has been made both with the KvN formalism and the differential geometry approach to CM. In particular, $\widetilde{\cal H}$ turns out to be the Lie derivative \cite{libromarsden} of the Hamiltonian flow associated to $H$. 

The reader may wonder why we want to do the CPI for a field theory. The reasons are basically two: the first one is that some {\it classical} solutions \cite{rebbi} (solitons, instantons, monopoles, etc.) seem to play an important role in several non-perturbative aspects of {\it quantum} field theory, like confinement, phase transitions in the early universe and similar \cite{nuclproc}. So a formalism like the CPI, which is classical but formally quite similar to the QPI, may help in unveiling further connections between {\it classical} solutions and {\it quantum} effects. The second reason for the CPI is that classical field theory seems to play a central role in heavy-ion collisions \cite{venugopalan}. The reason is that in these collisions quantum effects are damped, due to the very high occupation numbers. People \cite{mueller}-\cite{jeon} have realized that even the perturbation theory to be used in heavy-ion collisions should be the classical  and not the {\it quantum} one. Clearly the best tool to use in order to derive the Feynman diagrams for classical field theory is the CPI. This CPI perturbative approach was first pioneered in Ref. \cite{jeon} for a $\varphi^4$ scalar field theory. Of course the classical fields involved in heavy-ion collisions are gauge and fermion fields. So a CPI version of these fields should be developed. For fermions things were done in Ref. \cite{field}, while for gauge fields we shall work things out in this paper, where we will concentrate on the more {\it formal} aspects of the CPI for YM. In particular, we will study the role that the gauge-fixing and the Faddeev-Popov determinant play even at the classical level both in the Faddeev-Popov (FP) and in the Batalin-Fradkin-Vilkovisky (BFV) \cite{henneaux} approach. Like in the standard CPI for the point particle we have, due to the many auxiliary variables, a lot of symmetries (like supersymmetry, ghost charge invariance etc.) which are {\it universal} in the sense that they are present for every Hamiltonian $H$ entering (\ref{1.5bis}). In the CPI for Yang-Mills (YM) theories these symmetries will be even more numerous, due to the gauge invariance of the theory. These formal tools will be developed in this first paper. In a second one \cite{riccardo} we will implement the perturbation theory both at the scalar field theory \cite{jeon} and at the YM level. In particular, thanks to the supersymmetry \cite{martin} that we will work out in this first paper, we shall show \cite{riccardo} that the complicated \cite{jeon} perturbation theory of the CPI can be simplified using the analog of the superfields studied in \cite{dequantization}. This perturbation theory will turn out to be very similar to the quantum one because, once we replace fields with superfields, the QPI goes into the CPI \cite{dequantization}. We hope this simplified perturbation theory can help the nuclear physicists \cite{jeon} who are exploring the quark-gluon phase in the heavy-ion collisions. 

The supersymmetry studied in this first paper will be the central tool which we shall use also in the third paper \cite{third} of this series which we hope to dedicate to the topological field theory aspects of the CPI for YM. It was in fact proved in \cite{topological} that any CPI can be turned into a topological field theory by properly choosing the boundary conditions. There \cite{topological} it was shown that the ``topological" phase of the CPI helps in building objects like the Euler number of the phase space or the Maslov index of the Lagrangian Grassmannian associated with the phase space. In the CPI for YM the analog of the phase space will be the space of gauge orbits, so the associated topological field theory will help in building the analog of the Euler number or Maslov index of the space of gauge orbits. These indices will help in unveiling some of the geometrical features of the space of gauge orbits which are known to be at the heart of many important non-perturbative aspects of gauge theories. 

Before concluding this Introduction, we should also remind the reader of the lot of work which has been done since the early 1970's \cite{marsden} on classical field theory. In those papers the classical tools of symplectic differential geometry were put at work for classical field theory. Those tools are the same which enter our CPI. It was in fact shown in \cite{martin}-\cite{math} that all the several auxiliary variables of the CPI have a nice geometrical meaning. For example, some of the Grassmann variables can be interpreted as differential forms, the charge \cite{martin} associated to one of our universal symmetries turns out to be the exterior derivative and the Hamiltonian $\widetilde{\cal H}$ associated to $\widetilde{\cal L}$ is basically, as we already said, the Lie derivative  \cite{libromarsden} of the Hamiltonian flow. These common tools can help the reader familiar with \cite{libromarsden} to easily understand the CPI \cite{math}Ê and vice versa. 

The paper is organized as follows. In Sec. 2 we review the YM formalism from the constraints point of view and build the associated CPI. This first version of the CPI will be called the ``natural" one because it arises in a very natural manner from simple considerations. In Sec. 3 we review the BFV method \cite{henneaux} and build the associated CPI. In Sec. 4 we prove the equivalence between the ``natural" CPI introduced in Sec. 2 and the BFV one of Sec. 3. In Sec. 5 we derive all the universal symmetries present in any CPI but now adapted to YM. We also build the charges related to the gauge invariances which have a different expression than in the quantum case because of the many auxiliary fields present. In Sec. 6 we build the universal supersymmetry present in this formalism and put forward some ideas for further developments which will appear in Ref. \cite{third}. Several calculational details are confined to few appendices. Because of the nature of this Journal we have tried to present things in a review form reviewing some topics which are somehow scattered in the literature. We have also presented the calculations in details because this first paper should contain many of the technical tools which will be used in further development \cite{riccardo}-\cite{third} of this project.

\section{Yang-Mills Theories}

As we have seen in the previous section, the implementation of the CPI for a classical scalar field theory is straightforward. Things become much more difficult when we study YM theories because of the presence of constraints. First of all, let us review some aspects of YM theories, just to fix the notation we will use later on. YM theories generalize quantum electrodynamics to systems endowed with a non-Abelian gauge invariance. The generators of the symmetry transformations form a Lie group $\mathcal{G}$, so they obey a well-defined algebra. If with $T_a, \; a=1, \ldots, n$ we indicate the basis elements of this Lie algebra, we will write the composition law among them as: 
\begin{displaymath}
i[T_a,T_b]=-C_{ab}^{\;\;c}T_c,
\end{displaymath}
where $C_{ab}^{\;\;c}$ are the so called structure constants. The antisymmetry and the Jacobi identity of the commutator imply that also the structure constants are antisymmetric and satisfy the Jacobi identity: 
\begin{displaymath}
C_{ab}^{\; \; c}=-C_{ba}^{\; \; c}, \qquad \quad C_{ab}^{\; \; e}C_{ec}^{\; \; d}+C_{bc}^{\; \; e}C_{ea}^{\; \; d}+
C_{ca}^{\; \; e}C_{eb}^{\; \; d}=0.
\end{displaymath}
Next we can define the second rank tensor $g_{ab}=C_{ad}^{\; \; e}C_{be}^{\;Ê\;d}$, which can be used to raise and lower the color indices. Another convention that we must specify is related with the choice of the space-time metric. We will use the following one: 
\begin{displaymath}
\eta^{\mu\nu}=\begin{pmatrix} -1 & 0 & 0 & 0 \cr
                     0 & 1 & 0 & 0 \cr
                     0 & 0 & 1 & 0 \cr
                     0 & 0 & 0 & 1
\end{pmatrix}
\end{displaymath}
to raise and lower the position of the space indices. For example, on the four vector and on the space-time derivatives the action of the metric tensor is the following: 
\begin{eqnarray}
\displaystyle
x^{\mu} = (t,{\bf x}) & \Longrightarrow & x_{\mu}=\eta_{\mu\nu}x^{\nu}=(-t,{\bf x}), \nonumber \medskip \\
\partial^{\mu} = \biggl(-\frac{\partial}{\partial t},\frac{\partial}{\partial {\bf x}}\biggr)
& \Longrightarrow &
\partial_{\mu}=\eta_{\mu \nu}\partial^{\nu}=\biggl(\frac{\partial}{\partialÊt}, \frac{\partial}{\partial {\bf x}}\biggr). \nonumber
\end{eqnarray}
The gauge transformation on the field $A$ has the following form:
\begin{equation}
\displaystyle
\delta A_a^{\mu}(x)=A^{\prime \mu}_a(x)-A^{\mu}_a(x)=C_a^{\; bc}\epsilon_b(x)
A_c^{\mu}(x)+\frac{1}{g}\frac{\partial \epsilon_a(x)}{\partial x_{\mu}}, \label{trasfa}
\end{equation}
where $g$ is the coupling constant, which, from now on, we shall put equal to one. If we define the antisymmetric field strength: 
\begin{displaymath}
F_a^{\mu \nu}\equiv \partial^{\mu}A^{\nu}_a-\partial^{\nu}A^{\mu}_a-gC_{a}^{\; bc}A_b^{\mu}
A_c^{\nu}
\end{displaymath}
we can deduce immediately from Eq. (\ref{trasfa}) that the gauge transformations of this field are $\delta F_a^{\mu \nu}=C_a^{\; bc}\epsilon_bF_c^{\mu \nu}$. The standard gauge invariant Lagrangian density is:
\begin{displaymath}
\displaystyle {\mathcal L}=-\frac{1}{4}F_a^{\mu\nu}F^a_{\mu\nu}.
\end{displaymath}
From ${\mathcal L}$ we can derive the following YM equations:
\begin{displaymath}
\partial_{\nu}
F_a^{\mu\nu}-C^b_{\;ac}F_b^{\mu\nu}A_{\nu}^c=0
\end{displaymath}
which can be written in terms of the covariant derivative $D^b_{\; a \nu}\equiv \delta_a^b\partial_{\nu}-C^b_{\;ac}A^c_{\nu}$ as follows: 
\begin{equation}
\displaystyle D^b_{\; a\nu}F^{\mu\nu}_b=0. \label{ymeq}
\end{equation}
It is interesting to rewrite Eq. (\ref{ymeq}) in components: the component $\mu=0$ becomes: 
\begin{equation}
\displaystyle \partial_k E_a^k-C^b_{\; ac}E^k_bA^c_k=0, \label{gausslaw}
\end{equation}
where we have defined the analog of the electric field $E^k_a\equiv F^{0k}_a$.
Eq. (\ref{gausslaw}) is the analog of the Gauss law for electromagnetism. Since they do not contain time derivatives, Eqs. (\ref{gausslaw}) are just constraints\footnote{As we will prove in Appendix A, they are the equations of the secondary constraints.}. The space components of Eq. (\ref{ymeq}) are instead real equations of motion and are given by: 
\begin{displaymath}
\displaystyle
-\frac{\partial}{\partial t}\vec{E}_a+\vec{\nabla}\times\vec{B}_a+C^b_{\; ac}
\vec{E}_bA^c_{\scriptscriptstyle 0}+C^b_{\; ac}\vec{B}_b\times\vec{A}^c=0,
\end{displaymath}
where $\vec{B}$ is the analog of the magnetic field and its components are defined as $\displaystyle B_a^{i}\equiv \frac{1}{2}\epsilon^{ijk}F_a^{jk}$. 

Like for the scalar field theory, in order to implement the CPI we need to build from the gauge invariant Lagrangian density $\displaystyle {\mathcal L}= -\frac{1}{4}F_a^{\mu\nu}F^a_{\mu\nu}$ the associated Hamiltonian. In this case we must take into account the constraints of the theory. According to the calculations presented in Appendix A, the canonical action is given by \cite{henneaux}: 
\begin{equation}
\displaystyle
S_{\textrm c}=\int {\textrm d}x \, \Biggl\{\dot{\vec{A}}_a\cdot \vec{\pi}^{a}-\frac{1}{2}\vec{\pi}_a\cdot\vec{\pi}^{a}
-\frac{1}{2}\vec{B}_a\cdot\vec{B}^{a}-A^{a}_{\scriptscriptstyle 0}(-\vec{\nabla}\cdot\vec{\pi}_a
+C^b_{\; ac}\vec{\pi}_b\cdot \vec{A}^c)\Biggr\}, \label{azcan}
\end{equation}
where $\pi^k_a=-F^{0k}_a$ and the arrows above indicates the space components of the fields. 
In the previous formula, all the scalar products involve only the space components of the fields.
From (\ref{azcan}) we identify immediately the following YM Hamiltonian density:
\begin{displaymath}
\displaystyle
{\cal H}^0_{\scriptscriptstyle \textrm{YM}}\equiv \frac{1}{2}\biggl(\vec{\pi}_a\cdot\vec{\pi}^{a}
+\vec{B}_a\cdot\vec{B}^{a}\biggr)
\end{displaymath}
and the constraints:
\begin{equation}
\sigma_a\equiv -\vec{\nabla}\cdot\vec{\pi}_a+C^b_{\; ac}\vec{\pi}_b\cdot \vec{A}^c\approx 0.  \label{nabla}
\end{equation}
Note that the $A^a_{\scriptscriptstyle 0}$ in (\ref{azcan}) act as Lagrange multipliers which will be indicated later on with $\lambda^a$. The constraints $\sigma$ form a closed algebra, i.e.:
\begin{displaymath}
\{\sigma_a(t, {\bf x}), \sigma_b(t, {\bf x}^{\prime})\}=C^c_{\; ab}\sigma_c(t, {\bf x})\delta({\bf x}-{\bf x}^{\prime}).
\end{displaymath} 
The above equation tells us that the Poisson brackets among the constraints are zero on the constraint surface, which implies that, according to the definitions of Dirac \cite{dirac}, the constraints are first class. With similar techniques it is easy to prove that the YM Hamiltonian has zero Poisson brackets with the constraints:
\begin{displaymath}
\displaystyle \left\{ \sigma_a(t,{\bf x}^{\prime}), \int {\textrm{d}}{\bf x}  \, {\cal H}^0_{\scriptscriptstyle \textrm{YM}}(t,{\bf x})\right\}=0.
\end{displaymath}
This means that the $\sigma_a$ are automatically conserved in time and no other constraint is generated. 

From the canonical action (\ref{azcan}) we can derive the equations of motion. By considering the variation with respect to $A^a_k$ we get:
\begin{equation}
\displaystyle \frac{\delta S_{\textrm c}}{\delta A^{a}_k}=0 \Longrightarrow \dot{\pi}^k_a=-\partial_iF^{ki}_a-
\lambda^dC^b_{\;da}\pi^k_b+C^b_{\;ac}A^c_iF_b^{\;ki}. \label{eqpi}
\end{equation}
Similarly, considering the variation with respect to $\pi^k_a$, we get the equation of motion for $A^{a}_k$:
\begin{equation}
\dot{A}^{a}_k=\pi^{a}_k+\partial_k\lambda^{a}+C^{a}_{\;dc}\lambda^dA^c_k. \label{eqa}
\end{equation}
Now, if we identify from the canonical action the following Hamiltonian:
\begin{displaymath}
\displaystyle
H=\int {\textrm d}{\bf x}\Biggl\{\frac{1}{2}\vec{\pi}_a\cdot \vec{\pi}^{a}+\frac{1}{2}\vec{B}_a\cdot \vec{B}^a+\lambda^{a}(-\vec{\nabla}\cdot \vec{\pi}_a+C^b_{\;ac}\vec{\pi}_b\cdot{\vec{A}}^{c})\Biggr\}
\end{displaymath}
we can rewrite the equations of motion (\ref{eqpi}) and (\ref{eqa}) as:
\begin{equation}
\displaystyle \dot{\vec{A}}^{a}=\frac{\partial H}{\partial\vec{\pi}_a}, \qquad \quad \dot{\vec{\pi}}_a=-\frac{\partial H}{\partial \vec{A}^{a}} \label{compeq}.
\end{equation}
Collecting together all the fields in the variables $\phi^{\scriptscriptstyle A}\equiv (\vec{A}^{a},\vec{\pi}_a)$, and introducing the symplectic matrix $\omega^{\scriptscriptstyle AB}=\begin{pmatrix}
0 & 1\cr -1 & 0\cr \end{pmatrix}$, the equations of motion (\ref{compeq}) can be written in a compact form as:
\begin{displaymath}
\displaystyle
\dot{\phi}^{\scriptscriptstyle A}=\omega^{\scriptscriptstyle AB}\partial_{\scriptscriptstyle B}H,
\end{displaymath}
where $\partial_{\scriptscriptstyle B}$ stands for the functional derivative w.r.t. $\phi^{\scriptscriptstyle B}$. 

Let us now proceed to implement the CPI. In order to do that we must have a one-to-one correspondence between the equations of motion and their solutions, once the initial conditions are given. This is clear from formula (\ref{trasf}). Its RHS is equivalent to the LHS only if, given some initial conditions, a unique solution can be found for the equations of motion appearing in the Dirac delta on the RHS of (\ref{trasf}). If the solutions or the ``zeros" of the equations of motion of the RHS of (\ref{trasf}) were more than one, one should use the functional analog of the following formula:
\begin{equation}
\displaystyle \sum_i \frac{\delta(x-x_i)}{\displaystyle \left| \frac{\partial f}{\partial x} \right|_{x=x_i}}=\delta[f(x)],
\label{1.4bis}
\end{equation}
where the sum on the LHS is over all the zeros $x_i$ of the function $f(x)$. The sum present on the LHS in (\ref{1.4bis}) does not allow us to bring the determinant $\displaystyle \left| \frac{\partial f}{\partial x}\right|_{x=x_i}$ to the RHS, as it is needed in order to implement the CPI. A similar problem has already been faced in the case of the point particle with reparametrization invariance \cite{bill}. In the case of a theory with gauge invariance, like YM theories described by the action (\ref{azcan}), the number of solutions of the equations of motion with the same initial conditions is more than one. In fact, since the Lagrangian multipliers $\lambda^{a}=A^{a}_{\scriptscriptstyle 0}$ are undetermined, when we solve the equations of motion we get an entire family of solutions depending on the parameters $\lambda^a$, even if we fix the initial conditions. This is the reason why we cannot write the path integral, using the same formal expression of the scalar field (\ref{trasf}), as:
\begin{equation}
Z_{\scriptscriptstyle \textrm{CPI}}^{\scriptscriptstyle \textrm{YM}}=\int {\cal D}\phi^{\scriptscriptstyle A}\delta(\phi^{\scriptscriptstyle A}-\phi^{\scriptscriptstyle A}_{\textrm{cl}})=
\int {\cal D}\phi^{\scriptscriptstyle A} \delta(\dot{\phi}^{\scriptscriptstyle A}-\omega^{\scriptscriptstyle AB}\partial_{\scriptscriptstyle B}H){\textrm{det}}\left[\delta({\bf x}-{\bf y})\delta^{\scriptscriptstyle A}_{\scriptscriptstyle B}\partial_t -\omega^{\scriptscriptstyle AC}\frac{\partial^2 H}{\partial \phi^{\scriptscriptstyle C}({\bf x})\partial \phi^{\scriptscriptstyle B}({\bf y})}\right]. \label{trentasette}
\end{equation}
Before passing from the Dirac delta of the solutions to the one of the equations of motions, we need to determine the Lagrange multipliers. This is possible by fixing the gauge \cite{dirac}-\cite{bill}. Suppose we choose the Lorentz gauge $-\partial^{\mu}A^{a}_{\mu}=0$. To implement it
we can use the standard procedure \cite{abers} which consists in multiplying $Z_{\scriptscriptstyle \textrm{CPI}}^{\scriptscriptstyle \textrm{YM}}$ by ``one" and in rewriting this ``one" as:
\begin{displaymath}
\displaystyle 1= \int {\cal D}A_{\scriptscriptstyle 0}\;\delta[-\partial^{\mu}A_{\mu}]\Delta_{\scriptscriptstyle \textrm{FP}}(A),
\end{displaymath}
where $\Delta_{\scriptscriptstyle \textrm{FP}}(A)$ is the FP determinant associated with the gauge-fixing $-\partial^{\mu}A_{\mu}=0$. In this way, Eq. (\ref{trentasette}) becomes:
\begin{eqnarray}
Z_{\scriptscriptstyle \textrm{CPI}}^{\scriptscriptstyle \textrm{YM}}&=&
\int {\cal D}\phi^{\scriptscriptstyle A}{\cal D}A_{\scriptscriptstyle 0}\;\delta[-\partial^{\mu}A^{a}_{\mu}]\Delta_{\scriptscriptstyle \textrm{FP}}[A]\, \delta(\dot{\phi}^{\scriptscriptstyle A}-\omega^{\scriptscriptstyle AB}\partial_{\scriptscriptstyle B}H)\cdot \nonumber \\
&& \cdot \textrm{det}\left[\delta({\bf x}-{\bf y})\delta^{\scriptscriptstyle A}_{\scriptscriptstyle B}\partial_t -\omega^{\scriptscriptstyle AC}\frac{\partial^2 H}{\partial \phi^{\scriptscriptstyle C}({\bf x})\partial \phi^{\scriptscriptstyle B}({\bf y})}\right] \label{trentasette2}.
\end{eqnarray}
The appearance of the FP determinant in a {\it classical} context is not a surprise: in fact, it is possible to prove that such a determinant can be reduced, via a suitable canonical transformation, to the determinant appearing in the Dirac procedure for {\it classical} systems with constraints, see for example \cite{dirac} or \cite{sund}.

Following \cite{Lee}, we want to find the explicit form for the FP determinant associated with the choice of the Lorentz gauge. First of all, according to (\ref{trasfa}), the generic gauge transformation of the field $A$ is given by:
\begin{displaymath}
\delta A^{\mu}_a=C_a^{\;bc}\epsilon_bA^{\mu}_c+\partial^{\mu}\epsilon_a
\end{displaymath}
which, for a gauge transformation in a neighborhood of the identity, implies that the gauge-transformed $-\partial^{\mu}A^{a}_{\mu}$ is given by:
\begin{displaymath}
\displaystyle
\left[ -\partial^{\mu}A^{a}_{\mu}\right] ^{\prime}=-\partial^{\mu}A^{a}_{\mu}+\frac{\partial(-\partial_{\mu}A^{a\mu})}
{\partial A^{b \nu}}\delta A^{b \nu}=-\partial^{\mu}D^{ba}_{\mu}\epsilon_b.
\end{displaymath}
From this expression we get that the FP determinant is given by: 
\begin{displaymath}
\Delta_{\scriptscriptstyle \textrm{FP}}[A]\sim \textrm{det} [-\partial^{\mu}D^{ba}_{\mu}].
\end{displaymath}
This determinant can be exponentiated via a couple of Grassmann variables $\psi_a, \eta_a$:
\begin{displaymath}
\Delta_{\scriptscriptstyle \textrm{FP}}[A]=\int{\cal D}\psi_{a}{\cal D}\eta_{a}\exp \left[ \int {\textrm d}x \, \partial^{\mu}\psi_aD_{\mu}^{ba}\eta_b\right].
\end{displaymath}
We can exponentiate also the gauge-fixing via an auxiliary variable\footnote{The $\pi_a$ should not be confused with the $\vec{\pi}_a$ appearing in (\ref{azcan}) which carry a space index $\pi^i_a$.} $\pi_a$:
\begin{displaymath}
\delta(-\partial^{\mu}A_{\mu}^{a})=\int {\cal D}\pi_a \exp \left[ -i\int {\textrm d}x\, \pi_a\partial^{\mu}A_{\mu}^{a}\right].
\end{displaymath}
Like in the standard CPI procedure, also in this case the Dirac delta of the equations of motion, which appears on the RHS of (\ref{trentasette2}), can be Fourier transformed via some bosonic auxiliary variables $\Lambda_{\scriptscriptstyle A}$. The associated functional determinant can be exponentiated via a couple of Grassmann variables $\Gamma^{\scriptscriptstyle A}$ and $\overline{\Gamma}_{\scriptscriptstyle A}$. The final result is:
\begin{equation}
Z_{\scriptscriptstyle \textrm{CPI}}^{\scriptscriptstyle \textrm{YM(NAT)}}=\int {\cal D}\mu \;\exp \left[ i\int {\textrm d}x\, \widetilde{\cal L}^{\scriptscriptstyle \textrm{NAT}} \right],
\label{ymnat}
\end{equation}
where the functional measure ${\cal D}\mu$ is given by:
\begin{equation}
{\cal D}\mu\equiv {\cal D}\phi^{\scriptscriptstyle A}{\cal D}\Lambda_{\scriptscriptstyle A}{\cal D}\overline{\Gamma}_{\scriptscriptstyle A}
{\cal D}\Gamma^{\scriptscriptstyle A}{\cal D}A^{a}_{\scriptscriptstyle 0}{\cal D}\pi_a{\cal D}\psi_a{\cal D}\eta_a, \label{Lagrnaturale0}
\end{equation}
while the Lagrangian $\widetilde{\cal L}^{\scriptscriptstyle \textrm{NAT}}$ is:
\begin{equation}
\widetilde{\cal L}^{\scriptscriptstyle \textrm{NAT}}=-\pi_a\partial^{\mu}A^{a}_{\mu}-i\partial^{\mu}\psi_aD^{ba}_{\mu}\eta_b
+\Lambda_{\scriptscriptstyle A}(\dot{\phi}^{\scriptscriptstyle A}-\omega^{\scriptscriptstyle AB}\partial_{\scriptscriptstyle B}H)+i\overline{\Gamma}_{\scriptscriptstyle A}(\delta_{\scriptscriptstyle B}^{\scriptscriptstyle A}\partial_t
-\omega^{\scriptscriptstyle AC}\partial_{\scriptscriptstyle B}\partial_{\scriptscriptstyle C}H)\Gamma^{\scriptscriptstyle B}   \label{Lagrnaturale}. 
\end{equation}
What we have done up to now is the first, more {\it natural} procedure to derive a path integral for a classical field theory with a non-Abelian gauge invariance. This is the reason why we have put an acronym ``{NAT}" (for Natural) everywhere in the previous formulae. 

We want to conclude this section by underlying again that, in order to implement the CPI, it was crucial to fix the gauge and as a consequence to introduce the FP determinant. So a gauge-fixing procedure is needed even at the {\it classical} level and not just in the {\it quantum} case. 
In quantum mechanics the gauge-fixing was needed in order to get the quantum propagator, or equivalently to be able to invert the operator in the kinetic part of the Lagrangian. Somehow also at the classical level we need the gauge-fixing for a similar reason, i.e. to ``propagate" the initial conditions in a well defined manner, by choosing only one of the many different trajectories. Even the FP determinant enters the CPI for reasons identical to the quantum case. This is very clear in our formalism while in the usual Dirac \cite{dirac} procedure, the FP determinant makes its appearance via the ``C" matrix \cite{dirac} needed to build the Dirac brackets from the Poisson ones. 

In the next section, we will implement another path integral for classical YM theories, which consists in starting from the so called BFV Lagrangian and in exponentiating the associated classical equations of motion. Even if they seem to be completely different, the two CPIs, the one analyzed previously and the BFV one which we will study in the next section, are instead totally equivalent, as we shall show later.

\section{BFV Method}

In the FP method analyzed in the previous section we seem to lose the original gauge invariance of the theory because of the gauge-fixing procedure. Nevertheless,  the functional integral maintain a symmetry, which carries the memory of the original gauge invariance of the theory, and which is called BRST symmetry, see for example \cite{henneaux}. This symmetry is crucial because it guarantees that nothing observable depends on which gauge-fixing we have used. At the end of the Seventies, an alternative approach to the FP procedure was developed by Batalin, Fradkin and Vilkovisky (BFV). Their method gives automatically a gauge-fixed action with all the extra terms of the ``FP type" plus the BRST symmetry mentioned above \cite{henneaux}. 

In what follows we briefly review the BFV method, following Refs. \cite{henneaux}, \cite{gov}. The nice feature of this method is that it is entirely classical even if it is mostly used at the quantum level. It is classical in the sense that it basically produces in an {\it automatized way}
a gauge-fixing (GF) and the extra terms of the FP type which are things that are needed even at the classical level, as shown in the previous section. The BFV method starts by associating a couple of auxiliary variables $\overline{\cal P}^{a}$ and ${\cal P}_{a}$ to each first class constraint $\tilde{\sigma}_a$ and by defining the Poisson brackets among them:
\begin{displaymath}
\left\{{\cal P}_{a}, \overline{\cal P}^{b}\right\}=
\left\{\overline{\cal P}^{b}, {\cal P}_{a}\right\}=-\delta_{a}^{b}.
\end{displaymath}
All the Poisson brackets between the Grassmannian odd variables $(\overline{\cal P}, {\cal P})$ and the original phase space variables are instead identically equal to zero. We will assign to the Grassmannian odd variables a grading factor, called ghost number: in particular $\overline{\cal P}^a$ will be assigned ghost number $+1$, while ${\cal P}_a$ will be assigned ghost number $-1$. Furthermore, the Lagrangian multipliers $\lambda^{a}=A^a_{\scriptscriptstyle 0}$ which enter the definition of the canonical action (\ref{azcan}) must be considered as dynamical variables, so we associate to each of them a bosonic conjugate momentum $\pi_{a}$, which satisfies the following Poisson brackets:
\begin{equation}
\left\{\pi_{a},\lambda^{b}\right\}=-\delta_{a}^{b}. \label{poi}
\end{equation}
Of course, if we do not want to change the dynamical content of the theory, these momenta $\pi_a$ must be zero and this happens if $\dot{\lambda}_a$ does not appear in the Lagrangian. 
To summarize, if we indicate with $z_{\scriptscriptstyle A}$ the canonical variables describing the original gauge fields, the full extended BFV space is given by 
$\displaystyle (z_{\scriptscriptstyle A},\lambda^{a},\pi_{a},\overline{\cal P}^{a},{\cal P}_{a})$. It is possible to prove that in this enlarged BFV space there exists a real Grassmannian odd charge $\Omega$, which satisfies the following properties: 
\begin{eqnarray}
\displaystyle
\textrm{ghost number of} \, \Omega & = & +1  \medskip \\
\frac{\partial}{\partial\overline{\cal P}^a}\Omega_{\overline{\cal P}^{a}={\cal P}_a=0} & = & \tilde{\sigma}_a  \label{term0} \nonumber \medskip \\
\{\Omega,\Omega\} & = & 0. \nonumber
\end{eqnarray}
The charge which satisfies the previous properties is the so called BRST charge: 
\begin{displaymath}
\displaystyle
\Omega=\int \textrm{d}{\bf x} \left[ \overline{\cal P}^{a}\tilde{\sigma}_a-\frac{1}{2}\overline{\cal P}^b\overline{\cal P}^cC^{a}_{\; cb}{\cal P}_a \right].
\end{displaymath}
In Dirac's theory of constraints the {\it observables} are those functions of the original phase space which have zero Poisson brackets with the constraints, once we sit on the constraint surface. This means that the Poisson brackets of an observable $O$ with every constraint $\tilde{\sigma}_a$ must be a linear combination of the constraints themselves: $\{O, \tilde{\sigma}_a\}=W_a^{\; b}\tilde{\sigma}_b\approx 0$. We call ``observables" the $O$ because they remain invariant under the gauge transformations generated by the first class constraints $\tilde{\sigma}_a$. Furthermore, two observables $O$ and $O^{\prime}$ are equivalent if their difference is just an arbitrary linear combination of the constraints $\tilde{\sigma}_a$, i.e. $O^{\prime}=O+k^{a}\tilde{\sigma}_a$. Let us see which is the analog of these definitions in  the BFV method. First of all, a function is an observable if it commutes with the BRST charge. The procedure to construct the observables is to extend the functions $O$ to functions $\bar{O}$ of the BFV space by requiring that:
\begin{itemize}
\item the functions $\bar{O}$ reduce to $O$ when the ghosts $\overline{\cal P}^a$ and the antighosts ${\cal P}_a$ are put equal to zero; 
\item the functions $\bar{O}$ have ghost number zero;
\item $\bar{O}$ are real and Grassmannian even.
\item $\bar{O}$ are BRST invariant.
\end{itemize}
Now, if $\bar{O}$ is an extension of the function $O$ which commutes with the BRST charge and satisfies the previous properties, it is possible to consider an entire class of equivalent extensions $\bar{O}_{\theta}$ which differ from $\bar{O}$ by a pure BRST variation: $\bar{O}_{\theta}=\bar{O}-\{\theta,\Omega\}$. Since the Hamiltonian itself is a particular observable, it also admits an entire family of BRST invariant extensions: 
\begin{equation}
H_{\theta}=H-\{\theta,\Omega\}. \label{gaugeBFVuno}
\end{equation}
The function $\theta$ is somehow related to the gauge freedom of the theory: fixing the gauge means choosing a particular function $\theta$ and consequently a particular Hamiltonian among the $H_{\theta}$ of Eq. (\ref{gaugeBFVuno}). Clearly, all the physical quantities must be independent of the choice of this function $\theta$ and this is guaranteed by the so called Fradkin and Vilkovisky theorem. 

Let us now apply the BFV method to the case of YM theories. As we already said before, the Lagrange multipliers $\lambda^{a}$ can be identified with the time components of the fields $A_{\mu}^a$, i.e., $A^{a}_{\scriptscriptstyle 0}$, so the associated conjugate momenta must be identified with the time components of the momenta $\pi^{\mu}$. The equal-time Poisson brackets are: 
\begin{equation}
\{\pi_b(t,{\bf x}),\lambda^{a}(t,{\bf y})\}=-\delta_b^{a}\delta({\bf x}-{\bf y}).
\end{equation} 
From the canonical action (\ref{azcan}) we get the following constraints $\pi_a\approx 0, \; a=1,\ldots n$. 
We know from Appendix A that they are the primary constraints of the theory. From their consistency, i.e. the requirement that their time evolution leaves them invariant, we get just the $n$ secondary constraints of Eq. (\ref{nabla}), so the total number of constraints is $2n$. We will indicate them as 
\begin{equation}
\tilde{\sigma}_a=(\pi_b,\sigma_b), \;\; a=1,\ldots, 2n; \;\; b=1,\ldots n. \label{3.4bis}
\end{equation}
Following the BFV procedure, we shall associate to these constraints the ghosts ${\cal P}_a$ and $\overline{\cal P}^{a}$. All these auxiliary variables are Grassmannian odd and, as we have seen, they are conjugated in the sense that: $\{{\cal P}_a,\overline{\cal P}^b\}=-\delta_a^{b}.$
These ghosts can be divided in two blocks according to the type of constraints (\ref{3.4bis}) they refer to:
\begin{displaymath}
\overline{\cal P}^{a}=(-iP^b,C^b), \;Ê\quad {\cal P}_a=(i\overline{C}_b,\overline{P}_b)
\end{displaymath}
and their Poisson brackets are given by:
\begin{displaymath}
\{\overline{P}_a,C^b\}=-\delta_a^b=\{P^b,\overline{C}_a\}.
\end{displaymath}
Since the only non-zero structure constants are the ones among the secondary constraints, the BRST charge for the Yang-Mills theories becomes:  
\begin{eqnarray}
\displaystyle
\Omega &\hspace{-2mm}=&\hspace{-2mm} \int {\textrm d}{\bf x}\,\biggl[\overline{\cal P}^{a}\psi_a-\frac{1}{2}\overline{\cal P}^b\overline{\cal P}^cC^{a}_{\;cb}{\cal P}_a\biggr] \nonumber\\
&\hspace{-2mm}=&\hspace{-2mm} \int {\textrm d}{\bf x} \biggl[\sigma_aC^{a}-iP^{a}\pi_a+\frac{1}{2}\overline{P}_aC^{a}_{\;bc}C^bC^c \biggr] \label{caricaomega}.
\end{eqnarray}
The gauge freedom of the theory is entirely contained in the function $\theta$ which appears in (\ref{gaugeBFVuno}). Usually this arbitrariness is moved to a new function $\chi$ according to the definition:
\begin{displaymath}
\theta\equiv i\overline{C}_a\chi^{a}+\overline{P}_a\lambda^{a}. 
\end{displaymath}
This function $\chi$ plays the role of the GF \cite{henneaux}. For example,
the analog of the Coulomb gauge can be obtained by choosing $\chi^{a}=\vec{\nabla}\cdot\vec{A}^{a}$. With all these ingredients in our hands, we can now proceed to build the BFV action in the form given in Ref. \cite{henneaux}:
\begin{displaymath}
S^{\scriptscriptstyle \textrm{BFV}}=\int \textrm{d}x \biggl[\dot{\vec{A}}_a\cdot {\vec{\pi}}^{a}+\dot{\lambda}^{a}\pi_a+\dot{P}^{a}
\overline{C}_a+\dot{C}^{a}\overline{P}_a-{\cal H}^{\scriptscriptstyle 0}_{\scriptscriptstyle \textrm{YM}}+\{\theta,\Omega\}\biggr],
\end{displaymath}
where if $\theta$ is built out of the $\chi$ function associated with the ``Coulomb-like" gauge mentioned above then the commutator $\{\theta, \Omega \}$ is given by:
\begin{eqnarray}
\{\theta,\Omega\}&\hspace{-2mm}=&\hspace{-2mm}\int {\textrm d}{\bf x} \Bigl[-\pi_a\vec{\nabla}\cdot \vec{A}^{a}-\lambda^{a}\sigma_a-i\overline{P}_aP^{a}\nonumber\\
&&+\lambda^{a}\overline{P}_bC^b_{\; ac}C^c+i\overline{C}_a\vec{\nabla}\cdot(\vec{\nabla}C^{a}+C^{a}_{\;bc}
\vec{A}^cC^b) \Bigr]. \label{gaugeBFV} \label{thetaomega}
\end{eqnarray}
With this choice the specific form of $S^{\scriptscriptstyle \textrm{BFV}}$ is:
\begin{eqnarray}
\displaystyle
S^{\scriptscriptstyle \textrm{BFV}} &\hspace{-2mm} = &\hspace{-2mm} \int {\textrm d}x \Bigl\{\dot{\vec{A}}_a\cdot \vec{\pi}^{a}+\dot{\lambda}^{a}\pi_a+\dot{P}^{a}\overline{C}_a
+\dot{C}^{a}\overline{P}_a-\frac{1}{2}\vec{\pi}_a\cdot \vec{\pi}^{a}
-\frac{1}{4}F_a^{ij}F_{ij}^{a}-\pi_a\vec{\nabla}\cdot \vec{A}^{a} +\lambda^{a}\vec{\nabla}\cdot \vec{\pi}_a\nonumber\\
&&-\lambda^{a}C_{\; ac}^b\vec{\pi}_b\cdot\vec{A}^c-i\overline{P}_aP^{a}+\lambda^{a}\overline{P}_bC_{\; ac}^bC^c
+i\overline{C}_a\vec{\nabla}\cdot(\vec{\nabla}C^{a}
+C_{\; bc}^{a}\vec{A}^cC^b)\Bigr\}. \label{azionediBFV}
\end{eqnarray}
The previous expression is BRST invariant and from it we can read out the following BFV Hamiltonian: 
\begin{eqnarray}
\displaystyle
H^{\scriptscriptstyle \textrm{BFV}} &\hspace{-2mm} = &\hspace{-2mm}  \int {\textrm d}{\bf x} \,\Biggl\{\frac{1}{2}\vec{\pi}_a\cdot \vec{\pi}^{a}
+\frac{1}{4}F_a^{ij}F_{ij}^{a}+\pi_a\vec{\nabla}\cdot \vec{A}^{a}-\lambda^{a}\vec{\nabla}\cdot \vec{\pi}_a+\lambda^{a}C_{\;ac}^b\vec{\pi}_b\cdot \vec{A}^c \nonumber \\
&&+i\overline{P}_aP^{a}-\lambda^{a}\overline{P}_bC_{\; ac}^bC^c-i\overline{C}_a\vec{\nabla}\cdot (\vec{\nabla}C^{a}
+C_{\; bc}^{a}\vec{A}^cC^b)\Biggr\}. \label{BFVh} 
\end{eqnarray}
From the BFV action (\ref{azionediBFV}) we can derive, by variation with respect to all the fields of the theory, the following equations of motion:
\begin{eqnarray}
\dot{\vec{\pi}}_a &\hspace{-2mm} = &\hspace{-2mm} -(\vec{\nabla}\times \vec{B}_a) -C^b_{\;ac}(\vec{B}_b\times \vec{A}^c)+\vec{\nabla}\pi_a-\lambda^c C_{\; ca}^b\vec{\pi}_b-iC_{\; ba}^d(\vec{\nabla}\overline{C}_d)C^b  \label{eqBFVprima} \nonumber \\
\dot{\vec{A}}^{a} &\hspace{-2mm} = &\hspace{-2mm} \vec{\pi}^{a}+\vec{\nabla}\lambda^{a}+C_{\; dc}^{a}\lambda^d\vec{A}^c  \nonumber \\
\dot{\pi}_a &\hspace{-2mm} = &\hspace{-2mm} \vec{\nabla} \cdot \vec{\pi}_a-C_{\; ac}^b\vec{\pi}_b\cdot \vec{A}^c+\overline{P}_bC_{\; ac}^bC^c  \nonumber \\
\dot{\lambda}^{a} &\hspace{-2mm} = &\hspace{-2mm} \vec{\nabla}\cdot \vec{A}^{a}  \label{BFVeq} \\
\dot{\overline{C}}_a &\hspace{-2mm} = &\hspace{-2mm} i\overline{P}_a  \nonumber \\
\dot{P}^{a} &\hspace{-2mm} = &\hspace{-2mm} i\vec{\nabla}[(\vec{\nabla}\delta_b^{a}-C_{b}^{\; ac}\vec{A}_c)C^b]  \nonumber \\
\dot{\overline{P}}_a &\hspace{-2mm} = &\hspace{-2mm} -\lambda^b\overline{P}_cC_{\; ba}^c-i\vec{\nabla}\cdot \vec{\nabla} \overline{C}_a+iC_{\; ac}^d(\vec{\nabla}\overline{C}_d)\cdot \vec{A}^c  \nonumber \\
\dot{C}^{a} &\hspace{-2mm} = &\hspace{-2mm} C_{\; bc}^{a}\lambda^bC^c-iP^{a}.\nonumber 
\end{eqnarray}
If we introduce a compact notation indicating with $\xi^{\scriptscriptstyle A}$ all the fields above
\begin{displaymath}
\xi^{\scriptscriptstyle A}\equiv(\vec{A}^{a},\lambda^{a},\vec{\pi}_a,\pi_a,P^{a},C^{a},
\overline{C}_a, \overline{P}_a),
\end{displaymath}
then Eqs. (\ref{BFVeq}) can be written in the following form\footnote{In the equation of motion (\ref{tilda}), 
which involves also Grassmannian odd fields, it is important to distinguish between right $\overrightarrow{\partial}$ and left $\overleftarrow{\partial}$ derivatives. If we want to pass from right to left derivatives and vice versa, we have to use the following formula: 
$\overrightarrow{\partial}_{\scriptscriptstyle B}F=(-)^{\scriptscriptstyle [B]+ [B][F]}F\overleftarrow{\partial}_{\scriptscriptstyle B}$.}:
\begin{equation}
\dot{\xi}^{\scriptscriptstyle A}=\omega^{\scriptscriptstyle AB}\overrightarrow{\partial}_{\scriptscriptstyle B}H^{\scriptscriptstyle \textrm{BFV}}(\xi),
\label{tilda}
\end{equation}
where the index $A$ in $\xi^{\scriptscriptstyle A}$ runs over all the fields of the theory, and we have used the symplectic matrix $\omega^{\scriptscriptstyle AB}$, which has the following expression:
\begin{displaymath}
\omega^{\scriptscriptstyle AB}=\bordermatrix{&\vec{A}^{a} & \lambda^{a} & \vec{\pi}_a & \pi_a & P^{a} & C^{a} & \overline{C}_a & \overline{P}_a \cr
\vec{A}^{a} & 0 & 0 & 1 & 0 & 0 & 0 & 0 & 0 \cr
\lambda^{a} & 0 & 0 & 0 & 1 & 0 & 0 & 0 & 0 \cr
\vec{\pi}_a & -1 & 0 & 0 & 0 & 0 & 0 & 0 & 0 \cr
\pi_a & 0 & -1 & 0 & 0 & 0 & 0 & 0 & 0 \cr
P^{a} & 0 & 0 & 0 & 0 & 0 & 0 & -1 & 0 \cr
C^{a} & 0 & 0 & 0 & 0 & 0 & 0 & 0 & -1 \cr
\overline{C}_a & 0 & 0 & 0 & 0 & -1 & 0 & 0 & 0 \cr
\overline{P}_a & 0 & 0 & 0 & 0 & 0 & -1 & 0 & 0 \cr}.
\end{displaymath}
The matrix above has an antisymmetric block associated with the commuting fields, and a symmetric one associated with the BFV Grassmannian odd fields. So the symmetry properties of the matrix $\omega$ are: $\omega^{\scriptscriptstyle AB}=-(-)^{\scriptscriptstyle [A][B]}\omega^{\scriptscriptstyle BA}$, where $[A]$ indicates the Grassmannian parity of the field associated to $\xi^{\scriptscriptstyle A}$.

Indicating with $\omega_{\scriptscriptstyle AB}$ the inverse matrix of $\omega^{\scriptscriptstyle AB}$ we can write in a compact way also the kinetic part of the BFV action. In fact, up to surface terms, we get:
\begin{eqnarray}
\displaystyle
\frac{1}{2}\xi^{\scriptscriptstyle A}\omega_{\scriptscriptstyle AB}\dot{\xi}^{\scriptscriptstyle B} &\hspace{-2mm}=&\hspace{-2mm}
\frac{1}{2}\biggl(-\vec{A}^{a}\cdot \dot{\vec{\pi}}_a-\lambda^{a}\dot{\pi}_a
+\vec{\pi}_a\cdot \dot{\vec{A}}^{a}+\pi_a\dot{\lambda}^{a}-P^{a}\dot{\overline{C}}_a-C^{a}\dot{\overline{P}}_a-\overline{C}_a\dot{P}^{a}
-\overline{P}_a\dot{C}^{a}\biggr)\nonumber\\
&\hspace{-2mm}\simeq&\hspace{-2mm}\dot{\vec{A}}_a\cdot \vec{\pi}^{a}+\dot{\lambda}^{a}\pi_a+\dot{P}^{a}\overline{C}_a
+\dot{C}^{a}\overline{P}_a. \nonumber
\end{eqnarray}
Using the previous formula, the BFV action becomes:
\begin{displaymath}
S^{\scriptscriptstyle \textrm{BFV}}=\int {\textrm d}x\biggl(\frac{1}{2}\xi^{\scriptscriptstyle A}\omega_{\scriptscriptstyle AB}\dot{\xi}^{\scriptscriptstyle B}-H^{\scriptscriptstyle \textrm{BFV}}[\xi]\biggr).
\end{displaymath}
As we said previously, and as it is clear from Ref. \cite{henneaux}, this BFV procedure is a {\it classical} construction (and not necessarily a quantum one), which has automatically generated the GF and the analog of the FP terms. Because it is classical, we can think of using its equations of motion in the construction of the CPI. In this case, as the BFV action contains already the GF, we can explicitly solve the equations of motion without encountering the problem of the previous section. This means that we get one solution out of a definite initial condition, so the CPI procedure can go through without having to add the GF and FP in the measure as we had to do in the $Z^{\scriptscriptstyle \textrm{YM(NAT)}}_{\scriptscriptstyle \textrm{CPI}}$ of the previous section. 

So let us implement the CPI associated with the BFV action $S^{\scriptscriptstyle \textrm{BFV}}$. As usual, the initial  expression is given by a functional Dirac delta over the classical solutions of the equations of motion:
\begin{equation}
Z_{\scriptscriptstyle \textrm{CPI}}^{\scriptscriptstyle \textrm{YM(BFV)}}=\int {\cal D}\xi^{\scriptscriptstyle A}\, \delta(\xi^{\scriptscriptstyle A}-\xi^{\scriptscriptstyle A}_{\textrm{cl}}). \label{3.10}
\end{equation}
In the next step we have to pass from (\ref{3.10}) to the Dirac deltas containing the equations of motion. 
In doing that, we must take into account that we have both Grassmannian even ($x$) and Grassmann odd variables ($\theta$) among the fields $\xi^{\scriptscriptstyle A}$. Let us consider a generic change of variables 
\begin{displaymath}
\begin{pmatrix} 
x \cr \theta 
\end{pmatrix}=\begin{pmatrix}A & B\cr
         C & D\cr  \end{pmatrix}
\begin{pmatrix} x^{\prime} \cr \theta^{\prime} \end{pmatrix}
\end{displaymath}
where $\displaystyle
A=\frac{\partial x}{\partial x^{\prime}}, \; \; B=\frac{\partial x}{\partial \theta^{\prime}}, \; \; 
C=\frac{\partial \theta}{\partial x^{\prime}}, \; \; D=\frac{\partial \theta}{\partial \theta^{\prime}}$. The blocks $A$ and $D$ are Grassmannian even, while the blocks $B$ and $C$ are Grassmannian odd. 
The result of this change of variables in the integration measure is the following \cite{dewitt}: 
\begin{equation}
\displaystyle \int\prod {\textrm d}x_i\prod {\textrm d}\theta_{\mu}=
 \int\prod {\textrm d}x^{\prime}_{i}\prod \textrm{d}\theta^{\prime}_{\mu}\,
\textrm{sdet}\begin{pmatrix}A & B\cr
         C & D\cr \end{pmatrix}.   \label{sdet}
\end{equation}
The expression $\textrm{sdet}$ means ``superdeterminant" \cite{dewitt} and its definition is given by 
\begin{displaymath}
\textrm{sdet}\begin{pmatrix}A & B\cr
         C & D\cr \end{pmatrix}\equiv \textrm{det}(A-BD^{-1}C) \textrm{det}^{-1}(D).
\end{displaymath}
Let us notice that, when only Grassmannian even variables are present, the superdeterminant reduces to an ordinary determinant while, in presence of only Grassmannian odd variables, the superdeterminant becomes, as usual, the inverse of a determinant. In the case of our CPI we can say that, in going from the Dirac delta of the solutions to the Dirac delta of the equations of motion, a functional superdeterminant appears:
\begin{equation}
\delta \left(\xi^{\scriptscriptstyle A}-\xi^{\scriptscriptstyle A}_{\textrm{cl}}\right)=\delta \left(\dot{\xi}^{\scriptscriptstyle A}-\omega^{\scriptscriptstyle AB}\overrightarrow{\partial}_{\scriptscriptstyle B}H^{\scriptscriptstyle \textrm{BFV}}(\xi)\right)
\textrm{sdet}\left[\delta({\bf x}-{\bf y})\delta^{\scriptscriptstyle A}_{\scriptscriptstyle B}\partial_t -\omega^{\scriptscriptstyle AC}\frac{\overrightarrow{\partial}}{\partial \xi^{\scriptscriptstyle C}({\bf x})}H^{\scriptscriptstyle BFV}\frac{\overleftarrow{\partial}}{\partial \xi^{\scriptscriptstyle B}({\bf y})}\right].  \label{311bis}
\end{equation}
The next step is to exponentiate the equations of motion and the superdeterminant. The equations of motion can be exponentiated by using the following formula:
\begin{equation}
\widetilde{\delta}\left(\dot{\xi}^{\scriptscriptstyle A}-\omega^{\scriptscriptstyle AB}\overrightarrow{\partial}_{\scriptscriptstyle B}H^{\scriptscriptstyle \textrm{BFV}}\right)=\int {\cal D}\Lambda\,\exp\biggl[
i\int {\textrm d}x\, \Lambda_{\scriptscriptstyle A}\biggl(\dot{\xi}^{\scriptscriptstyle A}-\omega^{\scriptscriptstyle AB}\overrightarrow{\partial}_{\scriptscriptstyle B}H^{\scriptscriptstyle \textrm{BFV}}\biggr)\biggr], \label{311tris}
\end{equation}
where $\Lambda$ is a field with the same Grassmann parity of the field it refers to: $[\Lambda_{\scriptscriptstyle A}]=[\xi^{\scriptscriptstyle A}]=[A]$. Also the superdeterminant can be exponentiated by using a generalized Gaussian integral, see Appendix B for details:
\begin{eqnarray}
&& \textrm{sdet}\left[\delta({\bf x}-{\bf y})\delta^{\scriptscriptstyle A}_{\scriptscriptstyle B}\partial_t -\omega^{\scriptscriptstyle AC}\frac{\overrightarrow{\partial}}{\partial \xi^{\scriptscriptstyle C}({\bf x})}H^{\scriptscriptstyle BFV}\frac{\overleftarrow{\partial}}{\partial \xi^{\scriptscriptstyle B}({\bf y})}\right] \nonumber\\
&& \qquad
=\int {\cal D} \overline{\Gamma}_{\scriptscriptstyle A} {\cal D} \Gamma^{\scriptscriptstyle A}\exp\left[-\int \textrm{d}x \, \overline{\Gamma}_{\scriptscriptstyle A}(\partial_t
\delta_{\scriptscriptstyle B}^{\scriptscriptstyle A}-\omega^{\scriptscriptstyle AC}\overrightarrow{\partial}_{\scriptscriptstyle C}H\overleftarrow{\partial}_{\scriptscriptstyle B})\Gamma^{\scriptscriptstyle B}\right].
\label{312}
\end{eqnarray}
The auxiliary fields $\Gamma$ and $\overline{\Gamma}$ have a Grassmann parity which is opposite w.r.t. the one of the fields they refer to: $[\Gamma^{\scriptscriptstyle A}]=[\overline{\Gamma}_{\scriptscriptstyle A}]=[\xi^{\scriptscriptstyle A}]+1=[A]+1$.

Collecting all the terms in (\ref{3.10}), (\ref{311bis}), (\ref{311tris}) and (\ref{312}), we get:
\begin{equation}
Z_{\scriptscriptstyle \textrm{CPI}}^{\scriptscriptstyle \textrm{YM(BFV)}}=\int {\cal D} \Lambda_{\scriptscriptstyle A} {\cal D} \xi^{\scriptscriptstyle A} {\cal D} \overline{\Gamma}_{\scriptscriptstyle A} {\cal D} \Gamma^{\scriptscriptstyle A}
\exp\left[ i \int {\textrm d}x\, \tilde{\cal L}^{\scriptscriptstyle \textrm{BFV}}\right], \label{CPIYM}
\end{equation}
where the Lagrangian density is given by:
\begin{equation}
\widetilde{{\cal L}}^{\scriptscriptstyle \textrm{BFV}}= \Lambda_{\scriptscriptstyle A}\left[\dot{\xi}^{\scriptscriptstyle A}-\omega^{\scriptscriptstyle AB}\overrightarrow{\partial}_{\scriptscriptstyle B}H\right]
+i\overline{\Gamma}_{\scriptscriptstyle A}\left(\partial_t\delta_{\scriptscriptstyle B}^{\scriptscriptstyle A}-\omega^{\scriptscriptstyle AC}\overrightarrow{\partial}
_{\scriptscriptstyle C}H\overleftarrow{\partial}_{\scriptscriptstyle B}\right)\Gamma^{\scriptscriptstyle B}. \label{superlagr}
\end{equation}
Since the previous Lagrangian density contains only first order time derivatives, we can immediately obtain the following Hamiltonian density: 
\begin{equation}
\widetilde{\cal H}^{\scriptscriptstyle \textrm{BFV}}=\Lambda_{\scriptscriptstyle A}\omega^{\scriptscriptstyle AB}\overrightarrow{\partial}_{\scriptscriptstyle B}H+i\overline{\Gamma}_{\scriptscriptstyle A}\omega^{\scriptscriptstyle AC}\overrightarrow{\partial}
_{\scriptscriptstyle C}H\overleftarrow{\partial}_{\scriptscriptstyle B}\Gamma^{\scriptscriptstyle B}.  \label{sbfv}
\end{equation}
From $\widetilde{{\cal L}}^{\scriptscriptstyle \textrm{BFV}}$ of (\ref{superlagr}) we can derive the following equation for the fields $\Gamma$: 
\begin{equation}
\left(\partial_t\delta_{\scriptscriptstyle B}^{\scriptscriptstyle A}-\omega^{\scriptscriptstyle AC}\overrightarrow{\partial}
_{\scriptscriptstyle C}H\overleftarrow{\partial}_{\scriptscriptstyle B}\right)\Gamma^{\scriptscriptstyle B}=0. \label{Jacfields}
\end{equation}
Like for the point particle \cite{martin}, also in the case of YM theories (\ref{Jacfields}) is the equation of evolution of the Jacobi fields, i.e. the equation of evolution of the first variations of the theory, as we will prove in Appendix C.

The BFV Lagrangian density $\widetilde{\cal L}^{\scriptscriptstyle \textrm{BFV}}$ and Hamiltonian density $\widetilde{\cal H}^{\scriptscriptstyle \textrm{BFV}}$ assume a particularly simple form if we combine all the fields of our theory in a set of real superfields: 
\begin{equation}
\Xi^{\scriptscriptstyle A}\equiv \xi^{\scriptscriptstyle A}+(-)^{\scriptscriptstyle [A]}\theta \Gamma^{\scriptscriptstyle A}+(-)^{\scriptscriptstyle [A]}\bar{\theta}\omega^{\scriptscriptstyle AB}\overline{\Gamma}_{\scriptscriptstyle B}
+i(-)^{\scriptscriptstyle [A]}\bar{\theta}\theta \omega^{\scriptscriptstyle AB}\Lambda_{\scriptscriptstyle B},  \label{supcomp}
\end{equation} 
where $\theta$ and $\bar{\theta}$ are two anticommuting parameters and $(-)^{\scriptscriptstyle [A]}$ indicates the Grassmannian parity of the field $\xi^{\scriptscriptstyle A}$. The fact that such Grassmannian parity appears means that, differently from the point particle case, we have to distinguish between the even superfields defined as: 
\begin{displaymath}
\Xi_{\textrm{even}}^{\scriptscriptstyle A}\equiv \xi^{\scriptscriptstyle A}+\theta \Gamma^{\scriptscriptstyle A}+\bar{\theta}\omega^{\scriptscriptstyle AB}\overline{\Gamma}_{\scriptscriptstyle B}
+i\bar{\theta}\theta \omega^{\scriptscriptstyle AB}\Lambda_{\scriptscriptstyle B}  
\end{displaymath} 
and the odd superfields defined as:
\begin{displaymath}
\Xi_{\textrm{odd}}^{\scriptscriptstyle A}\equiv \xi^{\scriptscriptstyle A}-\theta \Gamma^{\scriptscriptstyle A}-\bar{\theta}\omega^{\scriptscriptstyle AB}\overline{\Gamma}_{\scriptscriptstyle B}
-i\bar{\theta}\theta \omega^{\scriptscriptstyle AB}\Lambda_{\scriptscriptstyle B}. 
\end{displaymath} 
Let us notice that each $\Xi^{\scriptscriptstyle A}$ has the same Grassmann parity of the field $\xi^{\scriptscriptstyle A}$ it is associated with. Using the definition (\ref{supcomp}) we can derive (up to surface terms for the Lagrangian) the following relations:
\begin{equation}
\displaystyle
\int \textrm{d}{\bf x} \, \widetilde{\cal H}^{\scriptscriptstyle \textrm{BFV}}=i\int {\textrm{d}}\theta \textrm{d}\bar{\theta}H^{\scriptscriptstyle \textrm{BFV}}(\Xi), \qquad \quad 
\int \textrm{d}{\bf x} \, \widetilde{\cal L}^{\scriptscriptstyle \textrm{BFV}}=i\int {\textrm{d}}\theta \textrm{d}\bar{\theta}L^{\scriptscriptstyle \textrm{BFV}}(\Xi). \label{lagham}
\end{equation}
The derivation of these formulae is not presented here in details because it is the same, modulo some grading factors, as the one contained in Ref. \cite{dequantization} for the point particle. 
So, in order to pass from the BFV Hamiltonian $H^{\scriptscriptstyle \textrm{BFV}}$ of Eq. (\ref{BFVh}) and Lagrangian $L^{\scriptscriptstyle \textrm{BFV}}$ of Eq. (\ref{azionediBFV}) to $\widetilde{\cal H}^{\scriptscriptstyle \textrm{BFV}}$ and $\widetilde{\cal L}^{\scriptscriptstyle \textrm{BFV}}$ respectively, it is sufficient to replace in $H^{\scriptscriptstyle \textrm{BFV}}$ and $L^{\scriptscriptstyle \textrm{BFV}}$ the fields $\xi^{\scriptscriptstyle A}$ with the superfields $\Xi^{\scriptscriptstyle A}$ and then perform the Grassmann integration $i\int {\textrm{d}}\theta {\textrm{d}}\bar{\theta}$. 

Since we will make an intensive use of these superfields later on, it can be useful to write their explicit form using the following symbols:
\begin{eqnarray}
\widetilde {A}^a_k&\hspace{-2mm}\equiv&\hspace{-2mm}A^a_k+\theta\Gamma^{A^a_k}+\bar{\theta}\overline{\Gamma}_{\pi^{k}_a}+i\bar{\theta}\theta\Lambda_{\pi^{k}_a}\nonumber\\
\widetilde {\pi}^k_a&\hspace{-2mm}\equiv&\hspace{-2mm}\pi^k_a+\theta\Gamma^{\pi^k_a}-\bar{\theta}\overline{\Gamma}_{A^{a}_k}-i\bar{\theta}\theta\Lambda_{A^{a}_k}\nonumber\\
\widetilde {\lambda}^a&\hspace{-2mm}\equiv&\hspace{-2mm}\lambda^a+\theta\Gamma^{\lambda^a}+\bar{\theta}\overline{\Gamma}_{\pi_a}+i\bar{\theta}\theta\Lambda_{\pi_a}\nonumber\\
\widetilde {\pi}_a&\hspace{-2mm}\equiv&\hspace{-2mm}\pi_a+\theta\Gamma^{\pi_a}-\bar{\theta}\overline{\Gamma}_{\lambda^{a}}-i\bar{\theta}\theta\Lambda_{\lambda^{a}}
\label{neretto}\\
\widetilde {C}^a&\hspace{-2mm}\equiv&\hspace{-2mm}C^a-\theta\Gamma^{C^a}+\bar{\theta}\overline{\Gamma}_{\overline{P}_a}+i\bar{\theta}\theta\Lambda_{\overline{P}_a}\nonumber\\
\widetilde {\overline{C}}_a&\hspace{-2mm}\equiv&\hspace{-2mm}\overline{C}_a-\theta\Gamma^{\overline{C}_a}+\bar{\theta}\overline{\Gamma}_{P^{a}}+i\bar{\theta}\theta\Lambda_{P^{a}}\nonumber\\
\widetilde {P}^a&\hspace{-2mm}\equiv&\hspace{-2mm}P^a-\theta\Gamma^{P^a}+\bar{\theta}\overline{\Gamma}_{\overline{C}_a}+i\bar{\theta}\theta\Lambda_{\overline{C}_a}\nonumber\\
\widetilde {\overline{P}}_a&\hspace{-2mm}\equiv&\hspace{-2mm}\overline{P}_a-\theta\Gamma^{\overline{P}_a}+\bar{\theta}\overline{\Gamma}_{C^{a}}+i\bar{\theta}\theta\Lambda_{C^{a}}. \nonumber
\end{eqnarray}

\section{Equivalence of the CPIs for Yang-Mills theories} 
 
In the two previous sections we have implemented two different CPIs for YM theories, the ``natural" and the BFV one. 
The presence of more than one path integral both at the classical and at the quantum level is due to the fact that in field theories there exist different procedures to fix the gauge, like the BFV and the FP ones. In this section we want to convince the reader that the two different functional integrals that we have obtained for the CPI are equivalent. 

First of all, let us prove that the equations of motion that can be derived from the standard FP effective action\footnote{We call ``FP effective action" the full action, with GF terms and FP ghosts that one gets in the quantum path integral if one uses the FP procedure \cite{henneaux},\cite{Lee}.}  and from the BFV action are basically the same. If we choose the Lorentz gauge $-\partial^{\mu}A_{\mu}^a=0$, the FP effective action becomes \cite{henneaux}, \cite{Lee}:
\begin{equation}
\displaystyle
L^{\scriptscriptstyle \textrm{FP}}=-\frac{1}{4}F^{a}_{\mu\nu}F^{\mu\nu}_a-\pi_a\partial^{\mu}A^{a}_{\mu}
-i\partial^{\mu}\psi_a\partial_{\mu}\eta^a+i\partial^{\mu}\psi_aC^{a}_{\;bc}\eta^cA^b_{\mu}, \label{Lagr.FP} 
\end{equation}
where $\pi_a$ are the auxiliary variables used to exponentiate the GF, $\psi_a$ and $\eta^a$ are the Grassmannian odd variables used to exponentiate the FP determinant and $C^{a}_{\;bc}$ are the structure constants of the gauge group under which $-\frac{1}{4}F^{a}_{\mu\nu}F^{\mu\nu}_a$  is invariant. In the previous section we have written the BFV equations of motion in a symplectic form and, if we want to compare the two procedures, i.e. the FP and the BFV one, we should write also the equations of motion that can be derived from (\ref{Lagr.FP}) in a symplectic form. To do this, we must derive from (\ref{Lagr.FP}) the associated Hamiltonian. We can start from the definition of the energy-momentum tensor containing  both Grassmannian even and odd variables:
\begin{displaymath}
\displaystyle
\Theta^{\mu\nu}=-\frac{\partial L^{\scriptscriptstyle \textrm{FP}}}{\partial(\partial_{\mu}A^{a}_{\lambda})}
\partial^{\nu}A^{a}_{\lambda}+\frac{\partial L^{\scriptscriptstyle \textrm{FP}}}{\partial(\partial_{\mu}\psi_a)}
\partial^{\nu}\psi_{a}+\frac{\partial L^{\scriptscriptstyle \textrm{FP}}}{\partial(\partial_{\mu}\eta^a)}
\partial^{\nu}\eta^{a}+\eta^{\mu\nu}L^{\scriptscriptstyle \textrm{FP}}. 
\end{displaymath}
In particular, the $00$ component gives:
\begin{eqnarray}
\displaystyle
\Theta^{\scriptscriptstyle 00}&\hspace{-2mm}=\hspace{-2mm}&\vec{\pi}_a\cdot \dot{\vec{A}}^a-i\dot{\eta}^a\dot{\psi}_a+\frac{1}{4}
F^{a}_{\mu\nu}F^{\mu\nu}_a \nonumber\\
&&+\pi_a\vec{\nabla}\cdot \vec{A}^{a}+i\vec{\nabla}\psi_a\cdot\vec{\nabla}\eta^a-i\vec{\nabla}\psi_a\cdot \vec{A}^b
C^{a}_{\;bc}\eta^c \label{alfa}.
\end{eqnarray}
From the definition of the conjugate momenta $\displaystyle \vec{\pi}_a=\frac{\partial L^{\scriptscriptstyle \textrm{FP}}}{\partial \dot{\vec{A}}^a}$ we get the following equations of motion for $\vec{A}$:
\begin{equation}
\dot{\vec{A}}_a=-\vec{\nabla}A^{\scriptscriptstyle 0}_a+C_a^{\;bc}A_{b0}\vec{A}_c+\vec{\pi}_{a}. \label{beta}
\end{equation}
Similarly, the momentum conjugated to $A^a_{\scriptscriptstyle 0}$ is given by:
\begin{equation}
\displaystyle
\pi_a=\frac{\partial L^{\scriptscriptstyle \textrm{FP}}}{\partial \dot{A}^{a}_{\scriptscriptstyle 0}}.  \label{sigmapii}
\end{equation}
Furthermore, by construction:
\begin{equation}
\displaystyle
\frac{1}{4}F^{a}_{\mu\nu}F^{\mu\nu}_a=\frac{1}{4}F^{a}_{ij}F^{ij}_a-\frac{1}{2}\vec{\pi}_a\cdot \vec{\pi}^{a}. \label{gamma}
\end{equation}
For what concerns the momenta conjugated to the Grassmann variables, we have:
\begin{displaymath}
\displaystyle
\pi_{\psi_a}\equiv \frac{\partial L^{\scriptscriptstyle \textrm{FP}}}{\partial \dot{\psi}_a}, \qquad \quad \pi_{\eta^a}\equiv \frac{\partial L^{\scriptscriptstyle \textrm{FP}}}{\partial \dot{\eta}^a}
\end{displaymath}
from which we easily derive the equations of motion for $\eta^a$ and $\psi_a$:
\begin{equation}
\dot{\eta}^a=-i\pi_{\psi_a}+C^{a}_{\;bc}A^{b}_{\scriptscriptstyle 0}\eta^{c}, \qquad \quad 
\dot{\psi}_a=i\pi_{\eta^a}. \label{delta}
\end{equation}
From the previous equations we note that the FP ghosts evolve just as the BFV ghosts:
\begin{displaymath}
\dot{C}^{a}=-iP^{a}+C^{a}_{\;bc}\lambda^bC^c, \qquad \quad 
\dot{\overline{C}}_a=i\overline{P}_a,
\end{displaymath}
so we can make the following identifications: 
\begin{equation}
C\leftrightarrow\eta,\;\;\qquad \overline{C}\leftrightarrow\psi. \label{iduno}
\end{equation}
Consequently, the variables $P$ and $\overline{P}$, which appear in the BFV formalism, can be identified with the momenta conjugated to the FP ghosts:
\begin{equation}
P\leftrightarrow\pi_{\psi},\;\;\qquad \overline{P}\leftrightarrow\pi_{\eta}. \label{iddue}
\end{equation}
Replacing (\ref{beta})-(\ref{delta}) into (\ref{alfa}), we have:
\begin{eqnarray}
\displaystyle
\Theta^{\scriptscriptstyle 00}&\hspace{-2mm}=\hspace{-2mm}&\vec{\pi}_a\cdot\vec{\nabla}\lambda^{a}+C^{a}_{\;bc} \lambda^b\vec{\pi}_a\cdot\vec{A}^c+\vec{\pi}_a\cdot\vec{\pi}^{a}
+\frac{1}{4}F^{a}_{ij}F^{ij}_a-\frac{1}{2}\vec{\pi}_a\cdot\vec{\pi}^{a}
+\pi_a\vec{\nabla}\cdot\vec{A}^{a}\nonumber\\
&&-(i\pi_{\psi_a}-C^{a}_{\;bc}\lambda^b\eta^c)\pi_{\eta^a}+
i\vec{\nabla}\psi_a\cdot \vec{\nabla}\eta^a-i\vec{\nabla}\psi_a\cdot \vec{A}^b C^{a}_{\;bc}\eta^c. \nonumber
\end{eqnarray}
Up to surface terms, we can identify the previous expression with the following FP Hamiltonian density:
\begin{eqnarray}
\displaystyle
H^{\scriptscriptstyle \textrm{FP}}&\hspace{-2mm}=&\hspace{-2mm} \frac{1}{2}\vec{\pi}_a\cdot\vec{\pi}^{a}+\frac{1}{4}F^{a}_{ij}
F^{ij}_a-\lambda^{a}\vec{\nabla}\cdot\vec{\pi}_a+\lambda^{a}C^b_{\;ac}\vec{\pi}_b\cdot\vec{A}^c+\pi_a\vec{\nabla}\cdot \vec{A}^{a} \nonumber\\
&&-i\pi_{\psi_a}\pi_{\eta^a}+\lambda^{a}C^b_{\;ac}\eta^{c}\pi_{\eta^b}
-i\psi_a\vec{\nabla}\cdot \left[(\vec{\nabla}\delta_b^{a}+C^{a}_{\;bc}\vec{A}^c)\eta^{b}\right], \nonumber
\end{eqnarray}
which coincides with the BFV Hamiltonian of Eq. (\ref{BFVh}) if we use the identifications (\ref{iduno})-(\ref{iddue}). This immediately implies that from the action 
\begin{displaymath}
S=\int {\textrm d}x\Bigl[\dot{\vec{A}}_a\cdot \vec{\pi}^{a}+\dot{\lambda}^{a}\pi_a+\dot{\psi}_a\pi_{\psi_a}+
\dot{\eta}^{a}\pi_{\eta^a}-H^{\scriptscriptstyle \textrm{FP}}\Bigr]
\end{displaymath}
we can derive, by variation with respect to the fields, the same equations of motion (\ref{BFVeq}) that we got via the BFV method. So we can say that the BFV method is somehow a canonical version of the FP method. In fact, either starting from the BFV or the effective FP Lagrangian, we get the same equations of motion in the symplectic form.
As a consequence from their ``exponentiation" we will get the same weight for the CPI, i.e., $\widetilde{\cal L}^{\scriptscriptstyle \textrm{BFV}}=\widetilde{
\cal L}^{\scriptscriptstyle \textrm{FP}}$. Note that $\widetilde{
\cal L}^{\scriptscriptstyle \textrm{FP}}$ is not the weight that appears in $Z^{\scriptscriptstyle \textrm{YM(NAT)}}_{\scriptscriptstyle \textrm{CPI}}$ of Eq. (\ref{ymnat}) because this last CPI was obtained not from the FP effective action but from the standard YM action. 

Finally, we want to prove that the CPI implemented using the BFV procedure
\begin{equation}
Z_{\scriptscriptstyle \textrm{CPI}}^{\scriptscriptstyle \textrm{YM(BFV)}}=\int {\cal D}\mu^{\prime}\,\exp \left[ i\int {\textrm d}x\, \widetilde{\cal L}^{\scriptscriptstyle \textrm{BFV}} \right],
\label{intuno}
\end{equation}
and the one obtained using the ``natural procedure"  $Z^{\scriptscriptstyle \textrm{YM(NAT)}}_{\scriptscriptstyle \textrm{CPI}}$ are equivalent. First of all, we shall prove in Appendix D that
\begin{equation}
\displaystyle L^{\scriptscriptstyle \textrm{BFV}}=L^{\scriptscriptstyle \textrm{FP}}.
\label{comparison}
\end{equation}
In this proof we made use of the equations of motion because in the CPI the system is always ``sitting" on the classical trajectories. Eq. (\ref{comparison}) can help us in giving a further proof of $\widetilde{\cal L}^{\scriptscriptstyle \textrm{BFV}}=\widetilde{
\cal L}^{\scriptscriptstyle \textrm{FP}}$. In fact, using the superfields $\Xi$ of Eq. (\ref{supcomp}), we can pass from the usual BFV Lagrangian density $L^{\scriptscriptstyle \textrm{BFV}}$ to the one which appears in the weight of the BFV classical path integral $\widetilde{\cal L}^{\scriptscriptstyle \textrm{BFV}}$ according to the following equation:
\begin{equation}
\widetilde{\cal L}^{\scriptscriptstyle \textrm{BFV}}=i\int {\textrm{d}}\theta {\textrm{d}}\bar{\theta} \, L^{\scriptscriptstyle \textrm{BFV}}(\Xi). \label{A}
\end{equation}
The same holds for the FP effective action:
\begin{equation}
\widetilde{\cal L}^{\scriptscriptstyle \textrm{FP}}=i\int {\textrm{d}}\theta {\textrm{d}}\bar{\theta} \,L^{\scriptscriptstyle \textrm{FP}}(\Xi). \label{B}
\end{equation}
So from $L^{\scriptscriptstyle \textrm{BFV}}=L^{\scriptscriptstyle \textrm{FP}}$ and Eqs. (\ref{A}) and (\ref{B}) we get again that $\widetilde{\cal L}^{\scriptscriptstyle \textrm{BFV}} = \widetilde{\cal L}^{\scriptscriptstyle \textrm{FP}}$. This $\widetilde{\cal L}^{\scriptscriptstyle \textrm{FP}}$ appears in the weight of $Z_{\scriptscriptstyle \textrm{CPI}}^{\scriptscriptstyle \textrm{YM(FP)}}$:
\begin{equation}
Z_{\scriptscriptstyle \textrm{CPI}}^{\scriptscriptstyle \textrm{YM(FP)}}=\int {\cal D}\mu^{\prime}\,\exp \left[ i\int {\textrm d}x\, \widetilde{\cal L}^{\scriptscriptstyle \textrm{FP}} \right], \label{dagger}
\end{equation}
where ${\cal D}\mu^{\prime}$ means the functional integral over all the fields present in the Lagrangian density $\widetilde{\cal L}^{\scriptscriptstyle \textrm{FP}}$.
Since $\widetilde{\cal L}^{\scriptscriptstyle \textrm{BFV}} = \widetilde{\cal L}^{\scriptscriptstyle \textrm{FP}}$ and since the fields in the measures can be identified, as done in (\ref{iduno}), (\ref{iddue}), we can say that 
$\displaystyle Z_{\scriptscriptstyle \textrm{CPI}}^{\scriptscriptstyle \textrm{YM(BFV)}}=
Z_{\scriptscriptstyle \textrm{CPI}}^{\scriptscriptstyle \textrm{YM(FP)}}$.
Because of this,  if we now want to show that the BFV classical path integral $Z_{\scriptscriptstyle \textrm{CPI}}^{\scriptscriptstyle \textrm{YM(BFV)}}$ is equivalent to the ``{\it natural}" one $Z_{\scriptscriptstyle \textrm{CPI}}^{\scriptscriptstyle \textrm{YM(NAT)}}$ it will be sufficient to prove that the functional integral $\displaystyle Z_{\scriptscriptstyle \textrm{CPI}}^{\scriptscriptstyle \textrm{YM(FP)}}$ in (\ref{intuno}) is equivalent to the one given by Eq. (\ref{ymnat}):
\begin{equation}
\displaystyle Z_{\scriptscriptstyle \textrm{CPI}}^{\scriptscriptstyle \textrm{YM(NAT)}}=\int {\cal D}\mu\,\exp \left[i\int {\textrm d}x \, \widetilde{\cal L}^{\scriptscriptstyle \textrm{NAT}}\right]. \label{intdue}
\end{equation}
The comparison is easier if we divide the weight of the FP classical path integral in three parts: the GF, the ghost and the bosonic part:
\begin{displaymath}
\widetilde{\cal L}^{\scriptscriptstyle \textrm{FP}}=\widetilde{\cal L}^{\scriptscriptstyle \textrm{FP}}_{\textrm{gf}}
+\widetilde{\cal L}^{\scriptscriptstyle \textrm{FP}}_{\textrm{gh}}
+\widetilde{\cal L}^{\scriptscriptstyle \textrm{FP}}_{\textrm{bos}}. 
\end{displaymath}
The three parts can be written in terms of the superfields as: 
\begin{eqnarray}
\displaystyle
\widetilde{\cal L}^{\scriptscriptstyle \textrm{FP}}_{\textrm{gf}}&\hspace{-2mm}=\hspace{-2mm}&i\int {\textrm{d}}\theta {\textrm{d}}\bar{\theta}\, \left\{-\widetilde{\pi}_a\partial^{\mu}
\widetilde{A}^{a}_{\mu}
\right\}\nonumber\\
\widetilde{\cal L}^{\scriptscriptstyle \textrm{FP}}_{\textrm{gh}}&\hspace{-2mm}=\hspace{-2mm}&i\int {\textrm{d}}\theta {\textrm{d}}\bar{\theta}\, \left\{-i\partial^{\mu}\widetilde{\overline{C}}_a
\widetilde{D}^{ba}_{\mu}\widetilde{C}_b\right\}\label{4.13bis} \\
\widetilde{\cal L}^{\scriptscriptstyle \textrm{FP}}_{\textrm{bos}}&\hspace{-2mm}=\hspace{-2mm}&i\int {\textrm{d}}\theta {\textrm{d}}\bar{\theta}\, \left\{-\frac{1}{4}\widetilde{F}^{a}_{\mu\nu}
\widetilde{F}^{\mu\nu}_a\right\} \nonumber.
\end{eqnarray}
As usual, we have indicated with a tilde sign the superfields of Eq. (\ref{neretto}). Similarly, we can divide the natural Lagrangian in the following three parts:
\begin{eqnarray}
\displaystyle
&&\widetilde{\cal L}^{\scriptscriptstyle \textrm{NAT}}_{\textrm{gf}}=-\pi_a\partial^{\mu}A^{a}_{\mu}\nonumber \bigskip \\
&&\widetilde{\cal L}^{\scriptscriptstyle \textrm{NAT}}_{\textrm{gh}}=-i\partial^{\mu}\psi_a D^{ba}_{\mu}\eta_b\nonumber \bigskip \\
&&\widetilde{\cal L}^{\scriptscriptstyle \textrm{NAT}}_{\textrm{bos}}=\Lambda_{\scriptscriptstyle A}(\dot{\phi}^{\scriptscriptstyle A}-\omega^{\scriptscriptstyle AB}\partial_{\scriptscriptstyle B}H)+i
\overline{\Gamma}_{\scriptscriptstyle A}(\delta_{\scriptscriptstyle B}^{\scriptscriptstyle A}\partial_t-\omega^{\scriptscriptstyle AC}\partial_{\scriptscriptstyle B}\partial_{\scriptscriptstyle C}H)\Gamma^{\scriptscriptstyle B}. \nonumber
\end{eqnarray} 
The main obstacle in proving that the two functional integrals 
(\ref{dagger}) and (\ref{intdue}) are equivalent is that the number of variables in $\widetilde{\cal L}^{\scriptscriptstyle \textrm{FP}}$ is larger than the number of variables in $\widetilde{\cal L}^{\scriptscriptstyle \textrm{NAT}}$, in particular for what concerns the GF and the ghost parts.
For example, the generating functional $Z_{\scriptscriptstyle \textrm{CPI}}^{
\scriptscriptstyle \textrm{YM(NAT)}}$ lacks all the functional integrals over the CPI ghosts $\Gamma$, $\overline{\Gamma}$ and $\Lambda$ associated with the BFV ghosts. However it is possible to prove that all the extra fields present in $Z_{\scriptscriptstyle \textrm{CPI}}^{\scriptscriptstyle \textrm{YM(FP)}}$ can be integrated away to get just $Z_{\scriptscriptstyle \textrm{CPI}}^{\scriptscriptstyle \textrm{YM(NAT)}}$. Since the proof is rather long, we have confined it in Appendix E. 

So we can conclude this section by saying that the two different CPIs for YM theories that we have implemented in the previous sections, the BFV and the ``{\it natural}" one, are equivalent.

\section{Universal symmetries of the CPI for Yang-Mills theories}

In the previous section, we have proved that the CPIs derived using the ``{\it natural}" and the BFV procedures are equivalent, so in the study of the symmetries of the YM CPI we can concentrate ourselves on (\ref{CPIYM}), which was obtained by implementing the BFV method. First of all, let us derive from (\ref{CPIYM}) the graded commutators of the theory. To do that we use the standard method developed by Feynman \cite{RMP} for the standard commutators of quantum mechanics. In our case from (\ref{CPIYM}) we get:
\begin{displaymath}
\displaystyle
\left[\xi^{\scriptscriptstyle B}(t,{\bf x}),\Lambda_{{\scriptscriptstyle A}}(t,{\bf y})\right]=i\delta_{\scriptscriptstyle A}^{\scriptscriptstyle B}\delta({\bf x}-{\bf y}), \qquad
\left[\Gamma^{{\scriptscriptstyle B}}(t,{\bf x}),\overline{\Gamma}_{{\scriptscriptstyle A}}(t,{\bf y})\right]=\delta_{\scriptscriptstyle A}^{\scriptscriptstyle B}\delta({\bf x}-{\bf y}) \label{comm2}.
\end{displaymath}
All the other commutators are zero. In particular, the commutator among the fields $\xi$ is zero: $[\xi^{\scriptscriptstyle A},\xi^{\scriptscriptstyle B}]=0$. We know that the CPI involves a number of fields which is larger than the one needed to describe the system, so, as it emerges from \cite{martin}, this redundancy in the description of the system implies the presence of a certain number of symmetries.
We can immediately generalize those symmetries from the CPI of the point particle to the one for YM theories. For example, the BRS charge, derived in \cite{martin} for the point particle, becomes:
\begin{equation}
Q(t)=i\int {\textrm d}{\bf x} \, \Gamma^{\scriptscriptstyle A}(t,{\bf x})\Lambda_{\scriptscriptstyle A}(t,{\bf x}). \label{caricaqu}
\end{equation}
As usual, $Q$ is Grassmannian odd and nilpotent and this is the reason why we called it a BRS charge. In spite of its name, it has nothing to do with the BRS charges \cite{becchi} of gauge theories. In fact the BRS charge of the CPI made its appearance even in the point particle case \cite{martin} where no gauge field was present. There  \cite{martin} we showed that the CPI-BRS charge could be identified with the exterior derivative \cite{libromarsden} on the phase space. In the case of YM theories the phase space is built out of the whole $\xi^{\scriptscriptstyle A}$ variables (which include the gauge ghosts besides the gauge fields), so the BRS charge (\ref{caricaqu}) becomes the exterior derivative on this enlarged space\footnote{We will later show that in the case of the CPI for YM we have, besides this universal BRS charge, another one related to gauge theories (see Sec. 5.1).}. In the point particle case there was another Grassmannian odd and nilpotent charge which was called \cite{martin} the anti-BRS charge and that in the CPI for YM theories becomes:
\begin{equation}
\overline{Q}(t)=-i\int {\textrm d}{\bf x}\, \Lambda_{\scriptscriptstyle A}(t,{\bf x})\omega^{\scriptscriptstyle AB}\overline{\Gamma}_{\scriptscriptstyle B}(t,{\bf x}).
\label{antiqu}
\end{equation}
The charges $Q$ and $\overline{Q}$ anticommute among themselves and commute with the Hamiltonian $\int \textrm{d}{\bf x}\, \widetilde{\cal H}^{\scriptscriptstyle \textrm{BFV}}$, i.e. they are conserved charges. 

In the CPI for a point particle there were three further conserved charges $K$, $\bar{K}$, $Q_{\textrm{f}}$, which were bilinear in the ghosts and that together with $Q$ and $\bar{Q}$ made an ISp(2) algebra. What can we say in the case of YM theories? The expression \cite{martin} of those charges must be slightly modified in the gauge case in order to preserve the ISp(2) algebra. They are: 
\begin{equation}
\displaystyle K = -\int {\textrm d}{\bf x}\, \frac{1}{2}\Gamma^{\scriptscriptstyle A}\omega_{\scriptscriptstyle BA}\Gamma^{\scriptscriptstyle B}, \quad
\overline{K} = \int {\textrm d}{\bf x}\, \frac{1}{2}\overline{\Gamma}_{\scriptscriptstyle A}\omega^{\scriptscriptstyle AB}\overline{\Gamma}_{\scriptscriptstyle B},  \quad 
Q_{\textrm{f}}  =  -\int {\textrm d}{\bf x}\, \overline{\Gamma}_{\scriptscriptstyle A}\Gamma^{\scriptscriptstyle A}. \label{othercharges}
\end{equation}
These three charges, modulo central charges, close on the Sp(2) algebra: 
\begin{equation}
[Q_{\textrm{f}}, K]=2K, \;\;\; [Q_{\textrm{f}},\overline{K}]=-2\overline{K}, \;\;\; 
[K,\overline{K}]=Q_{\textrm{f}} \label{spdue}
\end{equation}
and they are all conserved. Furthermore, their graded commutators with the BRS and anti-BRS charges reproduce the inhomogeneous part of the ISp(2) algebra: 
\begin{eqnarray}
&&[Q_{\textrm{f}},Q]=Q, \;\;\; [Q_{\textrm{f}},\overline{Q}]=-\overline{Q}, \;\;\; [K,Q] = 0, \medskip \nonumber\\
&&[K,\overline{Q}]=Q, \;\;\; [\overline{K},Q]=\overline{Q}, \;\;\; [\overline{K},\overline{Q}] = 0. \label{ispdue}
\end{eqnarray}
Like for the point particle, also in the YM case the symmetry transformations generated by the previous charges can be represented in a very compact way by introducing the superfield formalism. As we have seen in Eq. (\ref{supcomp}), the superfield associated with $\xi^{\scriptscriptstyle A}$ is given by: 
\begin{equation}
\Xi^{\scriptscriptstyle A}\equiv \xi^{\scriptscriptstyle A}+(-)^{\scriptscriptstyle [A]}\theta \Gamma^{\scriptscriptstyle A}+(-)^{\scriptscriptstyle [A]}\bar{\theta}\omega^{\scriptscriptstyle AB}\overline{\Gamma}_{\scriptscriptstyle B}
+i(-)^{\scriptscriptstyle [A]}\bar{\theta}\theta \omega^{\scriptscriptstyle AB}\Lambda_{\scriptscriptstyle B}. \label{sup}
\end{equation}
Besides connecting the weights of the classical and quantum path integrals according to Eq. (\ref{lagham}), the definition of the superfields (\ref{sup}) allows us to represent the generators of the ISp(2) algebra in terms of differential operators \cite{martin} which act on the superspace $(t,\theta,\bar{\theta})$ in the following way:
\begin{equation}
\displaystyle
\hat{Q}=-\frac{\partial}{\partial \theta}, \;\;\; \hat{\overline{Q}}
=\frac{\partial}{\partial\bar{\theta}}, \;\;\;
\hat{K}=\bar{\theta}\frac{\partial}{\partial\theta}, \;\;\;
\hat{\overline{K}}=\theta\frac{\partial}{\partial\bar{\theta}}, \;\;\;
\hat{Q}_{\textrm{f}}=\bar{\theta}\frac{\partial}{\partial\bar
{\theta}}-\theta\frac{\partial}{\partial\theta}. \label{opYM}
\end{equation}
The previous operators generate the correct transformations according to the standard rule:
\begin{displaymath}
\delta\Xi^{\scriptscriptstyle A}=-\epsilon\hat{\Omega}\Xi^{\scriptscriptstyle A}, 
\end{displaymath}
where $\hat{\Omega}$ is one of the operators of Eq. (\ref{opYM}). These operators satisfy the same algebra of the associated charges. Before going on, let us notice that the operatorial representation of the ISp(2) charges on the superspace in the YM case is the same as the one of the original point particle formalism \cite{martin}, provided we change the definition of the superfields by introducing the suitable grading factors of Eq. (\ref{sup}).

\subsection{The BRS-BFV charge}

In the previous sections we have used the superfield formalism to derive in a simple and elegant way $\widetilde{\cal H}^{\scriptscriptstyle \textrm{BFV}}$ and $\widetilde{\cal L}^{\scriptscriptstyle \textrm{BFV}}$ from $H^{\scriptscriptstyle \textrm{BFV}}$ and $L^{\scriptscriptstyle \textrm{BFV}}$, see Eq. (\ref{lagham}), and to write in a compact way an operatorial representation for the generators of the ISp(2) algebra, see Eq. (\ref{opYM}). The superfields have also a third useful property: they allow us to write immediately a second conserved BRS charge. This charge is very different from the universal charges (\ref{caricaqu})-(\ref{othercharges}) that we have considered up to now. It is in fact related to the original gauge invariance of the theory and it exists only for the CPI of gauge theories. 

As we have already mentioned, even after the gauge-fixing procedure, in the quantum field theory a memory of the original gauge invariance still remains under the form of a symmetry generated by a gauge BRS charge \cite{becchi}. This invariance is present at the QM level to guarantee that the theory is independent of the particular choice of the gauge-fixing we may have made. Something similar should happen also at the CPI level because also there we have a GF but at the same time we want everything physical be independent of which form of the GF we choose. To derive the form of the gauge BRS charge in the CPI formalism, let us remember Eq. (\ref{lagham}): there we have constructed the BFV-CPI Hamiltonian\footnote{From now on, we shall not indicate the dependence of the fields on ${\bf x}$ and the integral over ${\bf x}$. We will also omit the index BFV on $H$ and $\widetilde{\cal H}$.} $\widetilde{\cal H}$ out of $H$ by replacing the fields with the superfields and performing the integration over the Grassmann partners of time $\theta$ and $\bar{\theta}$. Applying the same procedure to the charge $\Omega$ of Eq. (\ref{caricaomega}), we get the following charge in the enlarged space of the CPI:
\begin{displaymath}
\widetilde{\Omega}=i\int {\textrm{d}}\theta {\textrm{d}}\bar{\theta}\;\Omega(\Xi). 
\end{displaymath}
With a calculation identical to the one performed for the Hamiltonian, we get:
\begin{equation}
\widetilde{\Omega}=\Lambda_{\scriptscriptstyle A} \omega^{\scriptscriptstyle AB}\overrightarrow{\partial}_{\scriptscriptstyle B}\Omega-i\overline{\Gamma}_{\scriptscriptstyle A}\omega^{\scriptscriptstyle AC}\overrightarrow{\partial}_{\scriptscriptstyle C}
\Omega\overleftarrow{\partial}_{\scriptscriptstyle B}\Gamma^{\scriptscriptstyle B}. \label{secondacaricaBRS}
\end{equation}
Let us notice that, due to the Grassmannian odd character of $\Omega$, in (\ref{secondacaricaBRS}) there appears a minus relative sign in the second piece. 

Now, let us consider two different generic functions, $O_{\scriptscriptstyle 1}$ and $O_{\scriptscriptstyle 2}$, of the standard BFV phase space $\xi^{\scriptscriptstyle A}$ with Poisson brackets\footnote{The Poisson brackets are defined as: 
$\displaystyle \left\{ O_{\scriptscriptstyle 1}, O_{\scriptscriptstyle 2} \right\}_{\scriptscriptstyle \textrm{pb}} =O_{\scriptscriptstyle 1} \frac{\overleftarrow{\partial}}{\partial \xi^{\scriptscriptstyle A}}\omega^{\scriptscriptstyle AB}\frac{\overrightarrow{\partial}}{\partial \xi^{\scriptscriptstyle B}}O_{\scriptscriptstyle 2}.$}
$\left\{O_{\scriptscriptstyle 1},O_{\scriptscriptstyle 2}\right\}_{\scriptscriptstyle \textrm{pb}}=O_{\scriptscriptstyle 3}$.
As we will prove in Appendix F, the associated functions in the enlarged space of the CPI \cite{dequantization}:
\begin{equation}
\displaystyle \widetilde{O}_i=\Lambda_{\scriptscriptstyle A}\omega^{\scriptscriptstyle AB}\overrightarrow{\partial}_{\scriptscriptstyle B}O_i+i\overline{\Gamma}_{\scriptscriptstyle A}\omega^{\scriptscriptstyle AB}\overrightarrow{\partial}_{\scriptscriptstyle B}O_i\overleftarrow{\partial}_{\scriptscriptstyle C}\Gamma^{\scriptscriptstyle C}(-1)^{\scriptscriptstyle [O_i]} \label{definitions}
\end{equation}
satisfy the commutator 
\begin{equation}
\bigl[\widetilde{O}_{\scriptscriptstyle 1},\widetilde{O}_{\scriptscriptstyle 2}\bigr]=i\widetilde{O}_{\scriptscriptstyle 3}. \label{extcom}
\end{equation}
These functions $\widetilde{O}_i$ are nothing else \cite{martin}-\cite{math} than the Lie derivative \cite{libromarsden} of the Hamiltonian flow generated by $O_i$. The procedure to pass from $O_i$ to $\widetilde{O}_i$ is analog to the momentum map \cite{marsden} or its inverse. 
As a particular case, let us come back to the BRS-BFV charge $\Omega$ of Eq. (\ref{caricaomega}). It is conserved in the standard BFV space because $\left \{\Omega, H \right \}=0$. From (\ref{extcom}) this immediately implies that the commutator between the associated Lie derivatives is zero:
\begin{displaymath}
\bigl[ \widetilde{\Omega},\widetilde{\cal H}\bigr]=0.
\end{displaymath}
Since $\widetilde{\cal H}$ generates the time evolution in the extended CPI space, we can conclude that $\widetilde{\Omega}$ is a conserved charge. With similar calculations, the nilpotency of $\Omega$ implies the nilpotency of $\widetilde{\Omega}$: 
\begin{displaymath}
\bigl\{\Omega,\Omega\bigr\}=0 \, \Rightarrow \, \left[\widetilde{\Omega},\widetilde{\Omega}\right]=0.
\end{displaymath}
$\widetilde{\Omega}$ will be indicated as the BRS-BFV charge, to distinguish it from the other BRS charges, like $Q$, $\bar{Q}$, present in the CPI of every model. We would like to stress that, without using the superfield formalism, it would have been very difficult to find complicated charges like (\ref{secondacaricaBRS}) and to understand that they are symmetries for the Hamiltonian $\widetilde{\cal H}$.

It is a long but straightforward calculation to prove that the graded commutator of $\widetilde{\Omega}$ with the ISp(2) charges $Q,\overline{Q},K,\overline{K},Q_{\textrm{f}}$ is zero. In particular, $Q$ commutes with $\widetilde{\Omega}$. This is crucial because \cite{martin} $Q$ can be interpreted as an exterior derivative and we know that the exterior derivative is coordinate free, so it must be also gauge invariant. This is guaranteed just by the commutator $\left[Q,\widetilde{\Omega}\right]=0$. The main steps of the proof are given in Appendix G.

Before going on, we want to study the different physical meaning of the two symmetry charges $Q$ and $\widetilde{\Omega}$. If we write in a table all the fields of the CPI obtained implementing the BFV method we get: 
\begin{center}
$\widetilde{\Omega}$ \\
\vspace{-0.2cm}
\begin{picture}(240,3)
\psline[]{->}(2.1, 0)(6.2, 0)
\end{picture}
\begin{picture}(240,3)
\psline[]{->}(1.15, -0.2)(1.15, -1.6)
\end{picture}
\end{center}
\vspace{-0.9cm} 
\begin{tabbing}
$\hspace{5cm}$ \= ${\bar{\Gamma}_{A^a_k}}$ \= ${\bar{\Gamma}_{\pi^k_a}}$ \=  $\bar{\Gamma}_{\lambda_a}$ \= $\bar{\Gamma}_{\pi_a}$ \=
$\bar{\Gamma}_{C^a}$ \= $\bar{\Gamma}_{\bar{C}_a}$\= $\bar{\Gamma}_{P_a}$ \= $\bar{\Gamma}_{\bar{P}_a}$  \kill 
$\hspace{5cm}$ \> ${A^a_k}$ \> ${\pi^k_a}$ \> ${\lambda^a}$ \> ${\pi_a}$ \>
 ${C^a}$ \> ${\bar{C}_a}$ \> ${P_a}$ \> ${\bar{P}_a}$ \\
 $\hspace{4cm} Q $ \> ${\Gamma^{A^a_k}}$ \> ${\Gamma^{\pi^k_a}}$ \> ${\Gamma^{\lambda_a}}$ \> ${\Gamma^{\pi_a}}$ \>
 ${\Gamma^{C^a}}$ \> ${\Gamma^{\bar{C}_a}}$ \> ${\Gamma^{P_a}}$ \> ${\Gamma^{\bar{P}_a}}$ \\
 $\hspace{5cm}$ \> ${\bar{\Gamma}_{A^a_k}}$ \> ${\bar{\Gamma}_{\pi^k_a}}$ \> ${\bar{\Gamma}_{\lambda_a}}$ \> 
 ${\bar{\Gamma}_{\pi_a}}$ \>
 ${\bar{\Gamma}_{C^a}}$ \> ${\bar{\Gamma}_{\bar{C}_a}}$ \> ${\bar{\Gamma}_{P_a}}$ \> 
 ${\bar{\Gamma}_{\bar{P}_a}}$ \\
$\hspace{5cm}$ \>  ${\bar{\Lambda}_{A^a_k}}$ \> ${\bar{\Lambda}_{\pi^k_a}}$ \> ${\bar{\Lambda}_{\lambda_a}}$ \> 
 ${\bar{\Lambda}_{\pi_a}}$ \>
 ${\bar{\Lambda}_{C^a}}$ \> ${\bar{\Lambda}_{\bar{C}_a}}$ \> ${\bar{\Lambda}_{P_a}}$ \> 
 ${\bar{\Lambda}_{\bar{P}_a}}$
 \end{tabbing}
The action of the extended BRS-BFV charge $\widetilde{\Omega}$ on the fields of the first row via the CPI commutators is identical to the action of the charge $\Omega$ via the BFV Poisson brackets. In fact, they both generate the gauge-BRS transformations which mix only the fields which appear in the first row of the table. The other BRS charge $Q$ instead turns the fields of the first row into the associated Jacobi fields $\Gamma$, which are placed in the second row of the table, and the fields of the third row into those of the 4th. The anti-BRS $\bar{Q}$ instead turns the variables of the first row into those of the third row and the variables of the second row into those of the fourth. So if the charge $\widetilde{\Omega}$ allows us to move horizontally throughout the table, the charge $Q$ moves us only in the vertical direction. We want to stress again that, even if we indicated both charges as BRS charges, the physical meaning of $\widetilde{\Omega}$ and $Q$ is completely different: one comes from the gauge invariance, while the other comes from the pure CPI construction. In the original BFV theory \cite{hwang} there were other conserved charges besides $\Omega$. By ``original BFV theory" we mean the phase space $\xi^{\scriptscriptstyle A}$ and the Hamiltonian $H^{\scriptscriptstyle \textrm{BFV}}$ with its associated Poisson brackets. Those charges were an anti BRS-BFV charge: 
\begin{equation}
\displaystyle \overline{\Omega}=i\overline{P}_a\pi_a+\sigma_a\overline{C}_a+\frac{1}{2}\overline{C}_a\overline{C}_bC_c^{\; ab}P^c-\overline{P}_bC_{\;\;\;c}^{ba}C^c\overline{C}_a+\pi_a C_b^{\;ac}\lambda^b\overline{C}_c, \label{primaditre}
\end{equation}
a ghost BFV charge
\begin{equation}
\displaystyle \Omega_{\textrm{g}}=iP^a\overline{C}_a+iC^a\overline{P}_a, \label{secondaditre}
\end{equation}
and the analog of $K$ and $\overline{K}$ which are given by:
\begin{equation}
\displaystyle {\cal K}=-iP^dC^d+\frac{1}{2}C^{\;\;\;d}_{ef}\lambda_dC^eC^f, \qquad  
\overline{\cal K}=-i\overline{P}_d\overline{C}_d-\frac{1}{2}C^{efd}\lambda_d\overline{C}_f\overline{C}_e. \label{ultimaditre}
\end{equation}
The five charges of Eqs. (\ref{caricaomega}), (\ref{primaditre})-(\ref{ultimaditre}) satisfy another ISp(2) algebra. Out of these five charges we can build  the associated CPI charges defined via Eq. (\ref{definitions}), let us call them $\widetilde{\Omega}$, $\widetilde{\overline{\Omega}}$, $\widetilde{\Omega}_{\textrm{g}}$, $\widetilde{\cal K}$ and $\widetilde{\overline{\cal K}}$. It is clear from Eq. (\ref{extcom}) that these five charges close on an ISp(2) algebra in the enlarged space of the CPI and that they are conserved under the $\widetilde{\cal H}$ of Eq. (\ref{sbfv}). In Ref. \cite{third}, in connection with the topological field theory, we will also work out the various cohomologies associated to our $Q$ and $\widetilde{\Omega}$ charges. 

\section{Supersymmetry in classical mechanics}

In the second paper of Ref. \cite{martin} it was proved that in the CPI for the point particle, besides the universal symmetries we presented so far, there was also a $N=2$ supersymmetry \cite{enrico}. The charges associated to this supersymmetry depend explicitly on $H$, differently than the ones in Eq. (\ref{ispdue}). We want to prove that it is possible to find the analog of these charges for YM theory provided we introduce suitable grading factors. First of all, let us notice that the following two quantities:
\begin{displaymath}
N\equiv \Gamma^{\scriptscriptstyle A}\overrightarrow{\partial}_{\scriptscriptstyle A}H(-)^{\scriptscriptstyle [A]}, \qquad \quad 
\overline{N}\equiv \overline{\Gamma}_{\scriptscriptstyle B}\, \omega^{\scriptscriptstyle BC}\overrightarrow{\partial}_{\scriptscriptstyle C}H
\end{displaymath}
are conserved because they commute with $\widetilde{\cal H}$. 

Let us now build the following linear combinations of $N$ and $\overline{N}$ with the BRS and anti-BRS charges $Q$ and $\overline{Q}$ of Eqs. (\ref{caricaqu})-(\ref{antiqu}):
\begin{equation}
Q_{\scriptscriptstyle H}=Q-\beta N, \qquad \quad
\overline{Q}_{\scriptscriptstyle H}=\overline{Q}+\beta\overline{N}, \label{susy}
\end{equation}
where $\beta$ is an arbitrary parameter. Being linear combination of conserved charges, $Q_{\scriptscriptstyle H}$ and $\overline{Q}_{\scriptscriptstyle H}$ turn out to be also  conserved. If $\beta$ is infinitesimal $Q_{\scriptscriptstyle H}$ and $\overline{Q}_{\scriptscriptstyle H}$ can be written in terms of $Q$ and $\overline{Q}$ as:
\begin{displaymath}
Q_{\scriptscriptstyle H}=e^{\beta H}\,Q\,e^{-\beta H},\qquad \quad \overline{Q}_{\scriptscriptstyle H}=e^{-\beta H}\,\overline{Q}\,e^{\beta H}.
\end{displaymath}
What is more important is that they satisfy the following anticommutator $\left[Q_{\scriptscriptstyle H},\overline{Q}_{\scriptscriptstyle H}\right]=2i\beta\widetilde{\cal H}$, so they can be considered as generators of non-relativistic supersymmetry transformations. Once we apply them on the Lagrangian, they generate the following transformations: 
\begin{displaymath}
\displaystyle
\delta_{Q_H}\widetilde{\cal L}=-i\epsilon\beta\frac{\textrm{d}}{\textrm{d}t}N, \qquad \quad
\displaystyle
\delta_{\overline{Q}_H}\widetilde{\cal L}=-i\bar{\epsilon}\frac{\textrm{d}}{\textrm{d}t}\overline{Q}
\end{displaymath}
and, as the Lagrangian changes by a total derivative, we can say that $Q_{\scriptscriptstyle H}$ and $\overline{Q}_{\scriptscriptstyle H}$ generate symmetries for  the action $\widetilde{S}=\int \textrm{d} t \, \widetilde{\cal L}$. Finally, it is possible to prove that $Q_{\scriptscriptstyle H}$ and $\overline{Q}_{\scriptscriptstyle H}$ are nilpotent: $
[Q_{\scriptscriptstyle H},Q_{\scriptscriptstyle H}]=[\overline{Q}_{\scriptscriptstyle H},
\overline{Q}_{\scriptscriptstyle H}]=0$. 

As we did before with the BRS charges, we can give to these charges a differential operator representation in the superspace. Their action is defined via the equation
$\delta\, \Xi^{\scriptscriptstyle A}=-\epsilon\hat{\Omega}\Xi^{\scriptscriptstyle A}$, where $\Xi^{\scriptscriptstyle A}$ is the YM superfield of Eq. (\ref{supcomp}). The result is \cite{enrico}:
\begin{displaymath}
\hat{Q}_{\scriptscriptstyle H}  =  -\frac{\partial}{\partial\theta}-
\beta\bar{\theta}\frac{\partial}{\partial t}, \qquad \quad \hat{\overline{Q}}_{\scriptscriptstyle H}  =  \frac{\partial}{\partial \bar{\theta}}+
\beta\theta\frac{\partial}{\partial t}.
\end{displaymath}
With the previous choice we have the following anticommutator $\displaystyle [\hat{Q}_{\scriptscriptstyle H},\hat{\overline{Q}}_{\scriptscriptstyle H}]=-2\beta\frac{\partial}{\partial t}$, i.e. the operators $\hat{Q}_{\scriptscriptstyle H}$ and $\hat{\overline{Q}}_{\scriptscriptstyle H}$ close on $\displaystyle \frac{\partial}{\partial t}$,  which is the operatorial representation of the Hamiltonian in the superspace.

Every supersymmetric field theory is characterized by the number $N$ of the independent supersymmetry charges whose square reproduces the Hamiltonian 
\begin{displaymath}
Q_i^2=\widetilde{\cal H}, \quad i=1,\cdots, N.
\end{displaymath}
We want to prove that in the CPI such number is given by $N=2$ \cite{enrico}. First of all, let us consider again the four charges 
\begin{eqnarray}
&& Q=i\Gamma^{\scriptscriptstyle A}\Lambda_{\scriptscriptstyle A}, \qquad\qquad\;\; \quad \overline{Q}=-i\Lambda_{\scriptscriptstyle A}\omega^{\scriptscriptstyle AB}\overline{\Gamma}_{\scriptscriptstyle B},\nonumber\\
&& N=\Gamma^{\scriptscriptstyle A}\overrightarrow{\partial}_{\scriptscriptstyle A}H(-)^{\scriptscriptstyle [A]}, \qquad  \overline{N}=\overline{\Gamma}_{\scriptscriptstyle A} \omega^{\scriptscriptstyle AB}\overrightarrow{\partial}_{\scriptscriptstyle B}H. \nonumber
\end{eqnarray}
Using their algebra:
\begin{displaymath}
[Q,N]=[\overline{Q},\overline{N}]=0, \qquad  [Q,\overline{N}]=i\widetilde{\cal H}, \qquad [\overline{Q},N]=-i\widetilde{\cal H}
\end{displaymath}
we can easily construct the ``square roots" of the Hamiltonian $\widetilde{\cal H}$, which must have one of the following forms: 
\begin{displaymath}
Q_{\alpha,\beta}=\alpha Q+\beta\overline{N}, \qquad \overline{Q}_{\bar{\alpha},\bar{\beta}}=\bar{\alpha}\overline{Q}+\bar{\beta} N,
\end{displaymath}
where $\alpha$ $\beta$, $\bar{\alpha}$ and $\bar{\beta}$ are arbitrary parameters, which will be fixed in a while. $Q_{\alpha,\beta}$ and $\overline{Q}_{\bar{\alpha},\bar{\beta}}$ anticommute and satisfy:
\begin{equation}
\left[Q_{\alpha,\beta},Q_{\alpha,\beta}\right]=2\alpha\beta i\widetilde{\cal H},
\qquad \left[\overline{Q}_{\bar{\alpha},\bar{\beta}},\overline{Q}_{\bar{\alpha},\bar{\beta}}\right]=-2\bar{\alpha}\bar{\beta} i\widetilde{\cal H}. \label{sqroot}
\end{equation}
On the RHSs of (\ref{sqroot}) we get just the Hamiltonian $\widetilde{\cal H}$ only if
$\alpha\beta=-\frac{i}{2}$ and $\bar{\alpha}\bar{\beta}= \frac{i}{2}$.
Each of the previous equations can fix only one of the two parameters, so we have infinite possible square roots of the Hamiltonian. If we require that the charges commute with each other, we can extract the following four operators in the superspace:
\begin{eqnarray}
&& \hat{D}_{\scriptscriptstyle 1}  =  \frac{1}{\sqrt{2}} \Biggl\{-\frac{\partial}{\partial\theta}-i\theta\frac{\partial}{\partial t} \Biggr\}, \qquad \; \hat{Q}_{\scriptscriptstyle 1}  =  \frac{i}{\sqrt{2}} \Biggl\{-\frac{\partial}{\partial\theta}+i\theta\frac{\partial}{\partial t} \Biggr\}, \nonumber\\
&& \hat{D}_{\scriptscriptstyle 2}  =  \frac{1}{\sqrt{2}} \Biggl\{ \frac{\partial}{\partial\bar{\theta}}+i\bar{\theta}\frac{\partial}{\partial t} \Biggr\},\qquad \quad \hat{Q}_{\scriptscriptstyle 2}  =  \frac{i}{\sqrt{2}} \Biggl\{ \frac{\partial}{\partial\bar{\theta}}-i\bar{\theta}\frac{\partial}{\partial t} \Biggr\}. \nonumber
\end{eqnarray}
They are all conserved since they commute with the Hamiltonian. Furthermore, they commute among themselves and they can be considered as square roots of the Hamiltonian, in the sense that $[\hat{Q}_i,\hat{Q}_i]=2\hat{Q}_i^2=\hat{\widetilde{\cal H}}$. This implies that our supersymmetry is a $N=2$ supersymmetry, where $\hat{Q}_1$ and $\hat{Q}_2$ can be identified with the supercharges and $\hat{D}_1$ and $\hat{D}_2$ with the covariant derivatives \cite{Freund}. 

It is also possible to find the covariant derivatives associated with the original supersymmetry charges $Q_{\scriptscriptstyle H}$ e $\overline{Q}_{\scriptscriptstyle H}$. In analogy with what we have seen before, we can define:
\begin{displaymath}
\displaystyle
\hat{D}_{\scriptscriptstyle H}\equiv \frac{\partial}{\partial \theta}-\beta\bar{\theta}\frac{\partial}{\partial t}, \qquad \quad \hat{\overline{D}}_{\scriptscriptstyle H}\equiv -\frac{\partial}{\partial \bar{\theta}}+\beta\theta\frac{\partial}{\partial t}.
\label{dercovsusy}
\end{displaymath}
The previous operators commute with the supersymmetry operators $\hat{Q}_{\scriptscriptstyle H}$ e $\hat{\overline{Q}}_{\scriptscriptstyle H}$, so they can be considered as covariant derivatives. 

In supersymmetric field theories it is possible to associate with each covariant derivative a chiral superfield which is the one annihilated by the covariant derivative. Also in our case we can define a chiral superfield $\Xi^{a}_{\textrm{ch}}$ such that:
$\displaystyle \hat{D}_{\scriptscriptstyle H}\Xi^{a}_{\textrm{ch}}=0$.
From the previous equation we immediately get:
\begin{equation}
\Xi^{\scriptscriptstyle A}_{\textrm{ch}}=\xi^{\scriptscriptstyle A}+(-)^{\scriptscriptstyle [A]}\omega^{\scriptscriptstyle AB}\bar\theta\overline{\Gamma}_{\scriptscriptstyle B}-\bar{\theta}\theta\beta\dot{\xi}^{\scriptscriptstyle A}.
\label{Robinson}
\end{equation}
An antichiral superfield $\bar\Xi^{a}_{\textrm{ch}}$ is instead defined by the following equation: $\hat{\overline{D}}_{\scriptscriptstyle H}\bar\Xi^{a}_{\textrm{ch}}=0$, which is satisfied by
\begin{displaymath}
\overline{\Xi}^{\scriptscriptstyle A}_{\textrm{ch}}=\xi^{\scriptscriptstyle A}+(-)^{\scriptscriptstyle [A]}\theta\Gamma^{\scriptscriptstyle A}+\bar{\theta}\theta\beta\dot{\xi}^{\scriptscriptstyle A}.
\end{displaymath}
The chiral and the antichiral superfield depend only on two types of fields. If we require that a superfield is both chiral and antichiral we get: $\Xi^{\scriptscriptstyle A}=\xi^{\scriptscriptstyle A}$, i.e., we reduce ourselves to consider only the fields $\xi^{\scriptscriptstyle A}$ of the standard BFV phase space. 

It is easy to show that if we apply a supersymmetry transformation on a chiral superfield, we get another chiral superfield. In fact let us indicate with $Q_i$ a supersymmetry charge, with $D_i$ the covariant derivative and with $\Xi^i_{\textrm{ch}}$ the associated chiral superfield. Then under a supersymmetry transformation the chiral superfield changes as follows:
\begin{displaymath}
\Xi^{\prime i}_{\textrm{ch}}=(1-\epsilon Q_i)\Xi^i_{\textrm{ch}},
\end{displaymath}
with $\epsilon$ anticommuting parameter. If we apply $D_i$ to the transformed chiral superfield and we use the fact that $[Q_i,D_i]=0$, we get:
\begin{displaymath}
D_i\Xi^{\prime i}_{\textrm{ch}}=D_i\Xi^i_{\textrm{ch}}-\epsilon Q_iD_i\Xi^i_{\textrm{ch}}=0,
\end{displaymath} 
i.e. also $\Xi^{\prime i}_{\textrm{ch}}$ is a chiral superfield. This means the space of chiral superfields carry a representation of the supersymmetry. Chiral superfields are important because they are made of fewer components than the standard superfields. This means that they carry a representation of supersymmetry of lower dimension than the one carried by the superfields. Moreover, it is possible to prove that their representation is an irreducible one \cite{susy}. 

The supersymmetry charges, as most of the elements present in the CPI, have a nice geometrical interpretation. It was proven in \cite{enrico} that, like the BRS is an exterior derivative on phase space, the supersymmetry charges are equivariant exterior derivatives. We will not dwell here on this issue which has been thoroughly explored in Ref. \cite{enrico}. The same geometrical meaning is for sure valid in the gauge case and this issue will be further explored in Ref. \cite{third}, in connection to topological field theory. 

The supersymmetry that we have found in the CPI of any field theory is a {\it universal} one. It strongly indicates that from any non-supersymmetric field theory one can build, via the CPI, a supersymmetric one. So somehow supersymmetry seems to be a feature not of particular systems but of any system. Unfortunately this supersymmetry is non-relativistic. Why is so? We feel that this is due to the fact that, to build the CPI, we started from the Hamiltonian version of the equations of motion, which is intrinsecally non-covariant. One idea we would like to explore in \cite{third} is to start from that covariant version of the equations of motion, known as the de Donder-Weyl formalism, for a review see \cite{kastrup}. We feel that the CPI which can be derived from it should have a relativistic supersymmetry stemming from the covariance of the original formalism. It would be very nice if this happens because it would indicate that the relativistic supersymmetry 
is a universal phenomenon. Moreover, because of the geometrical meaning of the $Q_{\scriptscriptstyle H}$ mentioned before, it would also indicate that the supersymmetry is present because of some deep geometrical reason. This reason is because we basically wanted to reformulate the original field theory in a coordinate-free form, like the Lie derivative is. This may also explain why the partner Dirac fermions, which in the de Donder-Weyl formalism may turn out to be the differential forms $\Gamma$ of the CPI for bosonic field theories, are not fundamental fields but objects derived from the original fields of the theory. This may also explain why we do not see them in nature as independent degrees of freedom. The same trick would apply if we start from a fermionic theory and do its associated CPI. The forms would be bosonic and we would not see them because they are just differential forms. 

\section{Conclusions}

In this paper we have extended the CPI formalism from the point particle \cite{martin} to a gauge field theory. We have shown in details that, like in the quantum case, a gauge-fixing and a FP determinant are needed also in the CPI . We have also proven that this approach to the CPI is equivalent to a different one built out of the BFV formalism. In this first paper we have concentrated in building rigorously the formalism and the various universal charges that are present. In later works \cite{riccardo}-\cite{third} we will apply the formalism in the directions outlined in the Introduction and in Sec. 6. We have preferred to divide the project in this way because the formalism of the CPI for YM is rather heavy and we could have run easily into problems if we had tried to do too many things without having laid the foundations properly. Last, but not least, the reader may wonder of what happens in the CPI to the ``minimal coupling" scheme which was needed at the quantum level to couple particle degrees of freedom to gauge fields. This issue has already been thoroughly worked out in Ref. \cite{minimal} to which we refer the reader for details.

\section*{Acknowledgments}

This work has been supported by grants from the University of Trieste, MIUR and INFN. Two of us (E.G. and D.M.) warmly acknowledge many helpful discussions with M. Reuter. 

\newpage 
\begin{center}
{\LARGE\bf Appendices}
\end{center}

\appendix
\makeatletter
\@addtoreset{equation}{section}
\makeatother
\renewcommand{\theequation}{\thesection\arabic{equation}}

\section{Construction of the canonical action of Eq. (\ref{azcan})}
\noindent 
In this Appendix we want to derive in detail the ``canonical action" (\ref{azcan}) from the Lagrangian density $\displaystyle {\cal L}=-\frac{1}{4}F^{\mu\nu}_aF^{a}_{\mu\nu}$. First of all, from ${\cal L}$ we construct the energy-momentum tensor:
\begin{eqnarray}
\displaystyle
 \Theta^{\mu\nu}&\hspace{-2mm}=\hspace{-2mm}&-\frac{\partial {\cal L}}{\partial(\partial_{\mu}A^{a}_{\lambda})}
\partial^{\nu}A^{a}_{\lambda}+\eta^{\mu\nu}{\cal L}\nonumber\\
&\hspace{-2mm}=\hspace{-2mm}& F_a^{\mu\lambda}\partial^{\nu}A^{a}_{\lambda}
-\frac{1}{4}\eta^{\mu\nu}F_b^{\rho\sigma}F^b_{\rho\sigma}. \nonumber
\end{eqnarray}
The Hamiltonian is the integral over the space variables of the component $\Theta^{00}$ of such tensor, which is given by:
\begin{eqnarray}
\displaystyle
\Theta^{00}&\hspace{-2mm}=&\hspace{-2mm} F_a^{0\lambda}(\partial^{0}A^{a}_{\lambda})+\frac{1}{4}F^{\rho\sigma}_b
F_{\rho\sigma}^b\nonumber\\
&\hspace{-2mm}=&\hspace{-2mm} \vec{\pi}_a\cdot \dot{\vec{A}}^{a}-\frac{1}{2}
\vec{\pi}_a\cdot\vec{\pi}^a+\frac{1}{2}\vec{B}_a\cdot\vec{B}^a,
\end{eqnarray}
where we have used the definition of the conjugate momentum $\pi_a^k=-F_a^{0k}$ and of the ``magnetic field" $\displaystyle B_a^i=\frac{1}{2}\epsilon^{ijk}F_a^{jk}$. So we get the following expression for the Hamiltonian:
\begin{equation}
\displaystyle
H_{\textrm c}=\int {\textrm d}{\bf x} \, \Theta^{00}=\int {\textrm d}{\bf x} \, \biggl(\vec{\pi}_a\cdot\dot{\vec{A}}^{a}-\frac{1}{2}
\vec{\pi}_a\cdot\vec{\pi}^a+\frac{1}{2}\vec{B}_a\cdot\vec{B}^a\biggr).  \label{Hamcan}
\end{equation}
Taking into account the definition of the momentum:
\begin{equation}
\pi_a^{\mu}=-F_a^{0\mu}=\dot{A}_a^{\mu}+\partial^{\mu}A^0_a+C_a^{\; bc}A^0_bA^{\mu}_c \label{sigmapi}
\end{equation}
we can easily derive the equations of motion for $\vec{A}_a$:
\begin{equation}
\dot{\vec{A}}_a= \vec{\pi}_a-\vec{\nabla}A^0_a+C_a^{\; bc}\vec{A}_bA^0_c. \label{equazA}
\end{equation}
Using the previous equation into (\ref{Hamcan}) we get:
\begin{displaymath}
H_{\textrm c}=\int {\textrm d}{\bf x}\biggl(\frac{1}{2}\vec{\pi}_a\cdot\vec{\pi}^{a}+\frac{1}{2}\vec{B}_a\cdot
\vec{B}^{a}-\vec{\pi}_a\cdot\vec{\nabla}A^{0a}+C^{abc}\vec{\pi}_a\cdot\vec{A}_bA^0_c\biggr).
\end{displaymath}
The canonical Hamiltonian $H_{\textrm c}$ is not uniquely determined because of the constraints derived from the gauge invariance of the theory. First of all, from
(\ref{sigmapi}) we have that the antisymmetry of the tensor $F_a^{\mu\nu}$ and the definition itself of conjugate momenta imply that 
\begin{equation}
\pi_a^0=0, \qquad a=1,...,n. \label{briciolina}
\end{equation} 
This means that we have a number of primary constraints, equal to the number of generators of the Lie algebra that we are considering. It is clear that we can add to $H_{\textrm c}$ an arbitrary linear combination of the constraints and consider the following function:
\begin{displaymath}
H^{\prime}=\int {\textrm d}{\bf x}\, \left(H_{\textrm c}+v_1^{a}\pi^0_a\right)
\end{displaymath}
as generator of the time evolution. The equation of evolution of the primary constraints produces the following relation:
\begin{equation}
\displaystyle
\dot{\pi}^0_a(x) = \{\pi^0_a(x),H^{\prime}\}=
\vec{\nabla}\cdot\vec{\pi}_a-C^b_{\;ac}\vec{\pi}_b\cdot\vec{A}^c. \label{a4bis}
\end{equation}
From the consistency condition, i.e., the requirement that the primary constraints are conserved in time, we get that the RHS of (\ref{a4bis}) must be put equal to zero:
\begin{equation}
\sigma_a=-\vec{\nabla}\cdot\vec{\pi}_a+C^b_{\;ac}\vec{\pi}_b\cdot\vec{A}^c\approx 0. \label{asix}
\end{equation}
The $n$ secondary constraints (\ref{asix}), together with the $n$ primary constraints (\ref{briciolina}), which we have obtained before, form a set of $2n$ first class constraints.\footnote{``First class" basically means that the Poisson brackets of two constraints produce a linear combination of the constraints themselves, so this combination is zero on the constraints surface.} An arbitrary combination of the secondary constraints can be added to $H^{\prime}$ to get the following total  Hamiltonian \cite{dirac}: 
\begin{equation}
\begin{array}{l}
\displaystyle
H_{\scriptscriptstyle \textrm{T}}  =  \int {\textrm d}{\bf x} \,\biggl(\frac{1}{2}\vec{\pi}_a\cdot\vec{\pi}^{a}+\frac{1}{2}\vec{B}_a\cdot
\vec{B}^{a}+(\vec{\nabla}\cdot\vec{\pi}_a)A^{0a}+C^{abc}\vec{\pi}_a\cdot\vec{A}_bA^0_c
+ v_1^{a}\pi^0_a-v_2^{a}\sigma_a \biggr) \medskip  \\
\displaystyle \quad \;\; = \int {\textrm d}{\bf x}\, \biggl(\frac{1}{2}\vec{\pi}_a\cdot\vec{\pi}^{a}+\frac{1}{2}\vec{B}_a\cdot
\vec{B}^{a} +v_1^{a}\pi^0_a+(A^{0a}+v^{a}_2)(\vec{\nabla}\cdot\vec{\pi}_a+C^{b}_{\;ca}\vec{\pi}_b\cdot\vec{A}^c)\biggr). \label{espressacca}
\end{array}
\end{equation}
Starting from $H_{\scriptscriptstyle \textrm T}$ we can derive the equations of motion for the field $A(x)$:
\begin{displaymath}
\left\{  
\begin{array}{l}
\dot{A}^0_b(x)=\left\{A^0_b(x),H_{\scriptscriptstyle {\textrm T}}\right\}=-v_{1b}(x) \nonumber \medskip \\
 \dot{\vec{A}}_b=\vec{\pi}_b+C_{ab}^{\;\;c}\left(A^{0a}+v_2^{a}\right)\vec{A}_c-\vec{\nabla}\left(A^0_b+v_{2b}\right).
\end{array}
\right.
\end{displaymath}
Comparing the result with (\ref{equazA}) we can put $v_{2a}$=0.
Replacing this result into (\ref{espressacca}) we get:
\begin{displaymath}
H_{\scriptscriptstyle \textrm{T}}=\int {\textrm d}{\bf x}\,\biggl(\frac{1}{2}\vec{\pi}_a\cdot\vec{\pi}^{a}+\frac{1}{2}\vec{B}_a\cdot
\vec{B}^{a}-\dot{A}^0_a\pi^{0a}+A^{0a}\left(\vec{\nabla}\cdot\vec{\pi}_a-C^b_{\;ac}\vec{\pi}_b\cdot \vec{A}^c\right)
\biggr)
\end{displaymath}
from which we derive \cite{henneaux} the canonical action of Eq. (\ref{azcan}):
\begin{eqnarray}
S_{\textrm c} &\hspace{-2mm}=\hspace{-2mm}& \int {\textrm d}x\biggl(\dot{A}_a^{\mu}\pi^{a}_{\mu}-H_{\scriptscriptstyle \textrm{T}}\biggr)\nonumber\\
&\hspace{-2mm}=\hspace{-2mm}&\int {\textrm d}x \Biggl[\dot{\vec{A}}_a\cdot\vec{\pi}^{a}
-\frac{1}{2}\vec{\pi}_a\cdot\vec{\pi}^{a}
-\frac{1}{2}\vec{B}_a\cdot\vec{B}^{a}-A^{a}_{\scriptscriptstyle 0}\left(-\vec{\nabla}\cdot\vec{\pi}_a
+C^b_{\; ac}\vec{\pi}_b\cdot\vec{A}^c\right)\Biggr]. \nonumber
\end{eqnarray}

\newpage

\section{Proof of Eq. (\ref{312})}


As we have seen in Sec. 3, in the case of YM theories we deal with mixed variables, so we have to consider the superdeterminant of the Jacobian of the transformation that brings us from the solutions of the equations of motion to the equations of motion themselves. 
How can we exponentiate such a superdeterminant? 
In the case of purely Grassmannian even variables one considers the Jacobian of the transformation, which can be exponentiated via a pair of Grassmannian odd variables:
\begin{displaymath}
\textrm{det}^{-1}K^{ab}=\int{\cal D}\pi_a{\cal D}\xi_b \exp \left[-\pi_aK^{ab}\xi_b\right].
\end{displaymath}
In the case of purely Grassmannian odd variables one considers instead the inverse of the Jacobian of the transformation, which can be exponentiated via a pair of Grassmannian even variables according to the following rule:
\begin{displaymath}
\textrm{det}\, M^{ab}=\int{\cal D}\bar{\theta}_a{\cal D}\theta_b \exp \left[ {-\bar{\theta}_aM^{ab}\theta_b} \right]. 
\end{displaymath}
It is quite natural to think about the possibility of exponentiating the superdeterminant via a generalized Gaussian integral. In particular, we expect to have
a number of Grassmannian even auxiliary variables which is twice the number of the Grassmannian odd equations of motion and a number of Grassmannian odd auxiliary variables which is twice the number of the Grassmannian even equations of motion. 
In this Appendix we want to prove the previous statement, i.e: 
\begin{equation}
\displaystyle
\textrm{sdet}\begin{pmatrix} A & B\cr
         C & D\cr \end{pmatrix}=\int \textrm{d}\bar{\theta}\textrm{d}\theta \textrm{d}\bar{x}\textrm{d}x \;
\exp\left[-(\bar{\theta}^t\bar{x}^t)
\begin{pmatrix} A & B\cr
         C & D\cr \end{pmatrix} \begin{pmatrix} \theta \cr x \end{pmatrix}
         \right], \label{espsupdet}
\end{equation}
where $\theta$ and $\bar{\theta}$ are Grassmannian odd variables while $x$ and $\bar{x}$ are Grassmannian even variables.
Let us define the following quantities:
\begin{displaymath}
a=\bar{\theta}^tA\theta,\;\;\; b=\bar{\theta}^tBx,\;\;\; c=\bar{x}^tC\theta,\;\;\; d=\bar{x}^tDx,
\label{abbreviazioni}
\end{displaymath}
which are Grassmannian even, since the blocks $B$ and $C$ are Grassmannian odd, 
while $A$ e $D$ are even. 
At this point we can expand the exponential of Eq. (\ref{espsupdet}) as follows:
\begin{displaymath}
\displaystyle
\exp \left[ -(a+b+c+d)\right]=\sum_m\frac{(-)^m}{{m!}}(a+b+c+d)^m
\end{displaymath}
and use the ``multinomial formula":
\begin{displaymath}
(a+b+c+d)^m=\sum_{i_3\leq i_2\leq i_1\leq m}\frac{m!}{(m-i_1)!(i_1-i_2)!(i_2-i_3)!i_3!}
a^{m-i_1}b^{i_1-i_2}c^{i_2-i_3}d^{i_3}.
\end{displaymath}
Because of the presence of Grassmannian odd variables, a lot of the previous terms turn out to be zero. Let us indicate with $n$ the number of variables $\theta$ which are needed for the exponentiation.
For $n=2$ the only terms which are non zero and give a contribution to the integral (\ref{espsupdet}) are $a^2$,
$abc$, $b^2c^2$, for $n=3$ the only terms which survive are $a^3$, $a^2bc$,
$ab^2c^2$, $b^3c^3$ and so on. On the other side $d$ can have an arbitrary exponent, since it is made up by purely Grassmannian even terms, 
so we can write:
\begin{eqnarray*}
\displaystyle
I&=&\int \textrm{d}{\bar \theta}\textrm{d}{\theta}\textrm{d}\bar{x}\textrm{d}x \exp\left[-(a+b+c+d)\right]=
\int \textrm{d}{\bar \theta}\textrm{d}{\theta}\textrm{d}\bar{x}\textrm{d}x\Biggl\{e^{-d}+(-a+bc)e^{-d}\nonumber\\
&& +\left[\frac{1}{2!}\left(\frac{2!}{2!}a^2\right)-\frac{1}{3!}
\left(\frac{3!}{1!1!1!}abc\right)+\frac{1}{4!}
\left(\frac{4!}{2!2!}b^2c^2\right)\right]e^{-d}\nonumber\\
&&+\left[-\frac{1}{3!}\left(\frac{3!}{3!}a^3\right)+\frac{1}{4!}
\left(\frac{4!}{2!1!1!}a^2bc\right)-\frac{1}{5!}
\left(\frac{5!}{2!2!}ab^2c^2\right)+\frac{1}{6!}
\left(\frac{6!}{3!3!}b^3c^3\right)\right]e^{-d} + \cdots \Biggr\}.
\end{eqnarray*}
When we perform the Grassamnnian integrals, the only non-zero contribution comes from the terms which contain $n$ variables
$\theta$ and $n$ variables $\bar{\theta}$. Such terms include:
\begin{eqnarray*}
&& \Biggl[\frac{(-)^{n}}{n!}\Biggl(\frac{n!}{n!}a^n\Biggr)+\frac{(-)^{n+1}}{(n+1)!}\Biggl(
\frac{(n+1)!}{(n-1)!1!1!}a^{n-1}bc\Biggr)+...\nonumber\\
&& \qquad +\frac{(-)^{2n-1}}{(2n-1)!}\Biggl(\frac{(2n-1)!}{1!(n-1)!(n-1)!}ab^{n-1}c^{n-1}\Biggr)
+\frac{(-)^{2n}}{(2n)!}\Biggl(\frac{(2n)!}{n!n!}b^nc^n\Biggr)\Biggr] e^{-d} \nonumber\\
&&=\frac{(-)^n}{n!}\Biggl[a^n-\frac{n!}{(n-1)!}a^{n-1}bc+\frac{n!}{(n-2)!2!2!}a^{n-2}
b^2c^2+...\nonumber\\
&& \qquad +(-)^{n-1}\frac{n!}{(n-1)!(n-1)!}ab^{n-1}c^{n-1}+(-)^n\frac{n!}{n!n!}b^nc^n\Biggr]e^{-d}.
\end{eqnarray*}
From the definitions of $a$, $b$ and $c$ we get:
\begin{eqnarray*}
\displaystyle
I&=&\int \textrm{d}{\bar \theta}\textrm{d}{\theta}\textrm{d}\bar{x}\textrm{d}x\frac{(-)^n}{n!}\Biggl[(\bar{\theta}^t
A\theta)^n-\frac{n!}{(n-1)!}(\bar{\theta}^tA\theta)^{n-1}(\bar{\theta}^tBx\bar{x}^tC\theta)+ \cdots \nonumber\\
&& +(-)^{n-1}\frac{n!}{(n-1)!(n-1)!}(\bar{\theta}^tA\theta)
(\bar{\theta}^tBx\bar{x}^tC\theta)^{n-1}+(-)^n\frac{n!}{n!n!}(\bar{\theta}^tBx\bar{x}^tC\theta)^n\Biggr]e^{-\bar{x}^tDx}.
\end{eqnarray*}

Now we have to evaluate all the integrals. First of all, let us consider:
\begin{displaymath}
(\bar{\theta}^tBx\bar{x}^tC\theta)^n=\bar{\theta}_{i_1}B_{i_1j_1}x_{j_1}\bar{x}_{m_1}C_{m_1o_1} \theta_{o_1} \cdots \bar{\theta}_{i_n}B_{i_nj_n}x_{j_n}\bar{x}_{m_n}C_{m_no_n}\theta_{o_n}.
\end{displaymath}
Since $B$, $C$, $\theta$ and $\bar{\theta}$ are all Grassmannian odd, every block which appears in the previous expression is Grassmannian even. This means that we can first of all fix the indices of the variables
$\bar{\theta}$ and then sum over all the permutations (even) of $n$ Grassmannian even blocks:
\begin{displaymath}
(\bar{\theta}^tBx\bar{x}^tC\theta)^n=n!\bar{\theta}_1B_{1j_1}x_{j_1}\bar{x}_{m_1}C_{m_1o_1}
\theta_{o_1} \cdots \bar{\theta}_nB_{nj_n}x_{j_n}\bar{x}_{m_n}C_{m_no_n}\theta_{o_n}.
\end{displaymath}
Then, if we want to put in ascending order also the indices of $\theta$, we have to change some of the anticommuting parameters and give the right sign to each permutation:
\begin{displaymath}
(\bar{\theta}^tBx\bar{x}^tC\theta)^n=n!\bar{\theta}_1\theta_1\cdots \bar{\theta}_n\theta_n
\sum_{\sigma}|\sigma|(Bx\bar{x}^tC)_{1\sigma(1)} \cdots (Bx\bar{x}^tC)_{n\sigma(n)}.
\end{displaymath}
Let us suppose we have brought $D$ to a diagonal form (we can always do it via a unitary transformation), and consider the bosonic integral:
\begin{equation}
\displaystyle
\int\prod_i \textrm{d}\bar{x}_i\textrm{d}x_i\sum_{\sigma}|\sigma|B_{1i_1}x_{i_1}\bar{x}_{j_1}C_{j_1\sigma(1)} \cdots
B_{ni_n}x_{i_n}\bar{x}_{j_n}C_{j_n\sigma(n)}\exp\left[-\sum_k\bar{x}^kD_{kk}x^k\right]. \label{integbos}
\end{equation}
Let us perform the following change of variables:
\begin{displaymath}
x_j=t_j+iv_j,      \;\;\;\;\;    \bar{x}_j=t_j-iv_j.
\end{displaymath}
The integration measure in (\ref{integbos}) becomes:
$ \displaystyle \prod_j 2\textrm{d}t_j\textrm{d}v_j$,
while the exponential takes the form:
$\exp \left[ -\sum_j(t_j^2+v_j^2)D_{jj}\right]$.
These expressions are both symmetric for the exchange $t\leftrightarrow v$. 
This means that, to give a non zero contribution to the integral, the coefficients of the exponential in (\ref{integbos}) must also be symmetric for this exchange which is equivalent to the following one between the original variables:
\begin{displaymath}
x_j\rightarrow i\bar{x}_j, \;\;\;\;\; \bar{x}_j\rightarrow -ix_j.
\end{displaymath}
From these considerations we deduce that only the terms of the type:
\begin{displaymath}
\displaystyle
\prod_i(x_i\bar{x}_i)^{n_i}\exp\left[-\sum_k\bar{x}^kD_{kk}x^k\right]
\end{displaymath}
can give a non zero contribution to the integral (\ref{integbos}). The associated integrals can be solved in the following way:
\begin{eqnarray*}
\displaystyle
&&\int\prod_j\textrm{d}\bar{x}_j\textrm{d}x_j\prod_i(x_i\bar{x}_i)^{n_i}\exp\left[-\sum_k\bar{x}^kD_{kk}x^k\right]\nonumber\\
&&\qquad =\prod_i\left(-\frac{\textrm{d}}{\textrm{d}D_{ii}}\right)^{n_i}\left(\frac{1}{D_{ii}}\right)=
\prod_i\frac{n_i!}
{D_{ii}^{n_i}}\textrm{det}(D^{-1}).  \label{risolinteg}
\end{eqnarray*}
Let us notice that if $D$ is a diagonal matrix then the following sum:
\begin{displaymath}
B_{1i_1}D_{i_1j_1}^{-1}C_{j_1\sigma(1)}B_{2i_2}D_{i_2j_2}^{-1}C_{j_2\sigma(2)}\cdots \textrm{det}(D^{-1})
\end{displaymath}
contains terms which are proportional to:
\begin{displaymath}
\prod_i\frac{n_i!}{D_{ii}^{n_i}}\textrm{det}(D^{-1}),
\end{displaymath}
where $n_i$ is the number of eigenvalues $D_{ii}$ of the matrix $D$. This is just the result of the integration (\ref{risolinteg}).
So we can write:
\begin{displaymath}
\displaystyle
\int \textrm{d}\bar{x}\textrm{d}x(\bar{\theta}^tBx\bar{x}^tC\theta)^ne^{-\bar{x}^tDx}=
n! \textrm{det}(D^{-1})(\bar{\theta}^tBD^{-1}C\theta)^n.
\end{displaymath}
Similarly we have:
\begin{displaymath}
\displaystyle
\int \textrm{d}\bar{x}\textrm{d}x(\bar{\theta}^tA\theta)^{n_1}
(\bar{\theta}^tBx\bar{x}^tC\theta)^{n_2}e^{-\bar{x}^tDx}=n_2! \textrm{det}(D^{-1})(\bar{\theta}^tA\theta)^{n_1}
(\bar{\theta}^tBD^{-1}C\theta)^{n_2}
\end{displaymath}
Finally, collecting all these results, our integral is given by:
\begin{eqnarray*}
\displaystyle
I&\hspace{-2mm}=&\hspace{-2mm} \int \textrm{d}\bar{\theta}\textrm{d}\theta\frac{(-)^n}{n!}\Biggl[(\bar{\theta}^tA\theta)^n-\frac{n!}{(n-1)!}
(\bar{\theta}^tA\theta)^{n-1}(\bar{\theta}^tBD^{-1}C\theta)+ \cdots \nonumber\\
&\hspace{-2mm}&\hspace{-2mm}+(-)^{n-1}\frac{n!}{(n-1)!}(\bar{\theta}^tA\theta)(\bar{\theta}^tBD^{-1}C\theta)^{n-1}
+(-)^n\frac{n!}{n!}(\bar{\theta}^tBD^{-1}C\theta)^{n}\Biggr] \textrm{det}(D^{-1}) \nonumber\\
&\hspace{-2mm}=&\hspace{-2mm} \int \textrm{d}\bar{\theta}\textrm{d}\theta\frac{(-)^n}
{n!}(\bar{\theta}^t(A-BD^{-1}C)\theta)^{n}\textrm{det}(D^{-1}).
\end{eqnarray*}
Remembering the standard rule of exponentiation of the determinant we get: 
\begin{displaymath}
I=\textrm{det}(A-BD^{-1}C)\textrm{det}(D^{-1}). \label{sdetespo}
\end{displaymath}
The last expression is just the definition of the superdeterminant. This completes our proof of Eq. ({\ref{312}).

\newpage 

\section{Equations of the Jacobi fields}

In this Appendix we would like to prove that Eqs. (\ref{Jacfields}) are identical to the equations of evolution of the first variations $\delta \xi^A$ of the fields $\xi^A$, i.e. to the equations of the Jacobi fields \cite{martin}. For convenience, we will split $\xi^A$ in two sets and we will indicate the anticommuting variables with lower capital Latin indices 
and the commuting ones with Greek indices. The first variation of the action is:
\begin{displaymath}
\displaystyle
\delta S =\int \textrm{d}x \biggl[\frac{1}{2} \delta \xi^{\mu} \omega_{\mu\nu}\dot{\xi}^{\nu}+\frac{1}{2}\xi^{\mu}
\omega_{\mu\nu}\delta\dot{\xi}^{\nu}-\delta\xi^{\mu}\partial_{\mu}H
+\frac{1}{2}\delta \xi^{a}\omega_{ab}\dot{\xi}^{b}+\frac{1}{2}\xi^{a}
\omega_{ab}\delta\dot{\xi}^{b}-\delta\xi^{a}\,\partial_{a}H\biggr]. 
\end{displaymath}
Modulo some integrations by parts, we get that the second variation of the action is given by:  
\begin{eqnarray}
\displaystyle
&&\delta^2 S=\int \textrm{d}x \biggl[\delta^2\xi^{\mu}
\left(\omega_{\mu\nu}\dot{\xi}^{\nu}-\partial_{\mu}H\right)+
\delta^2\xi^{a}\left(\omega_{ab}\dot{\xi}^{b}-\partial_{a}H\right)\nonumber \\
&&\qquad\quad  +\delta\xi^{\mu}\left(\omega_{\mu\nu}\delta\dot{\xi}^{\nu}-\delta\xi^{a}\partial_{\mu}\partial_aH-\delta\xi^{\nu}\partial_{\mu}\partial_{\nu}H\right)\nonumber\\
&&\qquad\quad  +\delta\xi^a\left(\omega_{ab}\delta\dot{\xi}^b-\delta\xi^{\mu}\partial_a\partial_{\mu}H
+\delta\xi^b\partial_a\partial_bH\right)\biggr], \nonumber
\end{eqnarray}
which gives the following equations for the Jacobi fields $\delta \xi^A$: 
\begin{displaymath}
\left\{
\begin{array}{l}
\omega_{\mu\nu}\delta\dot{\xi}^{\nu}-\delta\xi^{a}\partial_{\mu}\partial_aH
-\delta\xi^{\nu}\partial_{\mu}\partial_{\nu}H=0, \medskip \\
\omega_{ab}\delta\dot{\xi}^b-\delta\xi^{\mu}\partial_a\partial_{\mu}H
+\delta\xi^b\partial_a\partial_bH=0.
\end{array}
\right.
\end{displaymath}
Multiplying the first equation by $\omega^{\rho\mu}$ and the second one by $\omega^{ca}$ we get the following equation for the first variations $\delta \xi$:
\begin{displaymath}
\left\{
\begin{array}{l}
\delta\dot{\xi}^{\rho}=\omega^{\rho\mu}\partial_{\mu}\partial_{\nu}H\delta\xi^{\nu}-
\omega^{\rho\mu}\partial_{\mu}\partial_aH\delta\xi^{a}, \medskip \nonumber\\
\delta\dot{\xi}^c=\omega^{ca}\partial_a\partial_{\mu}H\delta\xi^{\mu}-
\omega^{ca}\partial_a\partial_bH\delta\xi^b.
\end{array}
\right.
\end{displaymath}
In the previous equations all the derivatives are right derivatives. Using the relations: 
\begin{displaymath}
\displaystyle \overrightarrow{\partial}_aH=-H\overleftarrow{\partial}_a, \qquad \overrightarrow{\partial}_{\mu}H=H\overleftarrow{\partial}_{\mu},
\end{displaymath} 
we can rewrite the equations of motion for $\delta\xi$ as:
\begin{displaymath}
\left\{
\begin{array}{l}
\displaystyle \delta\dot{\xi}^{\rho}=\omega^{\rho\mu}\overrightarrow{\partial}_{\mu}H\overleftarrow{\partial}_{\nu}\,\delta\xi^{\nu}+
\omega^{\rho\mu}\overrightarrow{\partial}_{\mu}H\overleftarrow{\partial}_b\,\delta\xi^b \medskip \\
\displaystyle \delta\dot{\xi}^c=\omega^{ca}\overrightarrow{\partial}_aH\overleftarrow{\partial}_{\nu}\,\delta\xi^{\nu}+
\omega^{ca}\overrightarrow{\partial}_aH\overleftarrow{\partial}_b\,\delta\xi^b. 
\end{array}
\right.
\end{displaymath}
By comparing this equation with (\ref{Jacfields}) we can conclude that also for YM theories the fields $\Gamma$ evolve in time just like the Jacobi fields $\delta \xi$.

\section{Proof of Eq. (\ref{comparison})}

In this Appendix, making explicit use of the equations of motion, we want to prove in detail that $L^{\scriptscriptstyle \textrm{BFV}}=L^{\scriptscriptstyle \textrm{FP}}$, Eq. (\ref{comparison}), i.e. that: 
\begin{equation}
\begin{array}{l}
\displaystyle
-\pi_a\vec{\nabla}\cdot \vec{A}^{a}+\dot{\lambda}^{a}\pi_a+\dot{\vec{A}}^{a}\cdot\vec{\pi}_a-\frac{1}{2}\vec{\pi}_a\cdot\pi^{a}
-\frac{1}{2}\vec{B}_a\cdot\vec{B}^{a}-\lambda^{a}\sigma_a+\dot{P}^{a}\overline{C}_a+\dot{C}^{a}\overline{P}_a \medskip \\
\qquad \quad +i\overline{C}_a\partial^kD_{kb}^{\;\;\;a}C^b
-i\overline{P}_aP^{a}+\lambda^{a}\overline{P}_bC^b_{\;ac}C^c \medskip \\
=-\pi_a\partial^{\mu}A^{a}_{\mu}-\frac{1}{4}F^{a}_{\mu\nu}F^{\mu\nu}_a-i\partial^{\mu}
\overline{C}_aD_{\mu b}^{\;\;\;a}C^{b}. \label{inf}
\end{array}
\end{equation}
First of all, let us consider the gauge-fixing (GF) parts. From the LHS of (\ref{inf}) we have that the GF part of $L^{\scriptscriptstyle \textrm{BFV}}$ is given by: 
$\displaystyle -\pi_a\vec{\nabla}\cdot \vec{A}^{a}+\dot{\lambda}^{a}\pi_a$.
So, using our conventions $\partial^{\mu}=(-\partial_t,\vec{\nabla})$, this GF part of the BFV Lagrangian density can be rewritten as:
\begin{eqnarray}
-\pi_a\vec{\nabla}\cdot\vec{A}^{a}+\dot{\lambda}^{a}\pi_a &\hspace{-2mm}=&\hspace{-2mm}-\pi_a\vec{\nabla}\cdot  \vec{A}^{a}+\pi_a\partial_tA^{a}_{\scriptscriptstyle 0} \nonumber\\
&\hspace{-2mm}=&\hspace{-2mm} -\pi_a\partial^{\mu}A^{a}_{\mu}, \label{last}
\end{eqnarray}
where we have used the fact that the Lagrange multipliers $\lambda^a$ are the time components of the fields $A^a$. What we have 
obtained as last term of (\ref{last}) is just the GF term of the RHS of Eq. (\ref{inf}).
Now let us compare the terms of the Lagrangian densities, $L^{\scriptscriptstyle \textrm{BFV}}$ and $L^{\scriptscriptstyle \textrm{FP}}$, associated with the bosonic fields. Here we can use the equations of motion of the fields $\vec{A}^{a}$:
\begin{displaymath}
\dot{\vec{A}}^{a}=\vec{\pi}^{a}+\vec{\nabla}\lambda^{a}+C^{a}_{\;dc}\lambda^d\vec{A}^{c}
\end{displaymath}
and the definition of the secondary constraints:
\begin{displaymath}
\sigma_a= -\vec{\nabla}\cdot \vec{\pi}_a+C^{b}_{\;ac}\vec{\pi}_b\cdot \vec{A}^c
\end{displaymath}
to get
\begin{eqnarray}
\displaystyle
&&\dot{\vec{A}}^{a}\cdot \vec{\pi}_a-\frac{1}{2}\vec{\pi}_a\cdot \vec{\pi}^{a}
-\frac{1}{2}\vec{B}_a\cdot \vec{B}^{a}-\lambda^{a}\sigma_a=
\vec{\pi}^{a}\cdot \vec{\pi}_a+\vec{\nabla}\lambda^{a}\cdot \vec{\pi}_a+C^{a}_{\;dc}\lambda^d\vec{A}^c\cdot\vec{\pi}_a\nonumber\\
&& \qquad \quad -\frac{1}{2}\vec{\pi}_a\cdot \vec{\pi}^{a}-\frac{1}{2}\vec{B}_a\cdot \vec{B}^{a}+\lambda^{a}\vec{\nabla}\cdot \vec{\pi}_a-\lambda^{a}C^b_{\;ac}\vec{\pi}_b\cdot \vec{A}^c. \nonumber 
\end{eqnarray}
Up to surface terms, the previous identity becomes: 
\begin{equation}
\displaystyle
\dot{\vec{A}}^{a}\cdot \vec{\pi}_a-\frac{1}{2}\vec{\pi}_a\cdot \vec{\pi}^{a}
-\frac{1}{2}\vec{B}_a\cdot\vec{B}^{a}-\lambda^{a}\sigma_a=\frac{1}{2}\left(\vec{\pi}_a\cdot \vec{\pi}^{a}-\vec{B}_a\cdot \vec{B}^{a}\right). \label{zero}
\end{equation}
From the definition of the generalized magnetic field $B^k_a$ we have
\begin{equation}
\displaystyle
B^k_a=\frac{1}{2}\epsilon^{kij}F_a^{ij}\, \Longrightarrow \, -\frac{1}{2}\vec{B}_a\cdot \vec{B}^a=-\frac{1}{4}F^{ij}_aF^{a}_{ij}. \label{prima}
\end{equation}
Similarly, from the definition of the conjugate momentum we get:
\begin{equation}
\displaystyle
\pi^k_a=-F^{0k}_a\Rightarrow \frac{1}{2}\vec{\pi}_a\cdot\vec{\pi}^{a}=-\frac{1}{2}F^{0k}_aF^{a}_{0k}
\label{seconda}
\end{equation}
Collecting together (\ref{prima}) and (\ref{seconda}) we have:
\begin{displaymath}
\displaystyle
\frac{1}{2}\Bigl(\vec{\pi}_a\cdot \vec{\pi}^{a}_k-\vec{B}_a\cdot\vec{B}^{a}\Bigr)=-\frac{1}{4}F^{\mu\nu}_aF^{a}_{\mu\nu}, 
\end{displaymath}
which, once it is replaced in (\ref{zero}), gives:
\begin{displaymath}
\displaystyle
\dot{\vec{A}}^{a}\cdot \vec{\pi}_a-\frac{1}{2}\vec{\pi}_a\cdot\vec{\pi}^{a}
-\frac{1}{2}\vec{B}_a\cdot\vec{B}^{a}-\lambda^{a}\sigma_a=-\frac{1}{4}F^{\mu\nu}_a
F^{a}_{\mu\nu}.
\end{displaymath}
This proves that the terms associated with the bosonic variables on the LHS and on the RHS of Eq. (\ref{inf}) are the same.

Finally, we have to prove that the terms which refer to the Grassmannian odd variables of $L^{\scriptscriptstyle \textrm{BFV}}$ and $L^{\scriptscriptstyle \textrm{FP}}$ are the same. Here we need the equations of motion of $\overline{C}$ and the definition of the covariant derivative, i.e.
\begin{displaymath}
\dot{\overline{C}}_a=i\overline{P}_a, \qquad \quad D_{b\nu}^{\;\;\;a}=\delta^a_b\partial_{\nu}-C^{a}_{\;cb}A^c_{\nu}.
\end{displaymath} 
Using them in the  $L^{\scriptscriptstyle \textrm{FP}}$ density we get for the Grassmannian odd part of $L^{\scriptscriptstyle \textrm{FP}}$:
\begin{equation}
-i\partial^{\mu}\overline{C}_aD_{\mu b}^{\;\;\;a}C^b=
i\dot{\overline{C}}_a\dot{C}^{a}-i\dot{\overline{C}}_aC^{a}_{\;bc}C^cA^b_{\scriptscriptstyle 0}
-i\partial^{k}\overline{C}_aD_{kb}^{\;\;\;a}C^b.  \label{fpgr}
\end{equation}
Up to surface terms, the correspondent terms in the $L^{\scriptscriptstyle \textrm{BFV}}$ of Eq. (\ref{azionediBFV}) are given by:
\begin{eqnarray}
&& i\overline{C}_a\partial^kD_{kb}^{\;\;\;a}C^b-i\overline{P}_aP^{a}+\lambda^{b}\overline{P}_a
C^{a}_{\;bc}C^c+\dot{P}^a\overline{C}_a+\dot{C}^{a}\overline{P}_a\nonumber\\
&& \qquad =-i\partial^k\overline{C}_aD_{kb}^{\;\;\;a}C^b-i\overline{P}_aP^{a}+\lambda^{b}\overline{P}_a
C^{a}_{\;bc}C^c-P^a\dot{\overline{C}}_a+\dot{C}^{a}\overline{P}_a \nonumber \\
&& \qquad =i\dot{\overline{C}}_a\dot{C}^{a}-i\lambda^b\dot{\overline{C}}_aC^{a}_{\;bc}C^c-i
\partial^k\overline{C}_aD_{kb}^{\;\;\;a}C^b. \label{bfvgr}
\end{eqnarray}
Expressions (\ref{fpgr}) and (\ref{bfvgr}) are equal. So, since we have proved that also the terms which refer to the Grassmannian odd fields are equal, we can conclude that on shell the BFV and the FP Lagrangian densities are equivalent:
\begin{displaymath}
L^{\scriptscriptstyle \textrm{BFV}}=L^{\scriptscriptstyle \textrm{FP}}.
\end{displaymath}

\newpage

\section{Equivalence of the FP and the natural CPIs}

In this Appendix we want to show that it is possible to integrate away the extra fields present in $Z_{\scriptscriptstyle \textrm{CPI}}^{\scriptscriptstyle \textrm{YM(FP)}}$ to get just $Z_{\scriptscriptstyle \textrm{CPI}}^{\scriptscriptstyle \textrm{YM(NAT)}}$ and this proves the equivalence of the two approaches. As we announced in the paper, we will divide the proof in three parts, showing separately the equivalence of the GF, the ghost and the bosonic parts.

\subsection{The ghost part}

First of all, we want to compare the ghost parts. By making explicitly the Grassmann integrations over $\theta$ and $\bar{\theta}$ in (\ref{4.13bis}) we get the following dependence of $\widetilde{\cal L}^{\scriptscriptstyle \textrm{FP}}_{\textrm{gh}}$ on the fields of the theory:
\begin{eqnarray}
\widetilde{\cal L}^{\scriptscriptstyle \textrm{FP}}_{\textrm{gh}}&\hspace{-2mm}=\hspace{-2mm}&
i\partial^{\mu}\overline{C}_aD_{\mu}^{ba}
\Lambda_{\overline{P}_b}+\partial^{\mu}\overline{C}_aC^{a}_{\;cb}\Gamma^{A^c_{\mu}}\overline{\Gamma}_{\overline{P}_b}+\partial^{\mu}\overline{C}_aC^{a}_{\;cb}\overline{\Gamma}_{\pi^{\mu}_c}\Gamma^{C^b}+i\partial^{\mu}
\overline{C}_aC^{a}_{\;cb}\Lambda_{\pi^{\mu}_c}C^{b}\nonumber\\
&& +\partial^{\mu}\Gamma^{\overline{C}_a}D_{\mu}^{ba}
\overline{\Gamma}_{\overline{P}_b}-\partial^{\mu}\Gamma^{\overline{C}_a}C^{a}_{\;cb}\overline{\Gamma}_{\pi^{\mu}_c}C^{b}
-\partial^{\mu}\overline{\Gamma}_{P^{a}}D_{\mu}^{ba}\Gamma^{C^b}
\label{lagghost} \\
&& -\partial^{\mu}\overline{\Gamma}_{P^{a}}C^{a}_{\;cb}\Gamma^{A^c_{\mu}}C^b +i\partial^{\mu}\Lambda_{P^{a}}D_{\mu}^{ba}C^b.
\nonumber
\end{eqnarray}
Going to the functional integral, we have that its ghost part is given by:
\begin{equation}
Z^{\scriptscriptstyle \textrm{YM(FP)}}_{\textrm{gh}}=\int {\cal D}\overline{C}{\cal D}\Lambda_{\overline{P}}{\cal D}C{\cal D}\Lambda_{P}
{\cal D}\Gamma^{\scriptscriptstyle C}{\cal D}\Gamma^{\overline{C}}{\cal D}\overline{\Gamma}_P{\cal D}\overline{\Gamma}_{\overline{P}}\,
\exp \left[ i \int {\textrm d}x \,\widetilde{\cal L}^{\scriptscriptstyle \textrm{FP}}_{\textrm{gh}}\right]. \label{E1bis}
\end{equation}
For the ``natural" CPI of Eq. (\ref{ymnat}) the ghost part is instead given by:
\begin{equation}
Z^{\scriptscriptstyle \textrm{YM(NAT)}}_{\textrm{gh}}=\int {\cal D}\psi{\cal D}\eta\exp \left[ \int {\textrm d}x\partial^{\mu}\psi_aD_{\mu}^{ba}\eta_b\right]. \label{E1tris}
\end{equation}
We already know from (\ref{iduno}) that $\psi$ can be identified with $\overline{C}$; if we identify $\Lambda_{\overline{P}}$ with $\eta$, we get:  
\begin{equation}
\int {\cal D}\psi{\cal D}\eta\, \exp\left[\int {\textrm d}x\,\partial^{\mu}\psi_aD_{\mu}^{ba}\eta_b\right]=
\int {\cal D}\overline{C}{\cal D}\Lambda_{\overline{P}}\exp\left[i\int {\textrm d}x \,i\partial^{\mu}\overline{C}_aD_{\mu}^{ba}
\Lambda_{\overline{P}_b}\right]. \label{E1quater}
\end{equation}
It is clear that in the path integral $Z^{\scriptscriptstyle \textrm{YM(FP)}}_{\textrm{gh}}$ of (\ref{E1bis}) there are six other functional integrals, in the variables $C$, $\Lambda_{\scriptscriptstyle P}$, $\Gamma^{\scriptscriptstyle C}$, $\Gamma^{\scriptscriptstyle \overline{C}}$, $\overline{\Gamma}_{\scriptscriptstyle P}$, $\overline{\Gamma}_{\scriptscriptstyle \overline{P}}$, which are not present in the associated ``natural" path integral of Eq. (\ref{E1tris}). In what follows we want to show that, performing explicitly these integrations, we get a quantity which is independent of the fields of the theory. Let us start by evaluating the integral (\ref{E1quater}) over $\Lambda_{\scriptscriptstyle P}$:
\begin{displaymath}
T=\int{\cal D}\Lambda_{\scriptscriptstyle P}\exp\left\{\int {\textrm d}x\,(-\partial^{\mu}\Lambda_{P^{a}}\partial_{\mu}C^{a}+
\partial^{\mu}\Lambda_{P^{a}}C_{\;cb}^{a}A^c_{\mu}C^{b})\right\}.
\end{displaymath}
After an integration by parts we get:
\begin{eqnarray}
T&\hspace{-2mm}=&\hspace{-2mm}\int {\cal D}\Lambda_{\scriptscriptstyle P}\exp\left\{\int \textrm{d}x\left[\Lambda_{P_a}\left(\partial^{\mu}\partial_{\mu} 
C^{a}-C_{\;cb}^{a}\partial^{\mu}(A^c_{\mu}C^b)\right)\right]\right\}\nonumber\\
&&\sim\delta\left((\partial^{\mu}\partial_{\mu}\delta_b^{a}-C_{\;cb}^{a}\partial^{\mu}A^c_{\mu}
-C_{\;cb}^{a}A^c_{\mu}\partial^{\mu})C^b\right) \nonumber.
\end{eqnarray}
Using the gauge-fixing condition $\partial^{\mu}A_{\mu}=0$ and the properties of the Dirac deltas, we can write 
\begin{equation}
T\sim\delta(C^b)\,\textrm{det}(\partial^{\mu}\partial_{\mu}\delta_b^{a} \label{camminare}
-C_{\;cb}^{a}A^c_{\mu}\partial^{\mu}).
\end{equation}
The Dirac delta $\delta(C^b)$ allows us to perform immediately the integral over $C$, putting $C=0$ in all the terms appearing in (\ref{lagghost}). Next we have only four other functional integrations to perform:
\begin{eqnarray}
\det(\partial^{\mu}\partial_{\mu}\delta_b^{a}-C^{a}_{\;cb}A^{c}_{\mu}\partial^{\mu})
\hspace{-6mm}&& \int {\cal D}\Gamma^{\scriptscriptstyle C}{\cal D}\Gamma^{\scriptscriptstyle \overline{C}}{\cal D}\overline{\Gamma}_{\scriptscriptstyle P}{\cal D}\overline{\Gamma}_{\scriptscriptstyle \overline{P}}\,
\exp\int {\textrm d}x \, \Bigl(i\partial^{\mu}\overline{C}_aC^{a}_{\;cb}\Gamma^{A^c_{\mu}}\overline{\Gamma}_{\overline{P}_b}
\nonumber\\
&&+i\partial^{\mu}\overline{C}_aC^{a}_{\;cb}\overline{\Gamma}_{\pi^{\mu}_c}\Gamma^{C^b}+i
\partial^{\mu}\Gamma^{\overline{C}_a}D_{\mu}^{ba}
\overline{\Gamma}_{\overline{P}_b}-i
\partial^{\mu}\overline{\Gamma}_{P^{a}}D_{\mu}^{ba}\Gamma^{C^b}\Bigr).  \label{forterumore}
\end{eqnarray}
Using the definition of covariant derivative and integrating by parts, we can rewrite one of the terms appearing in the weight of (\ref{forterumore}) as follows:
\begin{eqnarray}
\int {\textrm d}x\;\partial^{\mu}\Gamma^{\overline{C}_a}D_{\mu}^{ba}\overline{\Gamma}_{\overline{P}_b}
&\hspace{-2mm}=&\hspace{-2mm} \int {\textrm d}x\, \left(\partial^{\mu}\Gamma^{\overline{C}_a}\partial_{\mu}\overline{\Gamma}_{\overline{P}_a}
-\partial^{\mu}\Gamma^{\overline{C}_a}C^{a}_{\;cb}A^c_{\mu}\overline{\Gamma}_{\overline{P}_b}\right)\nonumber\\
&\hspace{-2mm}=&\hspace{-2mm}-\int {\textrm d}x\;\Gamma^{\overline{C}_a}\left(\partial^{\mu}\partial_{\mu}\delta_b^{a}-C_{\;cb}^{a}\partial^{\mu}A^c_{\mu}
-C_{\;cb}^{a}A^c_{\mu}\partial^{\mu}\right)\overline{\Gamma}_{\overline{P}_b}. \nonumber
\end{eqnarray}
Note that also the integral over $\Gamma^{\overline{C}_a}$ produces a Dirac delta: 
\begin{equation}
\delta\left[(\partial^{\mu}\partial_{\mu}\delta_b^{a}-
C_{\;cb}^{a}A^c_{\mu}\partial^{\mu})\overline{\Gamma}_{\overline{P}_b}\right]=
\textrm{det}^{-1}\left(\partial^{\mu}\partial_{\mu}\delta_b^{a}-
C_{\;cb}^{a}A^c_{\mu}\partial^{\mu}\right)\delta(\overline{\Gamma}_{\overline{P}_b}).\label{stellacometa}
\end{equation}
The presence of the inverse of a determinant is due to the fact that the CPI variables $\overline{\Gamma}$ associated with the BFV ghosts are bosonic. The Dirac delta $\delta(\overline{\Gamma}_{\overline{P}_b})$ present in (\ref{stellacometa}) allows us to perform the integral over $\overline{\Gamma}_{\overline{P}_b}$. The result is that we can put everywhere: $\overline{\Gamma}_{\scriptscriptstyle \overline{P}}=0$ in (\ref{forterumore}). 
The two determinants of Eqs. (\ref{camminare})-(\ref{stellacometa}) cancel each other and the result is: 
\begin{eqnarray}
&&\int {\cal D}\Gamma^{\scriptscriptstyle C}{\cal D}\overline{\Gamma}_{\scriptscriptstyle P}\;\exp\left\{i\int \textrm{d}{\bf x} \left(\partial^{\mu}\overline{C}_a
C^{a}_{\;cb}\overline{\Gamma}_{\pi^{\mu}_c}\Gamma^{C^b}-\partial^{\mu}\overline{\Gamma}_{P^{a}}
\partial_{\mu}\Gamma^{C^{a}}+\partial^{\mu}\overline{\Gamma}_{P_a}C^{a}_{\;cb}A^c_{\mu}
\Gamma^{C^b}\right)\right\}\nonumber\\
&&\qquad =\textrm{det}^{-1}(\partial_{\mu}\partial^{\mu}\delta_b^{a}-
C_{\;cb}^{a}A^c_{\mu}\partial^{\mu}). \label{determinante}
\end{eqnarray}
So the only difference between $Z^{\scriptscriptstyle \textrm{YM(FP)}}_{\textrm{gh}}$
and $Z^{\scriptscriptstyle
\textrm{YM(NAT)}}_{\textrm{gh}}$ is given by the determinant in (\ref{determinante}). We want to prove that such determinant does not depend on the fields $A^c_{\mu}$. A functional determinant must be considered as derived from a matrix which depends on the space-time variables $x$:
\begin{displaymath}
D\equiv \textrm{det}^{-1}\left(\partial_{\mu}\partial^{\mu}\delta_b^{a}\delta(x-x^{\prime})-
C_{\;cb}^{a}\delta(x-x^{\prime})A^c_{\mu}\partial^{\mu}\right). 
\end{displaymath}
Using the relation $\delta(x-x^{\prime})=\partial_{\mu}\theta(x-x^{\prime})$, we get:
\begin{displaymath}
D= \textrm{det}^{-1}(\partial_{\mu})\textrm{det}^{-1}\left[\delta_b^{a}\delta(x-x^{\prime})-
C_{\;cb}^{a}\theta(x-x^{\prime})A^c_{\mu}\right]\textrm{det}^{-1}(\partial^{\mu}).
\end{displaymath}
The terms $\textrm{det}^{-1}(\partial_{\mu})$ and $\textrm{det}^{-1}(\partial^{\mu})$ are independent of the fields $A$. So it remains only to prove the independence of the term:
\begin{eqnarray}
\textrm{det}^{-1}\left[\delta_b^{a}\delta(x-x^{\prime})-
C_{\;cb}^{a}\theta(x-x^{\prime})A^c_{\mu}\right]&\hspace{-2mm}=&\hspace{-2mm}\exp^{-1}\textrm{Tr}\,\textrm{ln}(1-A)\nonumber\\
&\hspace{-2mm}=&\hspace{-2mm}\exp^{-1}\textrm{Tr}\left[-A-A^2/2 \cdots\right], \label{trace}  
\end{eqnarray}
where $A=C^{a}_{\;cb}\theta(x-x^{\prime})A^c_\mu$. On the RHS of (\ref{trace}) we have only performed standard manipulations. The trace in (\ref{trace}) is zero, thanks to the antisymmetry of the structure constants $C^a_{\;bc}$. Consequently, the determinant appearing in (\ref{determinante}) is independent of the fields of the theory. This completes the proof that the ghost terms of the two functionals $Z^{\scriptscriptstyle \textrm{YM(NAT)}}$ and $Z^{\scriptscriptstyle \textrm{YM(FP)}}$ are equivalent. 

\subsection{The bosonic parts}

The equivalence of the bosonic parts is much easier to prove. In fact from Eq. (\ref{Lagrnaturale}) we have
\begin{displaymath}
\widetilde{\cal L}^{\scriptscriptstyle \textrm{NAT}}_{\textrm{bos}}=\Lambda_{\scriptscriptstyle A}(\dot{\phi}^{\scriptscriptstyle A}-\omega^{\scriptscriptstyle AB}\partial_{\scriptscriptstyle B}H)+
i\overline{\Gamma}_{\scriptscriptstyle A}(\delta_{\scriptscriptstyle B}^{\scriptscriptstyle A}\partial_t-\omega^{\scriptscriptstyle AC}\partial_{\scriptscriptstyle B}\partial_{\scriptscriptstyle C}H)\Gamma^{\scriptscriptstyle B}.
\end{displaymath}
The previous Lagrangian was obtained by exponentiating the equations of motion 
derived from the canonical Lagrangian (\ref{azcan}):
\begin{equation}
\displaystyle
L=\dot{\vec{A}}_a\cdot\vec{\pi}^{a}-\frac{1}{2}\vec{\pi}_a\cdot\vec{\pi}^{a}-\frac{1}{4}F^{ij}_bF^b_{ij}-
\lambda^{a}(-\vec{\nabla}\cdot\vec{\pi}_a+C^{b}_{\;ac}\vec{\pi}_b\cdot\vec{A}^c). \label{lagrcanon}
\end{equation}
Using in (\ref{lagrcanon}) the equations of motion:
\begin{displaymath}
\dot{\vec{A}}^{a}=\vec{\pi}^{a}+\vec{\nabla}\lambda^{a}+C^{a}_{\;dc}\lambda^{d}\vec{A}^{c} 
\end{displaymath}
we get
\begin{displaymath}
\displaystyle
L=\frac{1}{2}\vec{\pi}_a\cdot\vec{\pi}^{a}-\frac{1}{4}F^{ij}_bF^{b}_{ij}=-\frac{1}{4}F^{a}_{\mu\nu}F^{\mu\nu}_a,
\end{displaymath}
i.e., just the original Lagrangian for YM theories. Using this result and the usual connection with the superfields we can write, up to surface terms
\begin{displaymath}
\displaystyle
\widetilde{\cal L}_{\textrm{bos}}^{\scriptscriptstyle \textrm{NAT}}=
\Lambda_{\scriptscriptstyle A}(\dot{\phi}^{\scriptscriptstyle A}-\omega^{\scriptscriptstyle AB}\partial_{\scriptscriptstyle B}H)+
i\overline{\Gamma}_{\scriptscriptstyle A}(\delta_{\scriptscriptstyle B}^{\scriptscriptstyle A}\partial_t-\omega^{\scriptscriptstyle AC}\partial_{\scriptscriptstyle B}\partial_{\scriptscriptstyle C}H)\Gamma^{\scriptscriptstyle B}=i\int {\textrm{d}}\theta {\textrm{d}}\bar{\theta}\left(-\frac{1}{4}\widetilde{F}^{a}_{\mu\nu}
\widetilde{F}^{\mu\nu}_a\right).
\end{displaymath}
Note that the RHS of this equation is also the bosonic part we would get in $Z^{\scriptscriptstyle \textrm{YM(FP)}}_{\scriptscriptstyle \textrm{CPI}}$. So this proves that the bosonic parts of $Z^{\scriptscriptstyle \textrm{YM(FP)}}_{\scriptscriptstyle \textrm{CPI}}$ and $Z^{\scriptscriptstyle \textrm{YM(NAT)}}_{\scriptscriptstyle \textrm{CPI}}$ are the same.  

\subsection{The gauge-fixing parts}

Finally, we have to prove that the two gauge-fixing parts are equivalent. In $\widetilde{\cal L}^{\scriptscriptstyle \textrm{NAT}}_{\textrm{gf}}$ we have only one term: $-\pi_a\partial^{\mu}A^{a}_{\mu}$,
while, using the standard superfields, the gauge-fixing part of $\widetilde{\cal L}^{\scriptscriptstyle \textrm{FP}}$ is given by:
\begin{eqnarray}
\widetilde{\cal L}^{\scriptscriptstyle \textrm{FP}}_{\textrm{gf}}&\hspace{-2mm}=&\hspace{-2mm}i\int {\textrm{d}}\theta {\textrm{d}}\bar{\theta}[-\widetilde{\pi}_a\partial^{\mu}
\widetilde{A}^{a}_{\mu}]\nonumber\\
&\hspace{-2mm}=&\hspace{-2mm}\pi_a\partial^{\mu}\Lambda_{\pi^{\mu}_a}
+i\Gamma^{\pi_a}\partial^{\mu}\overline{\Gamma}_{\pi^{\mu}_a}-i\overline{\Gamma}_{\lambda^{a}}
\partial^{\mu}\Gamma^{A^{a}_{\mu}}+\Lambda_{\lambda^{a}}\partial^{\mu}A^{a}_{\mu}.
\label{appear}
\end{eqnarray}
We can now formally identify the fields $\pi_a$, which appear in $\widetilde{\cal L}^{\scriptscriptstyle \textrm{NAT}}_{\textrm{gf}}$, with the fields $\Lambda_{\lambda^{a}}$, which appear in (\ref{appear}). Then, integrating $Z^{\scriptscriptstyle \textrm{YM(FP)}}_{\textrm{gf}}$ over $\pi_a$, $\Gamma^{\pi^{a}}$ and $\overline{\Gamma}_{\lambda^{a}}$, we get a term proportional to: 
\begin{equation}
\delta(\partial^{\mu}\Lambda_{\pi^{\mu}_a})\delta(\partial^{\mu}\overline{\Gamma}_{\pi^{\mu}_a})
\delta(\partial^{\mu}\Gamma^{A^a_{\mu}})=\det(\partial^{\mu}\delta_b^{a})\delta(\Lambda_{\pi^{\mu}_b})\delta(\overline{\Gamma}_{\pi^{\mu}_b})
\delta(\Gamma^{A^b_{\mu}}). \label{deltaspurie}
\end{equation}
The reason why we obtain a determinant and not an inverse of a determinant in (\ref{deltaspurie}) is because there are two anticommuting fields $\overline{\Gamma}_{\pi^{\mu}_b}$ and $\Gamma^{A^b_{\mu}}$ for each bosonic field $\Lambda_{\pi^{\mu}_b}$. The deltas which appear in (\ref{deltaspurie}) are not totally painless to work out since they contain fields which enter the expression of $\widetilde{\cal L}^{\scriptscriptstyle \textrm{FP}}_{\textrm{bos}}$. In particular, these Dirac deltas make $\widetilde{\cal L}^{\scriptscriptstyle \textrm{FP}}_{\textrm{bos}}= 0$. This could have been foreseen: in fact, our approach is on-shell and, using the equations of motion, we have: 
\begin{displaymath}
\widetilde{\cal L}^{\scriptscriptstyle \textrm{FP}}_{\textrm{bos}}=\Lambda_{\scriptscriptstyle A}(\dot{\phi}^{\scriptscriptstyle A}-\omega^{\scriptscriptstyle AB}\partial_{\scriptscriptstyle B}H)+
i\overline{\Gamma}_{\scriptscriptstyle A}(\delta_{\scriptscriptstyle B}^{\scriptscriptstyle A}\partial_t-\omega^{\scriptscriptstyle AC}\partial_{\scriptscriptstyle B}\partial_{\scriptscriptstyle C}H)\Gamma^{\scriptscriptstyle B}=0.
\end{displaymath}  
This tells us that the Dirac deltas of Eq. (\ref{deltaspurie}) are somehow superfluous. 
For example, let us consider the Dirac delta over $\Gamma^{A^b_{\mu}}$. In the bosonic part of the natural Yang-Mills CPI $Z^{\scriptscriptstyle \textrm{YM(NAT)}}_{\textrm{bos}}$ the terms depending on $\Gamma^{A^b_{\mu}}$ are the following ones:
\begin{equation}
\begin{array}{l}
\displaystyle
\int {\cal D}\Gamma^{A_{\mu}}{\cal D}\overline{\Gamma}_{A_{\mu}}\exp\left[-\int \textrm{d}x \, \overline{\Gamma}_{A^{a}_{\mu}}
\left(\delta_b^{a}\partial_t-\omega^{ac}\frac{\partial}{\partial A^{b}_{\nu}}
\frac{\partial}{\partial A^c_{\mu}}H\right)\Gamma^{A^b_{\nu}}\right] \\
\displaystyle \sim\int {\cal D}\Gamma^{A_{\mu}}\,\delta\Biggl(\dot{\Gamma}^{A^{a}_{\mu}}-\omega^{ac}\frac{\partial}
{\partial A^b_{\nu}}\frac{\partial}{\partial A^c_{\mu}}H\Gamma^{A^b_{\nu}}\Biggr). \label{intdelta}
\end{array}
\end{equation}
If we call $\Gamma^{A_{\mu}}\equiv y$ then the integral in (\ref{intdelta}) can be written in a compact form as $\displaystyle \int {\textrm d}y\,\delta[f(y)]$, with $f(y)=0$ at $y=0$. As we have seen in (\ref{deltaspurie}), the GF part produces another Dirac delta $\delta(y)$. This Dirac delta is superfluous if the following equation holds:
\begin{equation}
\int {\textrm d}y\, \delta[f(y)]\delta(y) \sim \int {\textrm d}y\, \delta[f(y)]. \label{sim}
\end{equation}
The ``$\sim$" sign means that the only difference between the LHS and the RHS of (\ref{sim}) is given by terms independent of the fields. 

Using the following properties of the Dirac deltas: 
\begin{displaymath}
\delta(y-a)\delta(y-b)=\delta(y-a)\delta(a-b), \qquad \quad
\int {\textrm d}y \, \delta[f(y)]=\int {\textrm d}y \, \frac{\delta(y)}
{\frac{\textrm{d}f(y)}{\textrm{d}y}\Bigm\arrowvert_{y=0}
}\end{displaymath}
we can write:
\begin{eqnarray}
\int {\textrm d}y \, \delta[f(y)]\delta(y)&\hspace{-2mm}=&\hspace{-2mm}\int {\textrm d}y \frac{\delta(y)\delta(y)}{\frac{{\textrm d}f(y)}{\textrm{d}y}\Bigm\arrowvert_{y=0}}=
\delta(0)\int {\textrm d}y\,\frac{\delta(y)}{\frac{{\textrm d}f(y)}{{\textrm d}y}\Bigm\arrowvert_{y=0}}\nonumber\\
&\hspace{-2mm}=&\hspace{-2mm}\delta(0)\int {\textrm d}y \,\delta[f(y)]. \nonumber
\end{eqnarray}
So $\delta[f(y)]\delta(y)$ and $\delta[f(y)]$ differ by a $\delta(0)$, which is independent of the gauge fields. This concludes our proof. 

\newpage 

\section{Proof of (\ref{extcom})}

In this Appendix we want to prove that if two functions $O_{\scriptscriptstyle 1}$ and $O_{\scriptscriptstyle 2}$ have zero Poisson brackets in the standard phase space 
then the associated 
\begin{equation}
\displaystyle \widetilde{O}_i\equiv\Lambda_{\scriptscriptstyle A}\omega^{\scriptscriptstyle AB}\overrightarrow{\partial}_{\scriptscriptstyle B}O_i+i\overline{\Gamma}_{\scriptscriptstyle A}\omega^{\scriptscriptstyle AB}\overrightarrow{\partial}_{\scriptscriptstyle B}O_i\overleftarrow{\partial}_{\scriptscriptstyle C}\Gamma^{\scriptscriptstyle C}(-1)^{\scriptscriptstyle [O_i]} \label{definitions2}
\end{equation}
commute in the extended Hilbert space underlying the CPI. More generally, we will prove that if $\{O_{\scriptscriptstyle 1},O_{\scriptscriptstyle 2}\}_{\scriptscriptstyle \textrm{pb}}=O_{\scriptscriptstyle 3}$, then, using the commutators of the CPI and the definitions (\ref{definitions2}), we have $[ \widetilde{O}_{\scriptscriptstyle 1},\widetilde{O}_{\scriptscriptstyle 2}]=i \widetilde{O}_{\scriptscriptstyle 3}$.

Since each $\widetilde{O}_{i}$ is the sum of two different terms, the commutator between $\widetilde{O}_{\scriptscriptstyle 1}$ and $\widetilde{O}_{\scriptscriptstyle 2}$ will be given by the sum of four different commutators. Let us analyze them one by one. The one between the first terms is:
\begin{eqnarray}
&& \displaystyle \left[ \Lambda_{\scriptscriptstyle A}\omega^{\scriptscriptstyle AB}\overrightarrow{\partial}_{\scriptscriptstyle B}O_{\scriptscriptstyle 1},
 \Lambda_{\scriptscriptstyle C}\omega^{\scriptscriptstyle CD}\overrightarrow{\partial}_{\scriptscriptstyle D}O_{\scriptscriptstyle 2} \right] =i\Lambda_{\scriptscriptstyle A}\omega^{\scriptscriptstyle AB}\overrightarrow{\partial}_{\scriptscriptstyle B}O_{\scriptscriptstyle 1}\overleftarrow{\partial}_{\scriptscriptstyle D}\omega^{\scriptscriptstyle DE}\overrightarrow{\partial}_{\scriptscriptstyle E}O_{\scriptscriptstyle 2}\nonumber \\
&& \qquad \displaystyle -i \Lambda_{\scriptscriptstyle D}\omega^{\scriptscriptstyle DE}
\overrightarrow{\partial}_{\scriptscriptstyle A}\overrightarrow{\partial}_{\scriptscriptstyle E}
O_{\scriptscriptstyle 2}\overrightarrow{\partial}_{\scriptscriptstyle B}O_{\scriptscriptstyle 1}\omega^{\scriptscriptstyle AB} (-)^{\scriptscriptstyle BO_2+O_1O_2+AD+A},
\label{comp1}
\end{eqnarray}
where each letter in the exponent of the minus sign indicates the Grassmannian parity that we have to introduce to take into account the fact that the CPI fields have mixed Grassmannian parity. The term on the RHS of (\ref{comp1}) must be compared with the Liouvillian associated with the standard Poisson brackets between $O_{\scriptscriptstyle 1}$ and $O_{\scriptscriptstyle 2}$:
\begin{eqnarray*}
&& \Lambda_{\scriptscriptstyle A}\omega^{\scriptscriptstyle AB}\overrightarrow{\partial}_{\scriptscriptstyle B}\left[O_{\scriptscriptstyle 1}\overleftarrow{\partial}_{\scriptscriptstyle C}\omega^{\scriptscriptstyle CD}
\overrightarrow{\partial}_{\scriptscriptstyle D}O_{\scriptscriptstyle 2}\right] \nonumber\\
&& = \Lambda_{\scriptscriptstyle A}\omega^{\scriptscriptstyle AB}\overrightarrow{\partial}_{\scriptscriptstyle B}O_{\scriptscriptstyle 1}\overleftarrow{\partial}_{\scriptscriptstyle D}\omega^{\scriptscriptstyle DE}\overrightarrow{\partial}_{\scriptscriptstyle E}O_{\scriptscriptstyle 2}+
\Lambda_A\omega^{\scriptscriptstyle AB}O_{\scriptscriptstyle 1}\overleftarrow{\partial}_{\scriptscriptstyle C}\omega^{\scriptscriptstyle CD}\overrightarrow{\partial}_{\scriptscriptstyle B}\overrightarrow{\partial}_{\scriptscriptstyle D}O_{\scriptscriptstyle 2} (-)^{\scriptscriptstyle (O_1+C)B} \nonumber \\
&&= -i  \cdot \left( i\Lambda_{\scriptscriptstyle A}\omega^{\scriptscriptstyle AB}\overrightarrow{\partial}_{\scriptscriptstyle B}O_{\scriptscriptstyle 1}\overleftarrow{\partial}_{\scriptscriptstyle D}\omega^{\scriptscriptstyle DE}\overrightarrow{\partial}_{\scriptscriptstyle E}O_{\scriptscriptstyle 2} -i \Lambda_{\scriptscriptstyle D}\omega^{\scriptscriptstyle DE}
\overrightarrow{\partial}_{\scriptscriptstyle A}\overrightarrow{\partial}_{\scriptscriptstyle E}
O_{\scriptscriptstyle 2}\overrightarrow{\partial}_{\scriptscriptstyle B}O_{\scriptscriptstyle 1}\omega^{\scriptscriptstyle AB} (-)^{\scriptscriptstyle BO_2+O_1O_2+AD+A} \right),
\end{eqnarray*}
which is just $-i$ times the RHS of (\ref{comp1}). 
Similar relations hold also for the terms coming from the other commutators, i.e.:
\begin{eqnarray}
&& \left[ \Lambda_{\scriptscriptstyle A}\omega^{\scriptscriptstyle AB}\overrightarrow{\partial}_{\scriptscriptstyle B}O_{\scriptscriptstyle 1}, i \overline{\Gamma}_{\scriptscriptstyle D} \; \omega^{\scriptscriptstyle DE}\overrightarrow{\partial}_{\scriptscriptstyle E}O_{\scriptscriptstyle 2}\overleftarrow{\partial}_{\scriptscriptstyle F}\Gamma^{\scriptscriptstyle F}\right] (-)^{\scriptscriptstyle O_2} \nonumber \\
&& \qquad =\overline{\Gamma}_{\scriptscriptstyle D}\omega^{\scriptscriptstyle DE}\overrightarrow{\partial}_{\scriptscriptstyle A}\overrightarrow{\partial}_{\scriptscriptstyle E}O_{\scriptscriptstyle 2}\overleftarrow{\partial}_{\scriptscriptstyle F}\Gamma^{\scriptscriptstyle F}\omega^{\scriptscriptstyle AB}\overrightarrow{\partial}_{\scriptscriptstyle B}O_{\scriptscriptstyle 1} (-)^{\scriptscriptstyle AD+O_2+O_2B+O_2O_1}; \label{comp2}  \bigskip  \\
&& \left[i \overline{\Gamma}_{\scriptscriptstyle A} \omega^{\scriptscriptstyle AB}\overrightarrow{\partial}_{\scriptscriptstyle B}O_{\scriptscriptstyle 1}\overleftarrow{\partial}_{\scriptscriptstyle C}\Gamma^{\scriptscriptstyle C},
\Lambda_{\scriptscriptstyle D} \, \omega^{\scriptscriptstyle DE}\overrightarrow{\partial}_{\scriptscriptstyle E}O_{\scriptscriptstyle 2}\right] (-)^{\scriptscriptstyle O_1} \nonumber \\
&& \qquad = -\overline{\Gamma}_{\scriptscriptstyle A}\omega^{\scriptscriptstyle AB}\overrightarrow{\partial}_{\scriptscriptstyle B} O_{\scriptscriptstyle 1}\overleftarrow{\partial}_{\scriptscriptstyle C}\overleftarrow{\partial}_{\scriptscriptstyle D}\omega^{\scriptscriptstyle DE}\overrightarrow{\partial}_{\scriptscriptstyle E}O_{\scriptscriptstyle 2}\Gamma^{\scriptscriptstyle C}(-)^{\scriptscriptstyle O_2(C+1)+O_1};  \bigskip \label{comp3} \bigskip \\
&& \left[ i\overline{\Gamma}_{\scriptscriptstyle A}\omega^{\scriptscriptstyle AB}\overrightarrow{\partial}_{\scriptscriptstyle B}O_{\scriptscriptstyle 1}\overleftarrow{\partial}_{\scriptscriptstyle C}\Gamma^{\scriptscriptstyle C}, i\overline{\Gamma}_{\scriptscriptstyle D}\omega^{\scriptscriptstyle DE}\overrightarrow{\partial}_{\scriptscriptstyle E}O_{\scriptscriptstyle 2}\overleftarrow{\partial}_{\scriptscriptstyle F}\Gamma^{\scriptscriptstyle F}\right](-)^{\scriptscriptstyle O_1+O_2} \nonumber\\
&& \qquad =-\overline{\Gamma}_{\scriptscriptstyle A}\omega^{\scriptscriptstyle AB}\overrightarrow{\partial}_{\scriptscriptstyle B}O_{\scriptscriptstyle 1}\overleftarrow{\partial}_{\scriptscriptstyle C}\;\omega^{\scriptscriptstyle CE}\overrightarrow{\partial}_{\scriptscriptstyle E}O_{\scriptscriptstyle 2}\overleftarrow{\partial}_{\scriptscriptstyle F}\Gamma^{\scriptscriptstyle F} (-)^{\scriptscriptstyle O_1+O_2} \label{comp4} \\ 
&& \qquad \quad 
+\overline{\Gamma}_{\scriptscriptstyle D}\omega^{\scriptscriptstyle DE}\overrightarrow{\partial}_{\scriptscriptstyle E}O_{\scriptscriptstyle 2}\overleftarrow{\partial}_{\scriptscriptstyle A}\omega^{\scriptscriptstyle AB}\overrightarrow{\partial}_{\scriptscriptstyle B}O_{\scriptscriptstyle 1}\overleftarrow{\partial}_{\scriptscriptstyle C}\Gamma^{\scriptscriptstyle C}
(-)^{\scriptscriptstyle O_2O_1+O_1+O_2}. \nonumber
\end{eqnarray}
Let us consider the second term coming from the definition of $\widetilde{O}_{\scriptscriptstyle 3}$, i.e. 
\begin{displaymath}
\displaystyle i\overline{\Gamma}_{\scriptscriptstyle A}\, \omega^{\scriptscriptstyle AB}\overrightarrow{\partial}_{\scriptscriptstyle B}\left[O_{\scriptscriptstyle 1}\overleftarrow{\partial}_{\scriptscriptstyle C}\,\omega^{\scriptscriptstyle CD}\overrightarrow{\partial}_{\scriptscriptstyle D}O_{\scriptscriptstyle 2}\right] 
\overleftarrow{\partial}_{\scriptscriptstyle E}\Gamma^{\scriptscriptstyle E} (-)^{\scriptscriptstyle O_1+O_2}.
\end{displaymath}
From this equation we get four different terms because the derivatives w.r.t. $\xi^{\scriptscriptstyle B}$ and $\xi^{\scriptscriptstyle E}$ can be applied on both $O_{\scriptscriptstyle 1}$ and $O_{\scriptscriptstyle 2}$. Let us write explicitly these four terms with the correct grading factors. By applying $\overrightarrow{\partial}_{\scriptscriptstyle B}$ over $O_{\scriptscriptstyle 1}$ and $\overleftarrow{\partial}_{\scriptscriptstyle E}$ over $O_{\scriptscriptstyle 2}$ we get:
\begin{displaymath}
\displaystyle i\overline{\Gamma}_{\scriptscriptstyle A}\omega^{\scriptscriptstyle AB}\overrightarrow{\partial}_{\scriptscriptstyle B}O_{\scriptscriptstyle 1}\overleftarrow{\partial}_{\scriptscriptstyle C}\, \omega^{\scriptscriptstyle CD}\overrightarrow{\partial}_{\scriptscriptstyle D}O_{\scriptscriptstyle 2}\overleftarrow{\partial}_{\scriptscriptstyle E}\Gamma^{\scriptscriptstyle E} (-)^{\scriptscriptstyle O_1+O_2}
\end{displaymath}
which, modulo a factor $i$, is just the first term on the RHS of (\ref{comp4}). By applying the derivatives w.r.t. $\xi^{\scriptscriptstyle E}$ and $\xi^{\scriptscriptstyle B}$ over $O_{\scriptscriptstyle 2}$ we get
\begin{displaymath}
-i \overline{\Gamma}_{\scriptscriptstyle D}\,\omega^{\scriptscriptstyle DE}\overrightarrow{\partial}_{\scriptscriptstyle A}\overrightarrow{\partial}_{\scriptscriptstyle E}O_{\scriptscriptstyle 2}\overleftarrow{\partial}_{\scriptscriptstyle F}\Gamma^{\scriptscriptstyle F}\omega^{\scriptscriptstyle AB}\overrightarrow{\partial}_{\scriptscriptstyle B}O_{\scriptscriptstyle 1} (-)^{\scriptscriptstyle AD+O_2+O_2B+O_2O_1},
\end{displaymath}
which is just the RHS of Eq. (\ref{comp2}) multiplied by $-i$. If we apply both the derivatives over $O_{\scriptscriptstyle 1}$ we get:
\begin{displaymath}
i \overline{\Gamma}_{\scriptscriptstyle A}\omega^{\scriptscriptstyle AB}\overrightarrow{\partial}_{\scriptscriptstyle B} O_{\scriptscriptstyle 1}\overleftarrow{\partial}_{\scriptscriptstyle C}\overleftarrow{\partial}_{\scriptscriptstyle D}\,\omega^{\scriptscriptstyle DE}\overrightarrow{\partial}_{\scriptscriptstyle E}O_{\scriptscriptstyle 2}\Gamma^{\scriptscriptstyle C}(-)^{\scriptscriptstyle O_2(C+1)+O_1},
\end{displaymath}
which is $-i$ times the RHS of Eq. (\ref{comp3}). Finally, if we apply the derivative w.r.t. $\xi^{\scriptscriptstyle E}$ over $O_{\scriptscriptstyle 1}$ and the derivative w.r.t. $\xi^{\scriptscriptstyle B}$ over $O_{\scriptscriptstyle 2}$ we get:
\begin{displaymath}
i \overline{\Gamma}_{\scriptscriptstyle D}\omega^{\scriptscriptstyle DE}\overrightarrow{\partial}_{\scriptscriptstyle E}O_{\scriptscriptstyle 2}\overleftarrow{\partial}_{\scriptscriptstyle A}\omega^{\scriptscriptstyle AB}\overrightarrow{\partial}_{\scriptscriptstyle B}O_{\scriptscriptstyle 1}\overleftarrow{\partial}_{\scriptscriptstyle C}\Gamma^{\scriptscriptstyle C}
(-)^{\scriptscriptstyle O_2O_1+O_1+O_2+1},
\end{displaymath}
which is $-i$ times the second term on the RHS of (\ref{comp4}). 
In this way we have proved that if $\{O_{\scriptscriptstyle 1}, O_{\scriptscriptstyle 2} \}_{\scriptscriptstyle \textrm{pb}}=
O_{\scriptscriptstyle 3} $ then the associated Lie derivatives $\widetilde{O}_{\scriptscriptstyle 1}$ and $\widetilde{O}_{\scriptscriptstyle 2}$ satisfy
$\left[\widetilde{O}_{\scriptscriptstyle 1},\widetilde{O}_{\scriptscriptstyle 2}\right]=i\widetilde{O}_{\scriptscriptstyle 3}$.

\newpage 

\section{Proof that $\left[Q,\widetilde{\Omega}\right]=0$}

In this section we want to prove that the two BRS charges, the universal $Q$ of the CPI  and the one $\widetilde{\Omega}$, due to the gauge invariance of the theory, commute. First of all we need the explicit expression of the BRS-BFV charge, which from Eq. (\ref{secondacaricaBRS}), is:
\begin{eqnarray*}
\displaystyle \widetilde{\Omega}&=&(\partial_k\pi^k_a)\Lambda_{\overline{P}_a}-i(\partial_k \Gamma^{\pi^k_a})\overline{\Gamma}_{\overline{P}_a}+i(\partial_k\overline{\Gamma}_{A^a_k})\Gamma^{C^a}-(\partial_k\Lambda_{A^a_k})C^a \nonumber \\
\displaystyle && -C^b_{\;ac}\pi^k_bA^c_k\Lambda_{\overline{P}_a}
+iC^b_{\; ac}\pi^k_b\Gamma^{A^c_k}\overline{\Gamma}_{\overline{P}_a} +iC^b_{\; ac}
\pi^k_b\overline{\Gamma}_{\pi^k_c}\Gamma^{C^a}-C^b_{\;ac}\pi^k_b\Lambda_{\pi^k_c}C^a \nonumber\\
\displaystyle && +iC^b_{\;ac}\Gamma^{\pi^k_b}A^c_k\overline{\Gamma}_{\overline{P}_a}
+iC^b_{\;ac}\Gamma^{\pi^k_b}\overline{\Gamma}_{\pi^k_c}C^a-iC^b_{\;ac}\overline{\Gamma}_{A^b_k}A^c_k\Gamma^{C^a}+iC^b_{\;ac}\overline{\Gamma}_{A^b_k}\Gamma^{A^c_k}C^a \nonumber \\
\displaystyle && +C^b_{\; ac}\Lambda_{A^b_k}A^c_kC^a-iP^a\Lambda_{\lambda^a}
-\Gamma^{P^a}\overline{\Gamma}_{\lambda^a}+\overline{\Gamma}_{\overline{C}_a}\Gamma^{\pi^a}+i\Lambda_{\overline{C}_a}\pi_a+\frac{1}{2}i\overline{P}_aC^a_{\;bc}\Gamma^{C^b}\overline{\Gamma}_{\overline{P}_c} \nonumber \\
&& -\frac{1}{2}i\overline{P}_aC^a_{\;bc} \overline{\Gamma}_{\overline{P}_b}\Gamma^{C^c}-\frac{1}{2}i\Gamma^{\overline{P}_a}C^a_{\; bc}C^b\overline{\Gamma}_{\overline{P}_c}
+\frac{1}{2}i\Gamma^{\overline{P}_a}C^a_{\;bc}\overline{\Gamma}_{\overline{P}_b}C^c+
\frac{1}{2}i\overline{\Gamma}_{C^a}C^a_{\;bc}C^b\Gamma^{C^c} \nonumber\\
&& -\frac{1}{2}i\overline{\Gamma}_{C^a}C^a_{\; bc}\Gamma^{C^b}C^c-\frac{1}{2}\Lambda_{C^a}C^a_{\; bc} C^bC^c -\frac{1}{2}\overline{P}_aC^a_{\; bc}\Lambda_{\overline{P}_b}C^c-\frac{1}{2}\overline{P}_aC^a_{\; bc}C^b \Lambda_{\overline{P}_c}.
\end{eqnarray*}
The expression of the CPI BRS charge is much more simpler (\ref{caricaqu}):
\begin{eqnarray}
\displaystyle Q &=& i\Gamma^{\scriptscriptstyle A}\Lambda_{\scriptscriptstyle A} =
i\Gamma^{A^a_k}\Lambda_{A^a_k}+i\Gamma^{\pi^k_a}\Lambda_{\pi^k_a}+i\Gamma^{\lambda^a} \Lambda_{\lambda^a}+i\Gamma^{\pi_a}\Lambda_{\pi_a} \nonumber\\
&& +i\Gamma^{C^a}\Lambda_{C^a}+i\Gamma^{P^a}\Lambda_{P^a}+i\Gamma^{\overline{C}_a}\Lambda_{\overline{C}_a}+i\Gamma^{\overline{P}_a}\Lambda_{\overline{P}_a}. 
\end{eqnarray}
The non zero commutators of the theory are: 
\begin{displaymath}
\displaystyle \left[ \xi^{\scriptscriptstyle B},\Lambda_{\scriptscriptstyle A}\right]=i\delta_{\scriptscriptstyle A}^{\scriptscriptstyle B}, \qquad \left[ \Gamma^{\scriptscriptstyle B},\overline{\Gamma}_{\scriptscriptstyle A}\right]=\delta_{\scriptscriptstyle A}^{\scriptscriptstyle B}. 
\end{displaymath}
Let us start by rewriting the non zero commutators of the first four terms of $\widetilde{\Omega}$ with $Q$:
\begin{eqnarray*}
\left[ \left(\partial_k\pi^k_a\right) \Lambda_{\overline{P}_a}, i \Gamma^{\pi^j_b}\Lambda_{\pi^j_b}\right] &=& \left(\partial_j \Gamma^{\pi^j_b}\right)\Lambda_{\overline{P}_b}  \medskip \\
\left[ -i\left( \partial_k\Gamma^{\pi^k_a}\right) \overline{\Gamma}_{\overline{P}_a},
i\Gamma^{\overline{P}_b}\Lambda_{\overline{P}_b}\right] &=& -\left(\partial_k \Gamma^{\pi^k_a}\right)\Lambda_{\overline{P}_a} \medskip  \\
\left[ i\left(\partial_k \overline{\Gamma}_{A^a_k}\right)\Gamma^{C^a}, i\Gamma^{A^b_j}\Lambda_{A^b_j}\right] &=& -\Gamma^{C^a}\left(\partial_k\Lambda_{A^a_k}\right) \medskip \\
\left[ -\left(\partial_k \Lambda_{A^a_k}\right)C^a, i \Gamma^{C^b}\Lambda_{C^b}\right] &=& 
\left(\partial_k \Lambda_{A^a_k}\right) \Gamma^{C^a}. 
\end{eqnarray*}
The first and the second commutator are opposite, as well as the third and the fourth. Something similar happens for all the other commutators. A first set is the following:
\begin{eqnarray*}
\left[ -C^b_{\;ac}\pi^k_bA^c_k\Lambda_{\overline{P}_a}, i\Gamma^{\pi^j_d}\Lambda_{\pi^j_d}+i\Gamma^{A^d_j}\Lambda_{A^d_j}\right]&=&
-C^b_{\;ac}\Gamma^{\pi^j_b}A^c_j\Lambda_{\overline{P}_a}-C^b_{\;ac}\pi^j_b\Gamma^{A^c_j}\Lambda_{\overline{P}_a}, \medskip \\
 \left[iC^b_{\;ac} \pi^k_b\Gamma^{A^c_k}\overline{\Gamma}_{\overline{P}_a},
i\Gamma^{\overline{P}_d}\Lambda_{\overline{P}_d}+i\Gamma^{\pi^j_d}\Lambda_{\pi^j_d} \right] &=& C^b_{\; ac}\pi^k_b\Gamma^{A^c_k}\Lambda_{\overline{P}_a}
+iC^b_{\; ac}\Gamma^{\pi^j_b}\Gamma^{A^c_j}\overline{\Gamma}_{\overline{P}_a}, \medskip \\
\left[ iC^b_{\; ac}\pi^k_b \overline{\Gamma}_{\pi^k_c}\Gamma^{C^a}, i\Gamma^{\pi^j_d}\Lambda_{\pi^j_d} \right] &=& i\Gamma^{\pi^j_b}\overline{\Gamma}_{\pi^j_c}\Gamma^{C^a}C^b_{\;ac}-C^b_{\;ac}\pi^k_b\Gamma^{C^a}\Lambda_{\pi^k_c}, \medskip \\
\left[ -C^b_{\;ac}\pi^k_b\Lambda_{\pi^k_c}C^a, i\Gamma^{C^d}\Lambda_{C^d}+i\Gamma^{\pi^j_d}\Lambda_{\pi^j_d}\right] &=&
C^b_{\;ac}\pi^k_b\Lambda_{\pi^k_c}\Gamma^{C^a}-C^b_{\;ac}\Gamma^{\pi^j_b}\Lambda_{\pi^j_c}C^a, \medskip \\
\left[ iC^b_{\;ac}\Gamma^{\pi^k_b}A^c_k\overline{\Gamma}_{\overline{P}_a},
i\Gamma^{\overline{P}_d}\Lambda_{\overline{P}_d}+i\Gamma^{A^d_j}\Lambda_{A^d_j}\right] &=& C^b_{\; ac}\Gamma^{\pi^k_b}A^c_k\Lambda_{\overline{P}_a}-iC^b_{\;ac}
\Gamma^{\pi^k_b}\Gamma^{A^c_k}\overline{\Gamma}_{\overline{P}_a}, \medskip \\
\left[ iC^b_{\;ac}\Gamma^{\pi^k_b}\overline{\Gamma}_{\pi^k_c}C^a, i \Gamma^{\pi^j_d}\Lambda_{\pi^j_d}+i\Gamma^{C^d}\Lambda_{C^d}\right]
&=& C^b_{\;ac}\Gamma^{\pi^k_b}C^a\Lambda_{\pi^k_c}-iC^b_{\; ac}\Gamma^{\pi^k_b}\overline{\Gamma}_{\pi^k_c}\Gamma^{C^a}.
\end{eqnarray*}
In the previous formula each commutator appears together with its opposite, so they do not give any contribution to the commutator $[\widetilde{\Omega},Q]$. Let us go on with other commutators: 
\begin{eqnarray}
\displaystyle \left[ -iC^b_{\; ac}\overline{\Gamma}_{A^b_k}A^c_k\Gamma^{C^a}, i\Gamma^{A^d_j}\Lambda_{A^d_j}\right]&=& C^b_{\; ac}A^c_k\Gamma^{C^a}\Lambda_{A^b_k}+i C^b_{\; ac}\overline{\Gamma}_{A^b_k}
\Gamma^{A^c_k}\Gamma^{C^a}, \nonumber \medskip \\
\left[ iC^b_{\; ac}\overline{\Gamma}_{A^b_k}\Gamma^{A^c_k}C^a, i\Gamma^{A^d_j}
\Lambda_{A^d_j}+i\Gamma^{C^d}\Lambda_{C^d}\right] &=& 
-C^b_{\; ac}\Gamma^{A^c_k}C^a\Lambda_{A^b_k}-iC^b_{\;ac}\overline{\Gamma}_{A^b_k}\Gamma^{A^c_k}\Gamma^{C^a}, \label{tre} \medskip \\
\left[ C^b_{\; ac}\Lambda_{A^b_k}A^c_kC^a, i\Gamma^{C^d}\Lambda_{C^d}+i\Gamma^{A^d_j}\Lambda_{A^d_j}\right]
&=& -C^b_{\; ac}\Lambda_{A^b_k}A^c_k\Gamma^{C^a} +C^b_{\; ac}\Lambda_{A^b_k}\Gamma^{A^c_k}C^a. \nonumber
\end{eqnarray}
Also this block of three commutators gives a zero contribution to the commutator $[\widetilde{\Omega},Q]$ since all the terms in the RHSs of Eq. (\ref{tre}) are one the opposite of the other. Let us consider the next block of four commutators:
\begin{eqnarray*}
&& \left[ -iP^a\Lambda_{\lambda^a}, i\Gamma^{P^b}\Lambda_{P^b}\right]= i\Gamma^{P^a}\Lambda_{\lambda^a}, \qquad \left[-\Gamma^{P^a}\overline{\Gamma}_{\lambda^a}, i\Gamma^{\lambda^b}\Lambda_{\lambda^b}\right]=-i\Gamma^{P^a}\Lambda_{\lambda^a} \ \nonumber\\ 
&& \left[ \overline{\Gamma}_{\overline{C}_a}\Gamma^{\pi_a}, i\Gamma^{\overline{C}_b}\Lambda_{\overline{C}_b}\right]=-i \Gamma^{\pi_a}\Lambda_{\overline{C}_a}, \qquad \left[i\Lambda_{\overline{C}_a}\pi_a, i\Gamma^{\pi_b}\Lambda_{\pi_b}\right] = i\Gamma^{\pi_a}\Lambda_{\overline{C}_a}.
\end{eqnarray*}
Finally, we have the last block of three blocks of three commutators:
\begin{eqnarray*}
\displaystyle \left[ \frac{1}{2}i\overline{P}_aC^a_{\;bc}\Gamma^{C^b}\overline{\Gamma}_{\overline{P}_c}, i\Gamma^{\overline{P}_d}\Lambda_{\overline{P}_d}\right] &=& -\frac{1}{2}iC^a_{\;bc}\Gamma^{\overline{P}_a}\Gamma^{C^b}\overline{\Gamma}_{\overline{P}_c}-
\frac{1}{2}C^a_{\;bc}\overline{P}_a\Gamma^{C^b}\Lambda_{\overline{P}_c} \medskip  \\
\displaystyle \left[ -\frac{1}{2}iC^a_{\; bc}\Gamma^{\overline{P}_a}C^b\overline{\Gamma}_{\overline{P}_c}, i\Gamma^{\overline{P}_d}\Lambda_{\overline{P}_d}+i\Gamma^{C^d}\Lambda_{C^d}\right]&=&-\frac{1}{2}C^a_{\;bc}\Gamma^{\overline{P}_a}C^b\Lambda_{\overline{P}_c}+\frac{1}{2}iC^a_{\; bc}\Gamma^{\overline{P}_a}\overline{\Gamma}_{\overline{P}_c}\Gamma^{C^b} \medskip \\
\displaystyle \left[ -\frac{1}{2}\overline{P}_aC^a_{\;bc}C^b\Lambda_{\overline{P}_c},
i\Gamma^{\overline{P}_d}\Lambda_{\overline{P}_d}+i\Gamma^{C^d}\Lambda_{C^d}\right]&=& \frac{1}{2}C^a_{\; bc}\Gamma^{\overline{P}_a}C^b\Lambda_{\overline{P}_c}+
\frac{1}{2}C^a_{\; bc}\overline{P}_a\Gamma^{C^b}\Lambda_{\overline{P}_c} 
\end{eqnarray*}
\begin{eqnarray*}
\displaystyle \left[ -\frac{1}{2}i\overline{P}_aC^a_{\;bc}\overline{\Gamma}_{\overline{P}_b}\Gamma^{C^c}, i\Gamma^{\overline{P}_d}\Lambda_{\overline{P}_d}\right]& = &\frac{1}{2}iC^a_{\;bc}\Gamma^{\overline{P}_a}\overline{\Gamma}_{\overline{P}_b}\Gamma^{C^c}
-\frac{1}{2}C^a_{\;bc}\overline{P}_a\Gamma^{C^c}\Lambda_{\overline{P}_b} \medskip\\
\displaystyle \left[ \frac{1}{2}i\Gamma^{\overline{P}_a}C^a_{\;bc}\overline{\Gamma}_{\overline{P}_b}C^c,i\Gamma^{\overline{P}_d}\Lambda_{\overline{P}_d}+i\Gamma^{C^d}\Lambda_{C^d}\right] &=&\frac{1}{2}\Gamma^{\overline{P}_a}C^a_{\;bc}C^c\Lambda_{\overline{P}_b}-\frac{1}{2}iC^a_{\;bc}\Gamma^{\overline{P}_a}\overline{\Gamma}_{\overline{P}_b}\Gamma^{C^c} \medskip\\
 \left[ -\frac{1}{2}\overline{P}_aC^a_{\; bc}\Lambda_{\overline{P}_b}C^c, i\Gamma^{\overline{P}_d}\Lambda_{\overline{P}_d}+i\Gamma^{C^d}\Lambda_{C^d}\right] &=& \frac{1}{2}C^a_{\; bc}\Gamma^{\overline{P}_a}\Lambda_{\overline{P}_b}C^c+\frac{1}{2}C^a_{\;bc}\overline{P}_a\Lambda_{\overline{P}_b}\Gamma^{C^c}
\end{eqnarray*}
\begin{eqnarray*}
\displaystyle \left[ \frac{1}{2}i\overline{\Gamma}_{C^a}C^a_{\; bc}C^b\Gamma^{C^c},i\Gamma^{C^d}\Lambda_{C^d}\right]&=&\frac{1}{2}C^a_{\; bc}C^b \Gamma^{C^c}\Lambda_{C^a}-\frac{1}{2}iC^a_{\;bc}\overline{\Gamma}_{C^a}\Gamma^{C^c}\Gamma^{C^b} \medskip \\
\left[-\frac{1}{2}i\overline{\Gamma}_{C^a}C^a_{\; bc}\Gamma^{C^b}C^c, i\Gamma^{C^d}\Lambda_{C^d}\right] &=& -\frac{1}{2}C^a_{\; bc}\Gamma^{C^b}C^c\Lambda_{C^a}+\frac{1}{2}iC^a_{\;bc}\overline{\Gamma}_{C^a}\Gamma^{C^b}\Gamma^{C^c} \medskip \\
\left[-\frac{1}{2}\Lambda_{C^a}C^a_{\;bc}C^bC^c,i\Gamma^{C^d}\Lambda_{C^d}\right]
&=& -\frac{1}{2}\Lambda_{C^a}C^a_{\;bc}\Gamma^{C^b}C^c+\frac{1}{2}\Lambda_{C^a}C^a_{\;bc}\Gamma^{C^c}C^b.
\end{eqnarray*}
It is easy to realize that the RHSs of each of the three blocks of commutators contain terms which are one the opposite of the other. This means that the contributions to the commutator  $[\widetilde{\Omega},Q]$ of each of the three blocks above are zero. This concludes our proof that $\left[ Q,\widetilde{\Omega} \right]=0$.

\end{document}